\documentclass[useAMS,usenatbib]{mn2e}

\usepackage{graphicx}
\usepackage{bm,epsfig} 
\usepackage{amssymb}
\usepackage{textcomp}

\begin{document}

\title[Gravitational-waves bursts from vortex avalanches]{Gravitational-wave bursts and stochastic background from superfluid vortex avalanches during pulsar glitches}
\author[L. Warszawski and A. Melatos]{L. Warszawski$^{1}$\thanks{lila@unimelb.edu.au (LW)} and A. Melatos$^{1}$\\
$^{1}$School of Physics, University of Melbourne, Parkville, VIC 3010, Australia\\}

\date{\today}
\maketitle

\begin{abstract}
The current-quadrupole gravitational-wave signal emitted during the 
spin-up phase of a pulsar glitch is calculated from first principles 
by modeling the vortex dynamics observed in recent Gross-Pitaevskii 
simulations of pinned, decelerating quantum condensates. Homogeneous and
inhomogeneous unpinning geometries, representing creep- and avalanche-like
glitches, provide lower and upper bounds on the gravitational wave signal strength respectively. The signal arising from homogeneous glitches is found to scale with the square root of glitch size, whereas the signal from inhomogeneous glitches scales proportional to glitch size. The signal is also computed
as a function of vortex travel distance and stellar angular velocity.
Convenient amplitude scalings are derived as functions of these parameters.
For the typical astrophysical situation, where the glitch duration (in units
of the spin period) is large compared to the vortex travel distance (in units
of the stellar radius), an individual glitch from an object $1\,\rm{kpc}$ from Earth
generates a wave strain of $10^{-24}[(\Delta\omega/\omega)/10^{-7}](\omega/10^2\,\rm{rad\,s}^{-1})^3(\Delta r/10^{-2}\,\rm{m})^{-1}$,
where $\Delta r$ is the average distance travelled by a vortex during a glitch, $\Delta\omega/\omega$ is the fractional glitch size, and $\omega$ is the pulsar angular velocity. The non-detection of a signal from the 2006 Vela glitch in data from the fifth science run conducted by the Laser Interferometer Gravitational-Wave Observatory implies that the glitch duration exceeds $\sim 10^{-4}\,\rm{ms}$. This represents the first observational lower bound on glitch duration to be obtained.
\end{abstract}
\begin{keywords}
dense matter  gravitational waves  stars: neutron  pulsars: general  stars: rotation
\end{keywords}
\date{\today}

\section{\label{sec:intro}Introduction}
Glitches are stochastic spin-up events that interrupt the smooth electromagnetic spin down of a pulsar.  They are commonly attributed to the simultaneous, transitory unpinning and outward motion of between $10^7$ and $10^{15}$ quantised superfluid vortices \citep{Anderson:1975p84,Anderson:1982p227,Alpar:1988p80,CHENG:1988p180,Link:1996p136,Warszawski:2008p4510,Melatos:2009p4511}, which pin metastably to nuclei in the crystalline inner crust \citep{Pines:1980p6820} except during a glitch.  The internal superfluid reorganises its velocity field nonaxisymmetrically during such an event, driving a time-varying current quadrupole moment, which emits gravitational radiation.   

Pulsar glitches are sources of both burst and continuous gravitational waves.  Previous theoretical and observational studies identified the post-glitch recovery phase, during which the viscous fluid component comes into corotation with the crust, as a continuous-wave source, with particular attention paid to how the shear viscosity, charged fluid fraction and mutual friction parameter of neutron stars may be extracted from the prospective signal \citep{Andersson:2001p1942,Peralta06b,vanEysden:2008p1745,Bennett:2010recovery}.  A new search algorithm, based on frequency-time maps of cross-power between two spatially separated terrestrial detectors, was proposed recently to identify `long transient' events lasting from seconds to weeks like post-glitch relaxation \citep{Thrane:2010}, along with a multi-detector Bayesian algorithm using $\mathcal{F}$-statistic amplitude priors \citep{Prix:2011}. 

The burst signal may be split into two categories.  The first category includes radiation from inertial $r$-mode oscillations \citep{Glampedakis:2009p1937} or quasi-radial acoustic and inertial modes involving one or two fluid components \citep{Sidery:2009p5840}.  The strongest signal is achieved when the velocities of the neutron superfluid and the charged fluid oscillate in anti-phase. For a detector with a spectral noise density of $10^{-24}\,\rm{Hz}^{-1/2}$ at $103\,\rm{Hz}$, the maximum signal-to-noise ratio for a glitch from a pulsar at the distance of Vela is 0.9 \citep{Andersson:2001p1942}.  The second category of signal, resulting from sudden superfluid deceleration when many vortices unpin simultaneously and move radially outwards, is discussed in this paper.  

Many other theories have been proposed for how this happens.  Most glitch mechanisms appeal to the storage of angular momentum in the superfluid interior, followed by its rapid release to the crust in discrete events via the sudden unpinnig of superfluid vortices \citep{Anderson:1975p84,Alpar:1984p6781,Haskell:2011}. Many ideas have been proposed for how this happens, the main point of distinction being the unpinning trigger.  \cite{Andersson:2002p1785} suggested that an instability between the viscous and inviscid fluid components, governed by the two fluids' relative velocity, can trigger a glitch.  \cite{Link:1996p136} invoked a sudden change in the coupling of the two fluid components, following the injection of heat into the interior, for example after a sudden rupture in the star's crust, known as a star quake.  Star quakes, in which the crust fails and abruptly lowers its moment of inertia, are also invoked as a stand-alone glitch theory for some objects \citep{Ruderman:1969p37,Middleditch:2006p360}, although star quakes on their own cannot account for the regular, large glitches observed in pulsars like Vela \citep{Crawford:2003}.

Gravitational waves result from non-axisymmetric rearrangements of mass and/or momentum field, so different glitch theories predict gravitational wave signals of different form and strength.  Therefore it is not possible to predict the form of the glitch gravitational wave signal independently of a glitch model, even though one can place limits on its maximum strength from energy arguments \citep{Andersson:2001p1942}.  In this paper, we consider specifically the self-organised, catastrophic unpinning model.  The model is inspired by the statistics of glitches [scale-free (power-law) sizes and independent (Poissonian) waiting times], which point to an underlying process involving the collective behaviour of many individual elements \citep{Melatos:2008p204}.  

A Bayesian search technique \citep{Clark:2007PRD} was recently applied to data collected during the fifth science run of the Laser Interferometer
Gravitational-wave Observatory (LIGO\footnote{http://www.ligo.org}) to look for gravitational radiation from quasi-normal mode oscillations excited by a glitch in the Vela pulsar.  No signal was detected.  The glitch, detected electromagnetically on 12 August, 2006, at the Hartebeesthoek Radio Observatory in South Africa, occurred during a period of five and a half hours during which the two LIGO detectors at Hanford returned high-quality, contiguous data.  Data from the third detector at Livinginston were not included in the search due to seismic noise.  The search returned an upper bound on the intrinsic strain amplitude of between $6.3\times 10^{-21}$ and $1.4\times 10^{-20}$, depending on the wave number of the mode \citep{Clark:2010}. \cite{Hayama:2008} employed a Monte-Carlo method to study the detection efficiency of possible gravitational waves triggered by the 2006 Vela pulsar glitch.  Glitches in Vela recur quasi-periodically, at intervals of approximately three years, so there is a case for altering the observation schedule to ensure that a world-wide detector network is operational for the next glitch [the most recent Vela glitch occurred in August, 2006 \cite{Flanagan:2006}].  

Radio telescope timing experiments on pulsar glitches reveal power-law-distributed sizes and exponential waiting times.  These statistics point to an underlying collective process that transitions between metastable states via vortex unpinning avalanches \citep{Melatos:2008p204}.  Quantum mechanical simulations confirm local and global knock-on mechanisms for triggering and sustaining unpinning avalanches \citep{Warszawski:2012PRB}.  In many models, the avalanches are spatially localised, so the glitch process is intrinsically nonaxisymmetric.    

In this paper, we calculate the gravitational wave signal from vortex avalanches during the spin-up phase of a pulsar glitch.   Section~{sec:GPE} summarises the first-principles evidence from quantum mechanical simulations, that a broadband gravitational wave signal arises from non-axisymmetric superfluid vortex rearrangement.  The simulations consist of numerical solutions of the non-linear Schr\"{o}dinger equation, known as the Gross-Pitaevskii equation, which describes the dynamics of a Bose-Einstein condensate.  Such simulations are limited to small systems containing $\sim10^2$ vortices.  In Sec.~\ref{sec:current} onwards we incorporate the generic behaviour observed in the Gross-Pitaevskii simulations into an approximate analytic theory of vortex motion, which is valid for realistically large numbers of vortices.  Sections~\ref{sec:current} and \ref{sec:discrete} evaluate the current quadrupole moment generated by spasmodic vortex motion in an idealised neutron star geometry.  In Sec.~\ref{sec:single}, we calculate the gravitational wave signal from the motion of a single vortex. The signal arising from a vortex avalanche is calculated in Sec.~\ref{sec:multi}, including consideration of the avalanche opening angle (Sec.~\ref{subsec:geom}), vortex speed (Sec.~\ref{subsec:motionGW}) and glitch size (\ref{subsec:omega}).  Criteria are given for detecting a gravitational wave burst from an individual glitch.  In Sec.~\ref{sec:ch6:conc}, we summarise our results and discuss critically the underlying assumptions of the calculation.  For completeness, we also explain in the appendix how the theoretical machinery in Sec.~\ref{sec:current}\,--\,\ref{sec:multi} can be applied to calculate rigorously the glitch contribution to the stochastic gravitational-wave background. The signal-to-noise ratio is small for current- and next-generation detectors, if only Galactic pulsars are considered, but it may be significant when extragalactic objects are included, an interesting avenue for future work.

\section{Motivation:  quantum mechanical simulations}\label{sec:GPE}
\begin{figure*}
\begin{center}
\includegraphics[scale=0.22,angle=90]{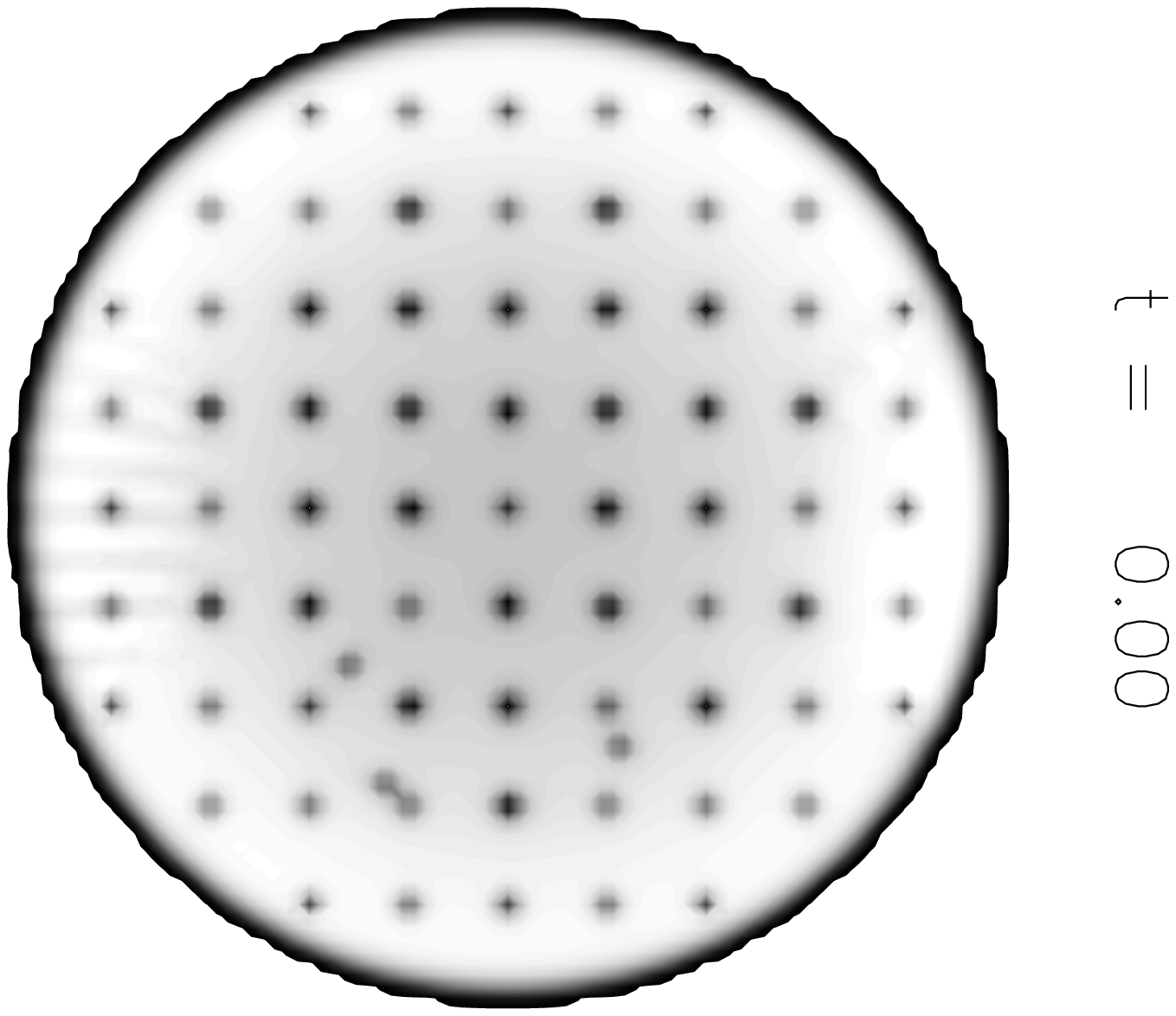}
\includegraphics[scale=0.22,angle=90]{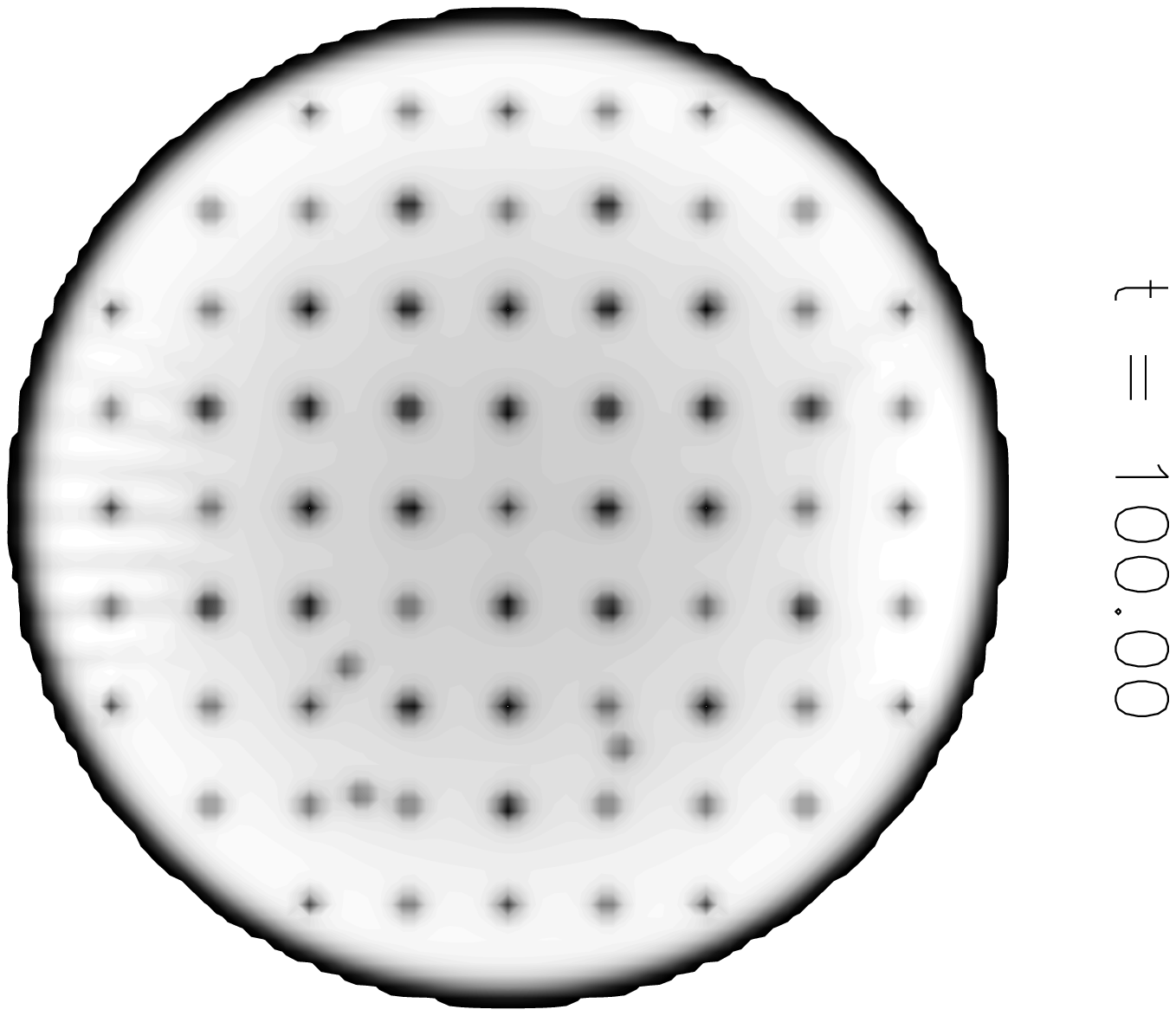}
\includegraphics[scale=0.22,angle=90]{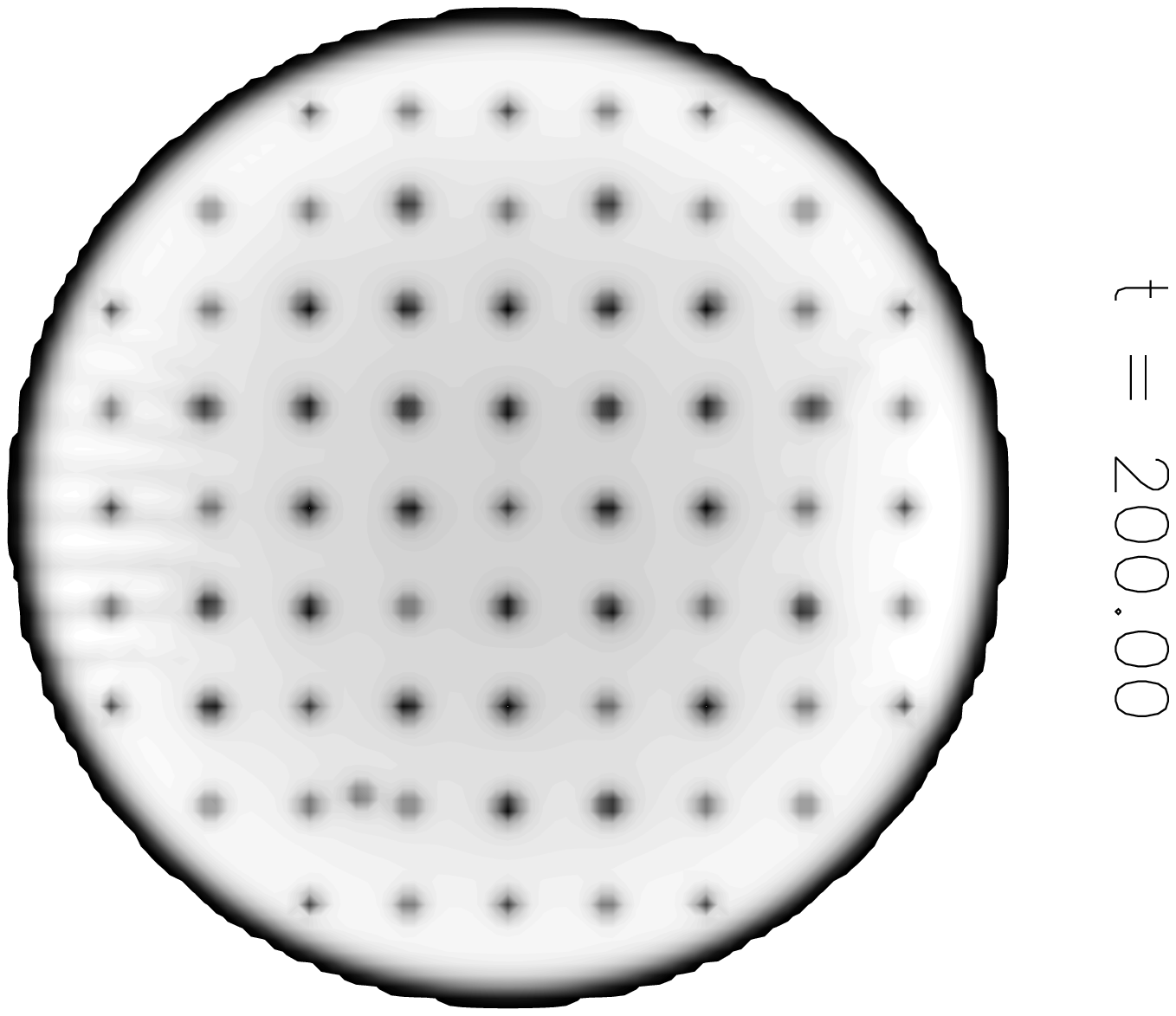}
\includegraphics[scale=0.22,angle=90]{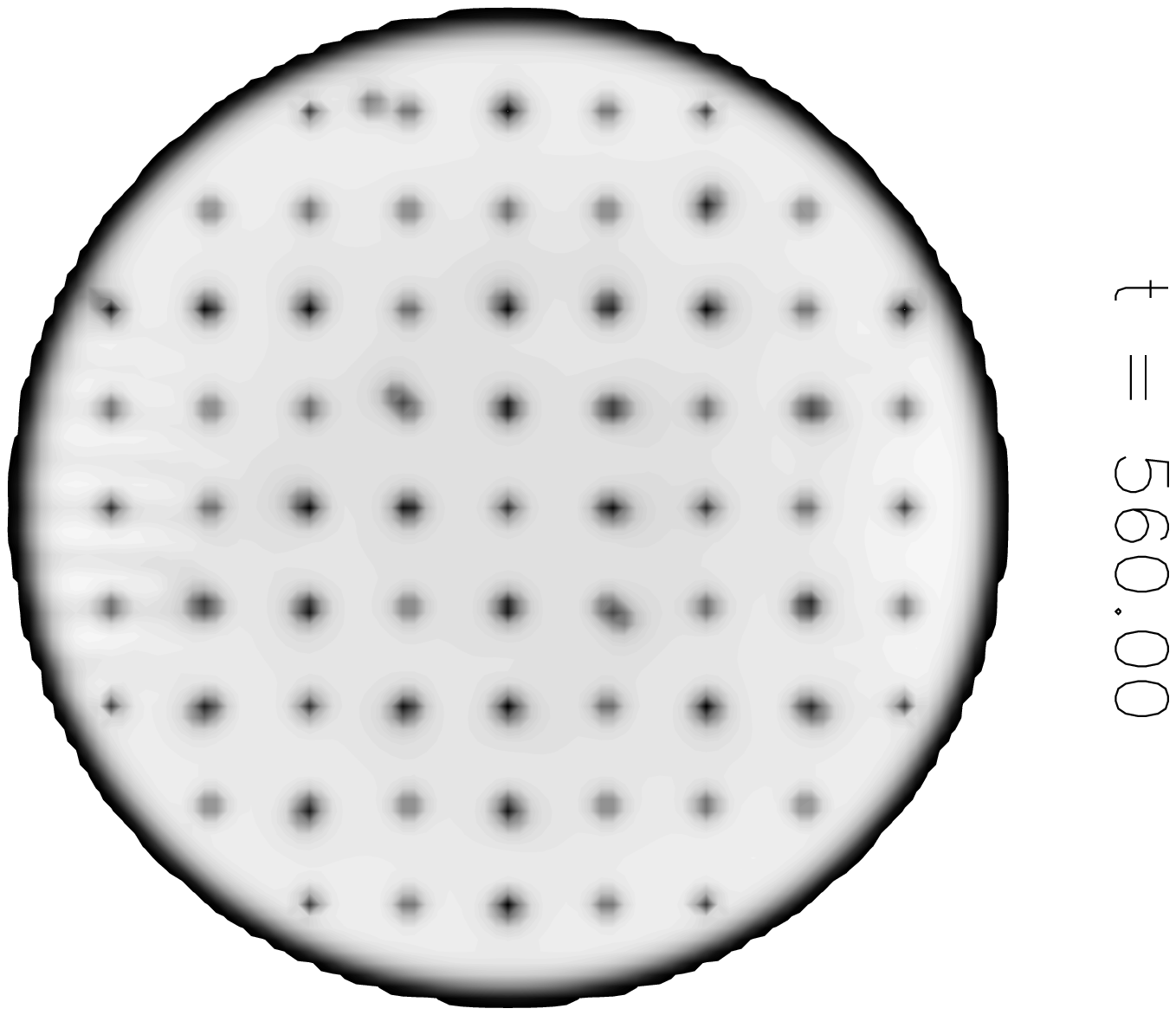}
\includegraphics[scale=0.22,angle=90]{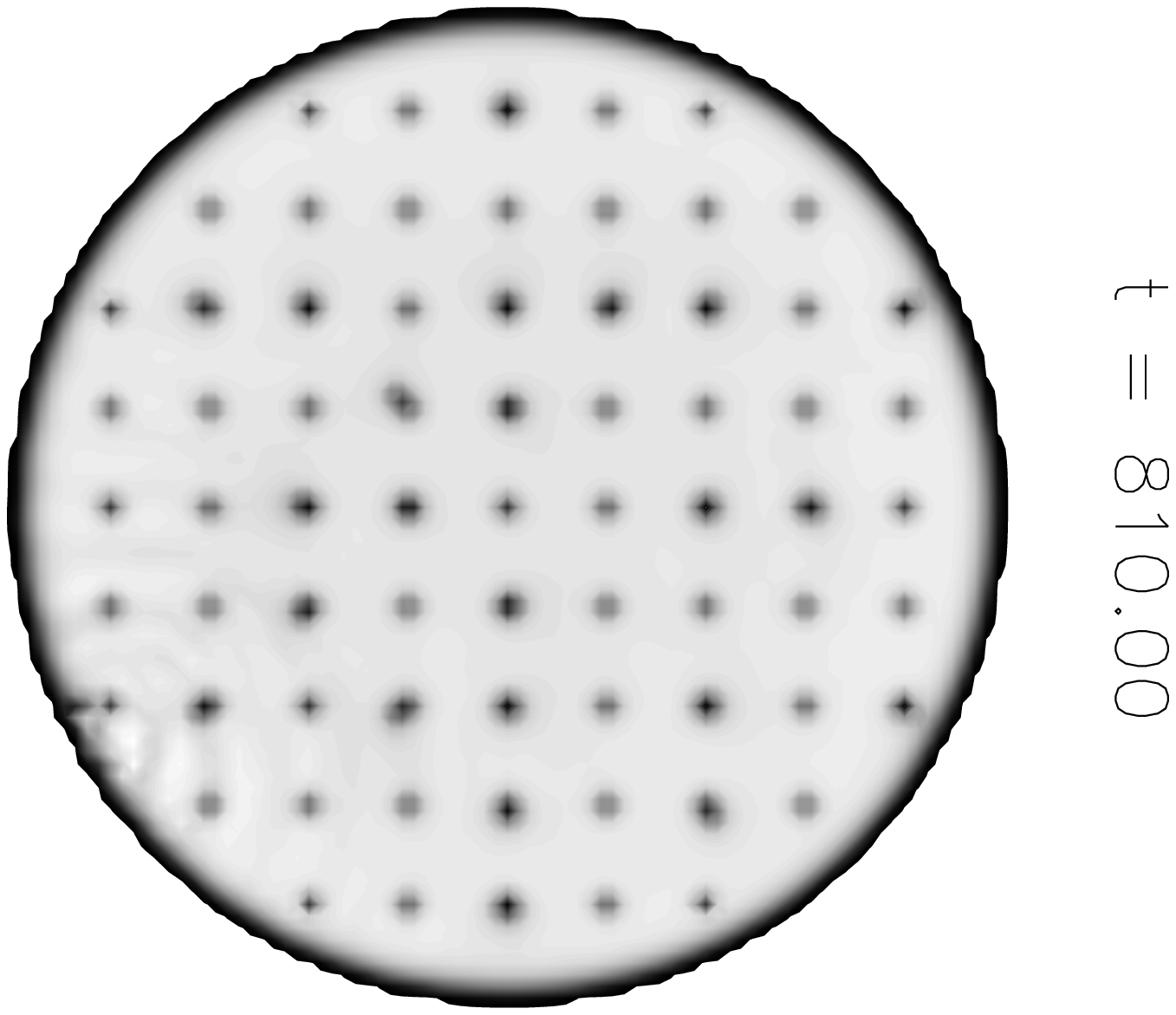}
\includegraphics[scale=0.55,angle=90]{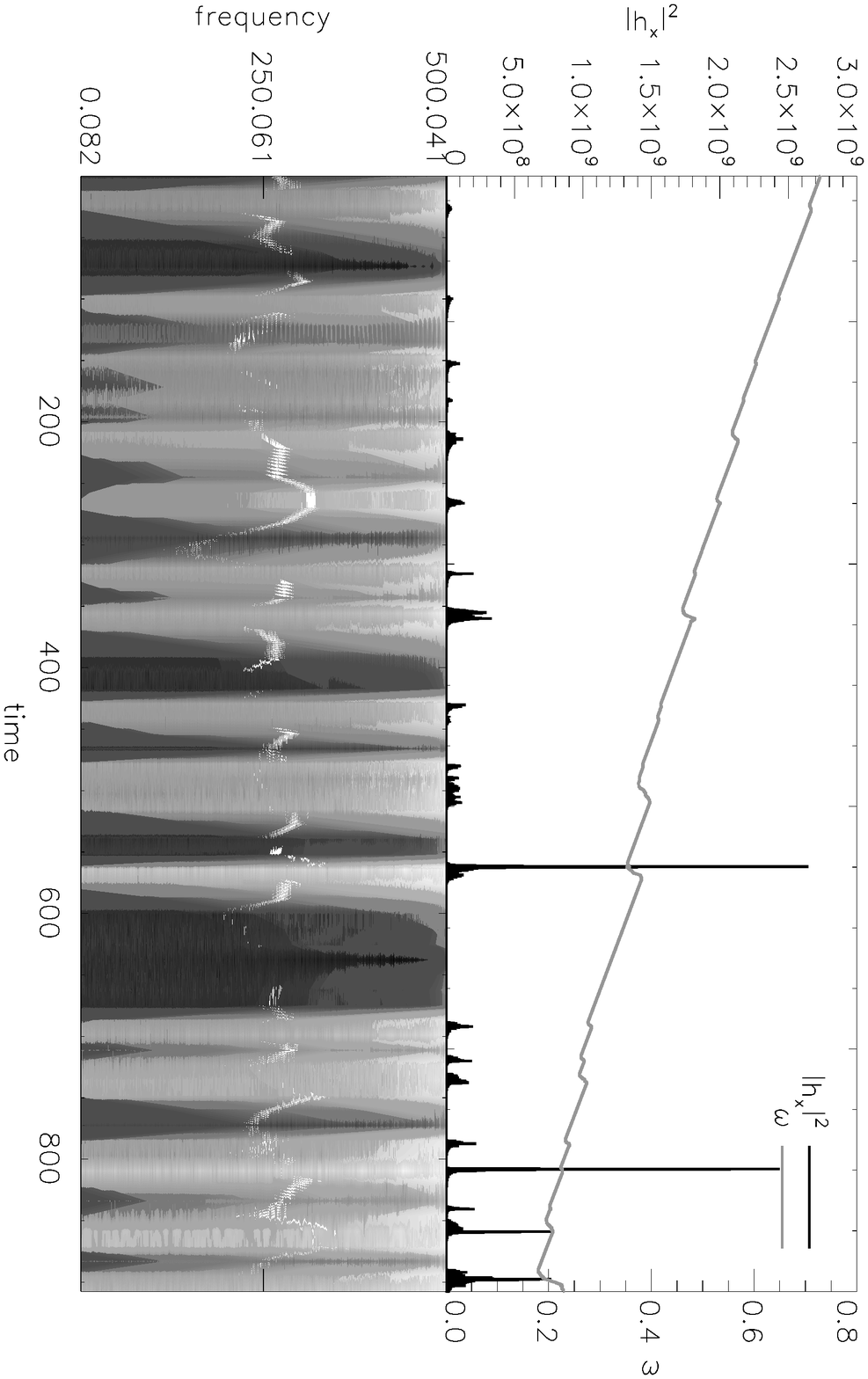}
\end{center}
\caption{\emph{Top}:  Greyscale snapshots of superfluid density $|\psi|^2$ (where $\psi$ is the two-dimensional complex order parameter) at $t=0$, 100, 200, 560 and 810 (in arbitrary units) from \emph{left} to \emph{right} respectively.  The \emph{darker dots}, and the \emph{lighter dots} that are not part of the rectangular array, are vortices; \emph{lighter dots} in the array are unoccupied pinning sites.  \emph{Middle}:  Squared modulus of the gravitational wave strain in the cross polarisation $|h_{\times}|^2$ as a function of time $t$ (\emph{black} curve plotted against \emph{left} vertical axis) and angular velocity $\omega(t)$ of the crust as a function of time (\emph{grey} curve plotted against \emph{right} vertical axis) for the simulations described in Sec.~\ref{sec:GPE} \citep[further detail is given in ][]{Warszawski:2010pulsar}.  \emph{Bottom}:  Frequency-time spectrogram of $|h_{\times}|^2$ calculated in time windows of width $12.3$.  Simulation parameters:  $N_{\rm{c}}/I_{\rm{c}}=10^{-3.0}$, $V_0=16.6$, $\eta=1$, $\Delta V_i/V_0=0.0$, $R=12.5$, $\Delta x=0.15$, $\Delta t=0.0025$, $\omega_0 = 0.8$ \citep[definitions of these quantities appear in ][]{Warszawski:2010pulsar}.}
\label{fig:ch6:GPE_uni}
\end{figure*}

\begin{figure*}
\begin{center}
\includegraphics[scale=0.45,angle=90]{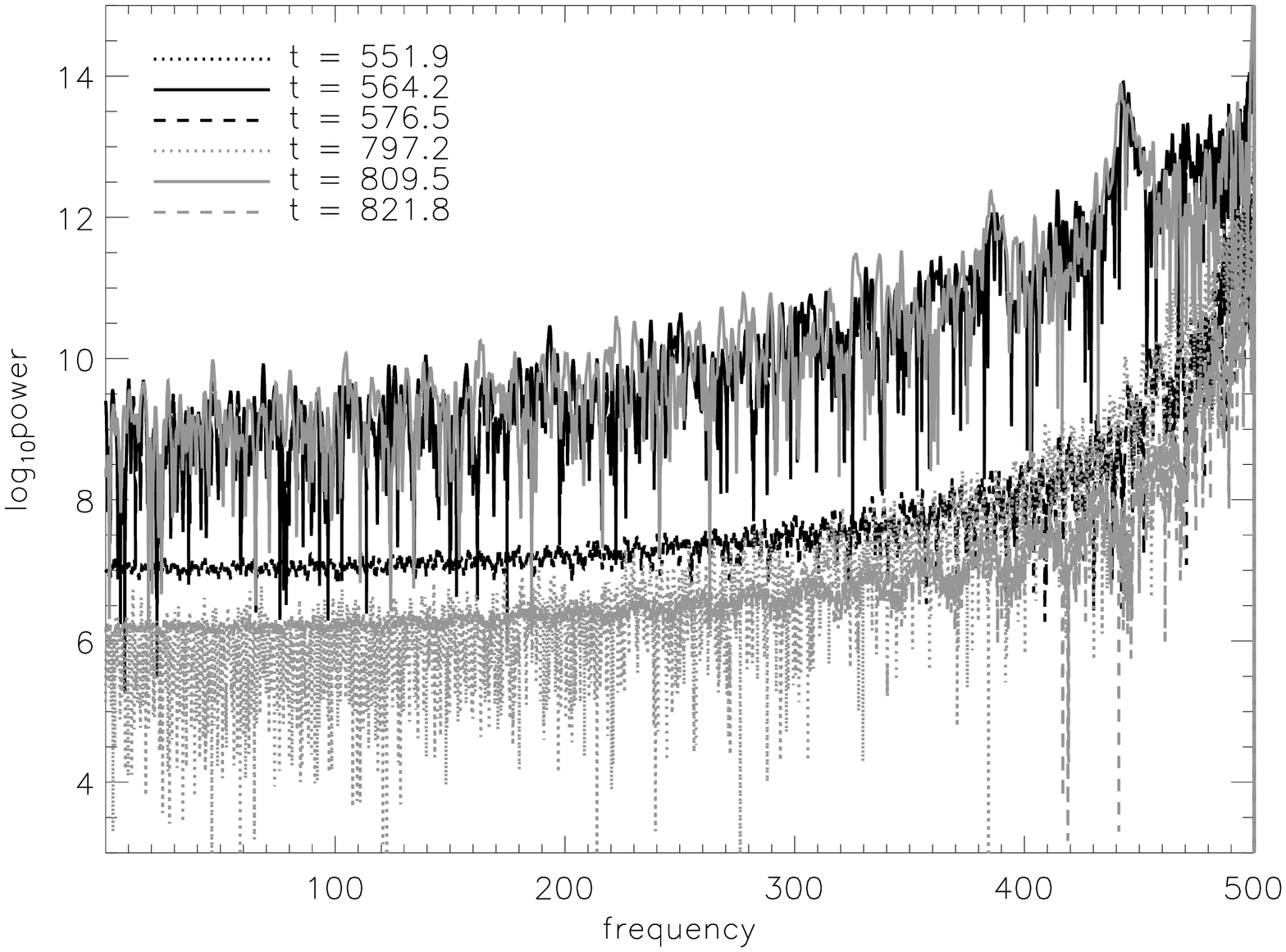}
\end{center}
\caption{Constant-$t$ cross-sections of the spectrogram of $|h_{\times}|^2$ graphed in Fig.~\ref{fig:ch6:GPE_uni}, at $t=551.9$, 564.2, 576.5, 797.2, 809.5 and 821.8 (\emph{black dotted}, \emph{black solid}, \emph{black dashed}, \emph{grey dotted}, \emph{grey solid} and \emph{grey dashed} curves respectively), showing spectral power as a function of frequency on log-linear axes (arbitrary units). }
\label{fig:ch6:GPE_uni_spectra}
\end{figure*}
Let us begin by confirming that, at the microscopic level, a quantum condensate (such as the neutron superfluid in the pulsar interior) in a decelerating container undergoes spasmodic vortex reorganisation, which emits a bursty broadband gravitational wave signal.  Such systems are traditionally studied by solving the Gross-Pitaevskii equation (GPE), a non-linear Schr\"{o}dinger equation of the form
\begin{equation}
i\hbar\frac{\partial \psi}{\partial t} = -\frac{\hbar^2}{2m}\nabla^2\psi +V\psi+ |\psi|^2\psi\,,
\end{equation}
where $\psi$ is the two-dimensional order parameter of a zero-temperature Bose-Einstein condensate, from which the condensate density, $|\psi|^2$, is derived, and $|\psi|^2\psi$ is the quantum pressure term which governs the structure of vortex cores.  $V$ is the potential, which comprises the container that confines the condensate, as well as a grid of spikes, which act as pinning sites for vortices, since the overlap of the empty vortex core with the region emptied of condensate by the high potential results in a low-energy configuration \citep{Avogadro:2008p29}. $V$ also contains a term proportional to the chemical potential. Initially, when the container is accelerated, vortices form and pin to the pinning sites.  The initial pinning positions depend on the number of vortices and the number and strength of pinning sites \citep{Sato:2007p8103,Mink:2008p1580,Goldbaum:2009p8052,Warszawski:2010pulsar} and exhibit hysteresis \citep{Jackson:2006p3196}.   

Previous numerical studies of a vortex lattice, pinned to a decelerating container reveal that vortices hop spasmodically between pinning sites as they move towards the container edge \citep{Warszawski:2010pulsar}.  Computational expense limits the system size to tens of vortices, so the results do not translate directly to a realistic pulsar.  Instead, we use the qualitative behaviour observed as a basis for analytic extrapolation in later sections, when calculating the gravitational radiation from real glitches.


Our simulations solve the Gross-Pitaevskii equation (GPE) for a zero-temperature condensate in a rotating container. For simplicity, we model an infinite vertical column of superfluid (strictly, its two-dimensional cross-section) in the presence of a rectangular grid of pinning sites.  A summary of the technical details is given in \cite{Warszawski:2012PRB}.  The initial state of the system, depicted in the \emph{top left} image of Fig.~\ref{fig:ch6:GPE_uni} (a greyscale snapshot of superfluid density), comprises 12 pinned vortices (\emph{darker} dots) and 38 unoccupied pinning sites (\emph{lighter dots}).   An external spin-down torque is then applied to the container, imitating the electromagnetic spin-down torque acting on a pulsar.  Pinning prevents vortices from moving radially outwards, fostering an angular velocity mismatch between the superfluid and container.  In response, vortices hop between pinning sites in spasmodic bursts, releasing discrete parcels of angular momentum to the container.  The images in the \emph{top} panel of Fig.~\ref{fig:ch6:GPE_uni} depict the superfluid density at $t=0$, 100, 200, 560 and 810 (the external torque is turned on at $t=0$; time is quoted in arbitrary simulation units).  We can see that vortices move from their initial positions.  For example, the interstitial vortex in the \emph{bottom right} quadrant at $t=0$ annihilates against the container wall by $t=560$.

From movies of the time evolution of the superfluid density, we see that the typical vortex trajectory is not simple.  As the vortex prepares to unpin, it migrates slowly to the outer edge of its pinning site.  It then accelerates off the pinning site, before spiralling into the new pinning site if and when it repins.  In later sections, we approximate the radial speed as a parabolic function of time to facilitate analytic calculations.  The motion of individual vortices is described in detail in \cite{Warszawski:2012PRB}.

Each time a vortex moves, the superfluid velocity field adjusts in response.  Each vortex generates a solenoidal velocity field, whose magnitude is inversely proportional to distance from the vortex core.  The total velocity, $\mathbf{v}_{\rm{s}}$, at any point is the vector sum of the contributions from each vortex.  Time-varying nonaxisymmetries in $\mathbf{v}_{\rm{s}}$ generate a time-varying current quadrupole moment, resulting in gravitational waves \citep{Thorne:1980,Melatos:2010turb}.

We graph the amplitude squared of the wave strain in the cross polarisation, $|h_{\times}|^2$ (in arbitrary units), whose mathematical definition is formalised in Sec.~\ref{sec:current}, as a \emph{black} curve in the \emph{middle} panel of Fig.~\ref{fig:ch6:GPE_uni}.  The \emph{grey} curve in the same plot tracks the angular velocity of the container, $\omega$.  Note that all calculations are performed in the reference frame that co-rotates with the container and pinning grid.  Episodes of non-zero $|h_{\times}|^2$ are clearly accompanied by crustal spin-up events, \emph{i.e}. glitches, when $\omega$ increases abruptly. Notably, the maximum $|h_{\times}|^2$ during a glitch is not a monotonic function of glitch size (measured as $\Delta\omega/\omega$, the fractional, impulsive change in $\omega$ during a glitch).  This is an important reminder that the superfluid non-axisymmetry and hence the gravitational wave strain depend on both the number of vortices that move and the degree of asymmetry in their positions within the star.  We discuss this further in the context of real pulsar glitches in Sec.~\ref{sec:multi}.  

The \emph{bottom} panel of Fig.~\ref{fig:ch6:GPE_uni} is a spectrogram of $|h_{\times}|^2$, calculated by finding the power spectrum of $|h_{\times}|^2$ in time windows of width $\Delta t=12.3$.  Since $|h_{\times}|^2$ is calculated in the co-rotating frame, a peak at $\omega$ is absent from the homodyned power spectrum. Once again, bright features in the spectrogram (e.g. at $t\approx 560$) coincide with rotational glitches.  The peak signal associated with each glitch occurs at the minimum preceding the glitch-induced spin up, for example, where the spikes in $|h_{\times}|^2$ intersect the $\omega(t)$ curve in the \emph{middle} panel.  In Fig.~\ref{fig:ch6:GPE_uni_spectra}, we graph instantaneous power spectra from instants before, during, and after the glitches at $t\approx 564$ and $t\approx 809$ (\emph{black} and \emph{grey} curves respectively).  All six curves exhibit a broadband burst signal, which increases with frequency, while the overall amplitude scales with $|h_{\times}|^2$.  The two \emph{solid} curves exhibit peaks at frequency values $390$ and $440$, which are absent from the pre-glitch (\emph{dotted}) and post-glitch (\emph{dashed}) curves, suggesting that the peaks are associated with glitch-induced radiation.

The results presented in this section confirm that the motion of vortices through a grid of pinning sites results in a bursty gravitational wave signal.  However, we emphasise that a systematic study remains to be done of how the system size and range of pinning site strengths, amongst many other properties, change the gravitational wave signal.  Recent results show that once the ratio of pinning sites to vortices exceeds unity, the statistics of spin-up events do not change \citep{Warszawski:2010pulsar} .  However, when vortices outnumber pinning sites, pinning no longer dominates the spin-down dynamics of the condensate, resulting in fewer and smaller glitches.   An important unanswered physics question is, does a system containing $\sim 10^{19}$ vortices exhibit unpinning avalanches like systems with $\sim 10^2$ vortices?  Semi-analytic studies \citep{Melatos:2009p4511} suggest that avalanche dynamics are still possible in large systems, especially when unpinned vortices can trigger further unpinnings in a domino-like effect \citep{Warszawski:2012PRB}.

\section{Current quadrupole moment}\label{sec:current}
Drawing inspiration from the vortex dynamics observed in recent studies \citep{Warszawski:2012PRB} and the burst gravitational wave signal computed in Fig.~\ref{fig:ch6:GPE_uni}, we now develop an analytic formalism that facilitates calculation of a gravitational wave signal from glitches involving realistic numbers ($\gtrsim 10^7$) of vortices.  We follow the methodology introduced by \cite{Thorne:1980} and employed recently in studies of f-mode and turbulence-driven emission \citep{Wasserman:2008p70,Melatos:2010turb,Bennett:2010recovery,Sidery:2009p5840}.  

The far-field metric perturbation (wave strain) generated by a superposition of current multipole moments $S^{lm}$ can be written as 
\begin{equation}
\label{eq:h}
h^{TT}_{jk} = \frac{G}{Dc^5}\sum_{l=2}^{\infty}\sum_{m=-l}^{l}T^{B2,lm}_{jk}\frac{\partial^lS^{lm}}{\partial t^l}~,
\end{equation}
in the transverse, traceless gauge, where $t$ is the retarded time, $D$ is the distance from source to observer, and $T^{B2,lm}_{jk}$ is the beam pattern, which is a function of the observer's orientation relative to the source. The physics of the source is housed in $S^{lm}$.  In general, $h^{TT}_{jk}$ also includes contributions from mass multipoles $I^{lm}$, which may dominate, but they are neglected in this paper, where we model the matter distribution inside a pulsar as incompressible and axisymmetric.

The ($l$,$m$)-th multipole moment, $S^{lm}(t)$ (units: $\rm{kg~m}\rm{s}^{-1}$), is given by \citep{Thorne:1980,Melatos:2010turb}
\begin{equation}
\label{eq:slm}
S^{lm}=c_{l}\int d^3 x Y^{*}_{lm}r^l \mathbf{x}\cdot \nabla\times \left(\rho\mathbf{v}_{\rm{s}}\right)~,
\end{equation}
where the integral is taken over the entire volume of the source, $\mathbf{v}_{\rm{s}}$ and $\rho$ are the fluid velocity and density respectively, $Y^{*}_{lm}$ is the complex conjugate of the scalar spherical harmonic 
\begin{equation}
 Y_{lm}=\sqrt{\frac{(2l+1)(l-m)!}{4\pi(l+m)!}}e^{im\phi}P^{m}_{l}(\cos\theta )~,
\end{equation}
$P^{m}_{l}(\cos\theta )$ is the associated Legendre function, and one has
\begin{equation}
 c_l = -\frac{32\pi}{\left( 2l+1\right)!!}\sqrt{\frac{l+2}{2l\left(l-1\right)\left(l+1\right)}}~.
\end{equation}
In our application, the leading term in the multipole expansion is $l=2$, for which only the $m=\pm 1$ terms are nonzero [$P^1_2(x)=-3x(1-x^2)^{1/2}$].  We make the simplifying assumption that the flow is purely azimuthal\footnote{\cite{Peralta06a} showed that this is inaccurate at high Reynolds numbers, when meridional circulation and Kolmogorov-like turbulence set in \citep{Peralta:2009,Melatos:2010turb}.}, such that the mass current term $\mathbf{x}\cdot\nabla\times (\rho\mathbf{v}_{\rm{s}})$ in Eq.~(\ref{eq:slm}) reduces to $z\left[\mathbf{\nabla}\times (\rho\mathbf{v}_{\rm{s}})\right]_{z}$.  It is convenient to switch to cylindrical coordinates ($R^{\prime},\phi^{\prime},z^{\prime}$), such that, for a sphere of radius $R_{\rm{s}}$, we have
\begin{eqnarray}
\label{eq:slm_sph}
 S^{21}&=&\sqrt{\frac{128\pi}{45}}\int_0^{2\pi} d\phi^{\prime}e^{i\phi^{\prime}}\nonumber\\
 & &\times \int_0^{R_{\rm{s}}} dR^{\prime} R^{\prime}\left(R_{\rm{s}}^2-R^{\prime 2}\right)^{3/2}\left[\nabla\times\rho\mathbf{v}_{\rm{s}}(R^{\prime})\right]_z ~.
\end{eqnarray}  
In Sec.~\ref{sec:discrete} onwards, vortices are treated as discrete carriers of quantised circulation.  It is then straight-forward to evaluate $\rho\mathbf{v}_{\rm{s}}$ and hence Eq.~(\ref{eq:slm_sph}) as a sum over vortices.

Denoting the inclination angle between the rotation axis of the star and the observer's line of sight by $\iota$, and the azimuth of the line of sight by $\zeta$ (relative to some reference plane), the polarised components of the wave strain are 
\begin{eqnarray}
h_+&=& -\frac{G}{2 c^5 D \left(\cos^2\iota + \cos^2\zeta \sin^2\iota\right)^2} 
 \sqrt{\frac{5}{2 \pi}} \nonumber\\& &\times \left[(\cos^2\iota \sin^2\iota \sin^2\iota) + 
    \sin^4\iota (\cos^4\zeta + 
       \cos^2\zeta \sin^2\zeta \right.\nonumber\\
& &~~~~\left.+ \cos^2\zeta \cot^2\iota) \cos^4\iota (1 + \cos^2\zeta \tan^2\iota)\right]\nonumber\\
& &\times \mathcal{R}e\left\lbrace\frac{\partial^2 S^{21}}{\partial t^2} \sin\iota \left[i e^{i \zeta} \cos^2\zeta \sin^2\iota+ie^{-i\zeta}\cos^2\iota \right]\right\rbrace \\
h_{\times}&=& -\frac{G\sqrt{\cos^2\iota + \cos^2\zeta\sin^2\iota}}{16 c^5 D \left(\cos^2\iota + \cos^2\zeta \sin^2\iota\right)^2} 
 \sqrt{\frac{5}{2 \pi}} \nonumber\\
& &\times \sec\iota\left[\cos^2\iota\sin^2\iota\sin^2\zeta+\sin^4\iota\cos^2\zeta\left(1+\cot^2\iota\right)\right.\nonumber\\
& &~~~~\left.+\cos^4\iota\left(1+\cos^2\zeta\tan^2\iota\right)\right]^{1/2}\nonumber\\
& &\times \mathcal{R}e\left\lbrace\frac{\partial^2 S^{21}}{\partial t^2} e^{i \zeta}\sin\iota  \left(5+3\cos 2\iota-2e^{4i\zeta}\sin^2\iota\right)\right\rbrace\,.
 \label{eq:hpol}
\end{eqnarray}
The angles $\iota$ and $\zeta$ enter through $T^{B2}_{jk}$.  For definiteness, we choose $\iota = \pi /3$ and $\zeta = 0$ in most of the figures and applications to follow.  In Sec.~\ref{sec:discrete}--\ref{sec:multi}, wave-strain estimates are presented mostly for $h_{\times}$, as one has $h_{+}\sim h_{\times}$, except for certain special (and unlikely) observer orientations.  In App.~\ref{sec:stoch}, when calculating the stochastic background, we average over $\iota\in[0,\pi]$ and set $\zeta=0$ without loss of generality.  

\section{Discrete vortex motion}\label{sec:discrete}

\begin{figure*}
\begin{center}
\includegraphics[scale=0.4]{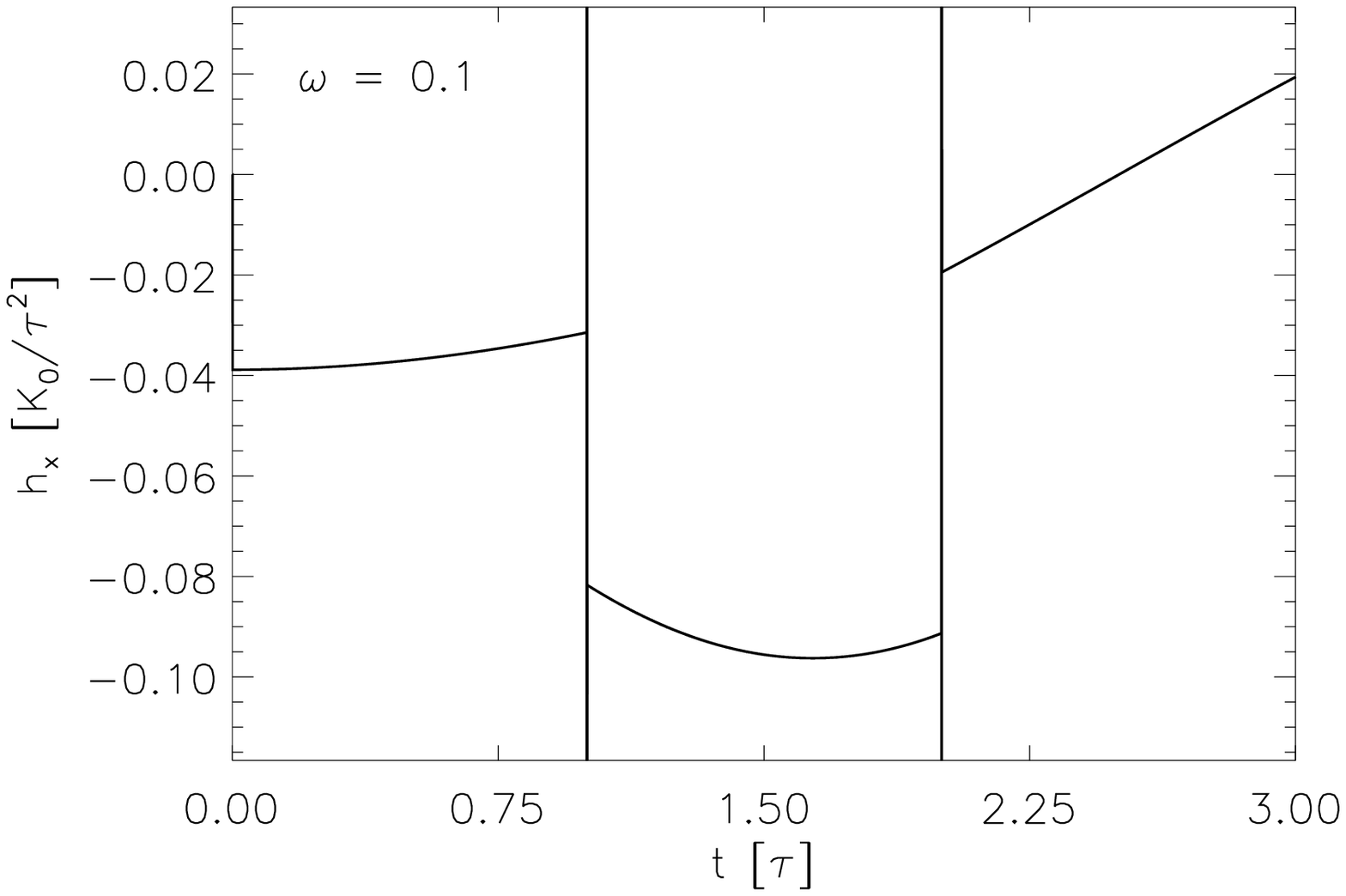}
\includegraphics[scale=0.4]{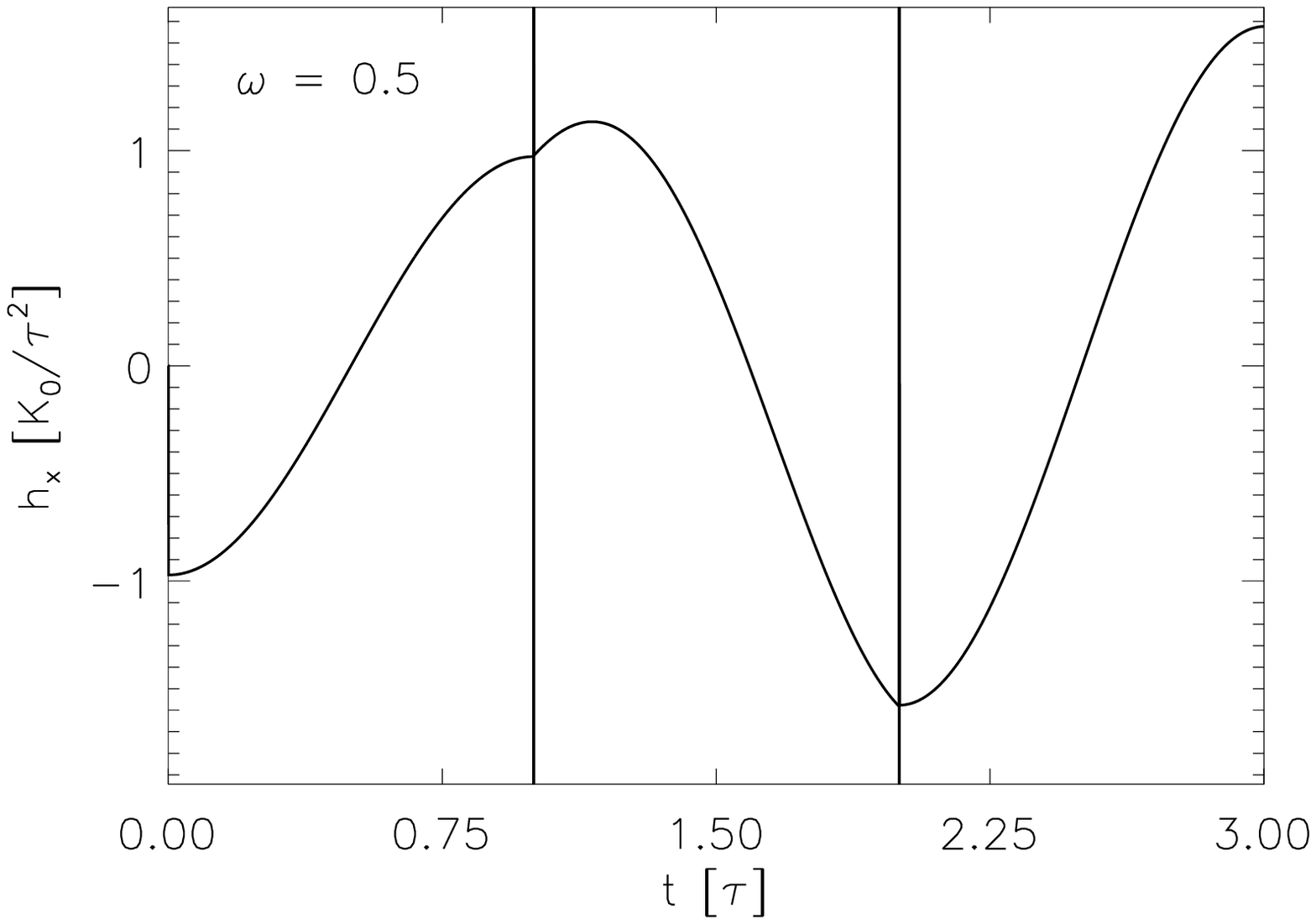}
\includegraphics[scale=0.4]{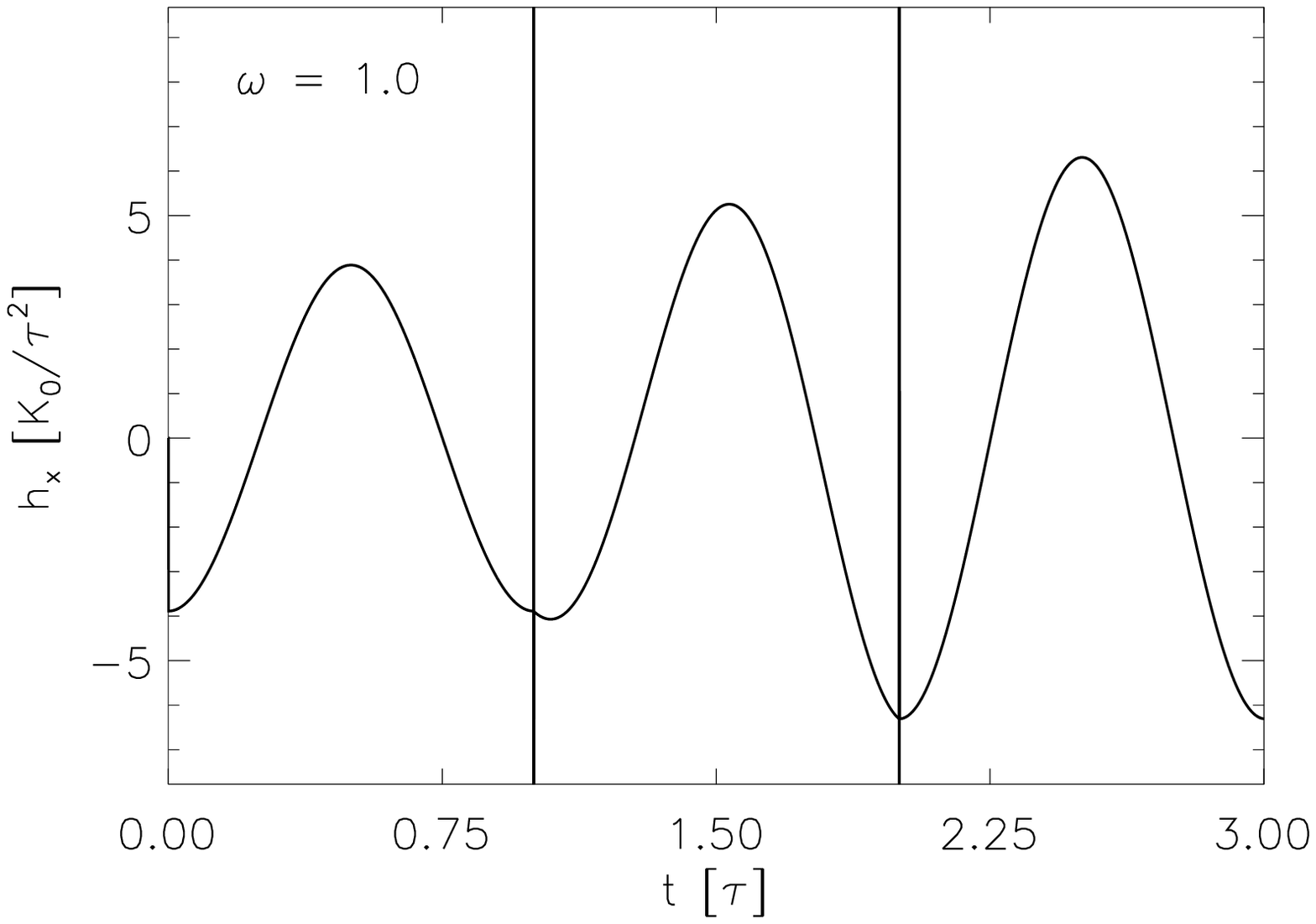}
\includegraphics[scale=0.4]{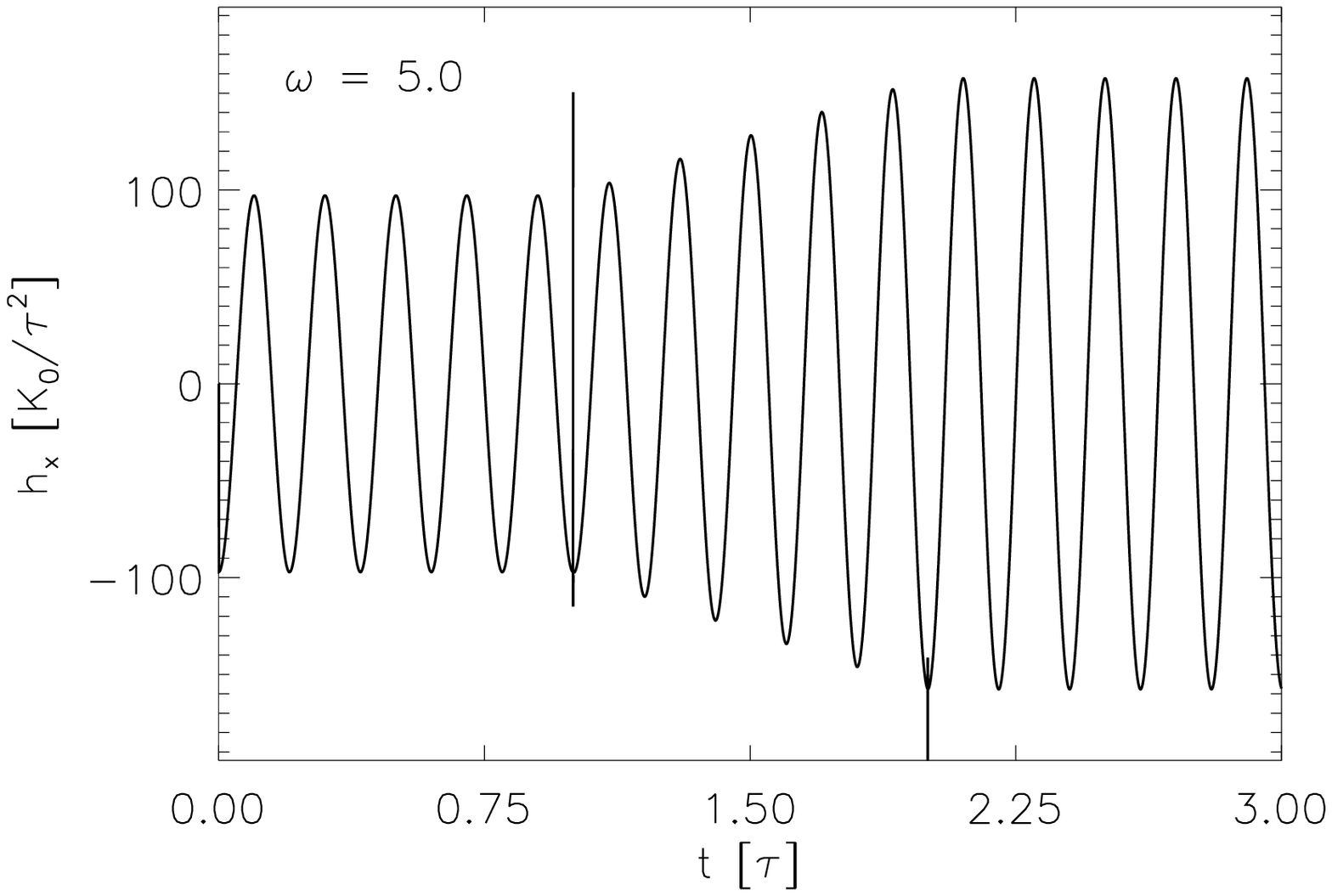}
\end{center}
\caption{Gravitational wave strain as a function of time in the cross polarisation $h_{\times}(t)$ in units of $K_0/\tau^2$ for a single vortex with initial position $\tilde{R}_0=0.1$, which moves a distance $\Delta\tilde{r}=0.2$, for angular velocity $\tilde{\omega}/(2\pi)=0.1$, 0.5, 1.0 and 5.0 (\emph{top left}, \emph{top right}, \emph{bottom left} and \emph{bottom right} panel respectively).  The vortex unpins at $\tilde{t}=1$ and repins at $\tilde{t}=2$ [bracketed by the \emph{vertical} spikes in $h_{\times}(t)$].}
\label{fig:ch6:CA_single_om}
\end{figure*}

\begin{figure*}
\begin{center}
\includegraphics[scale=0.4]{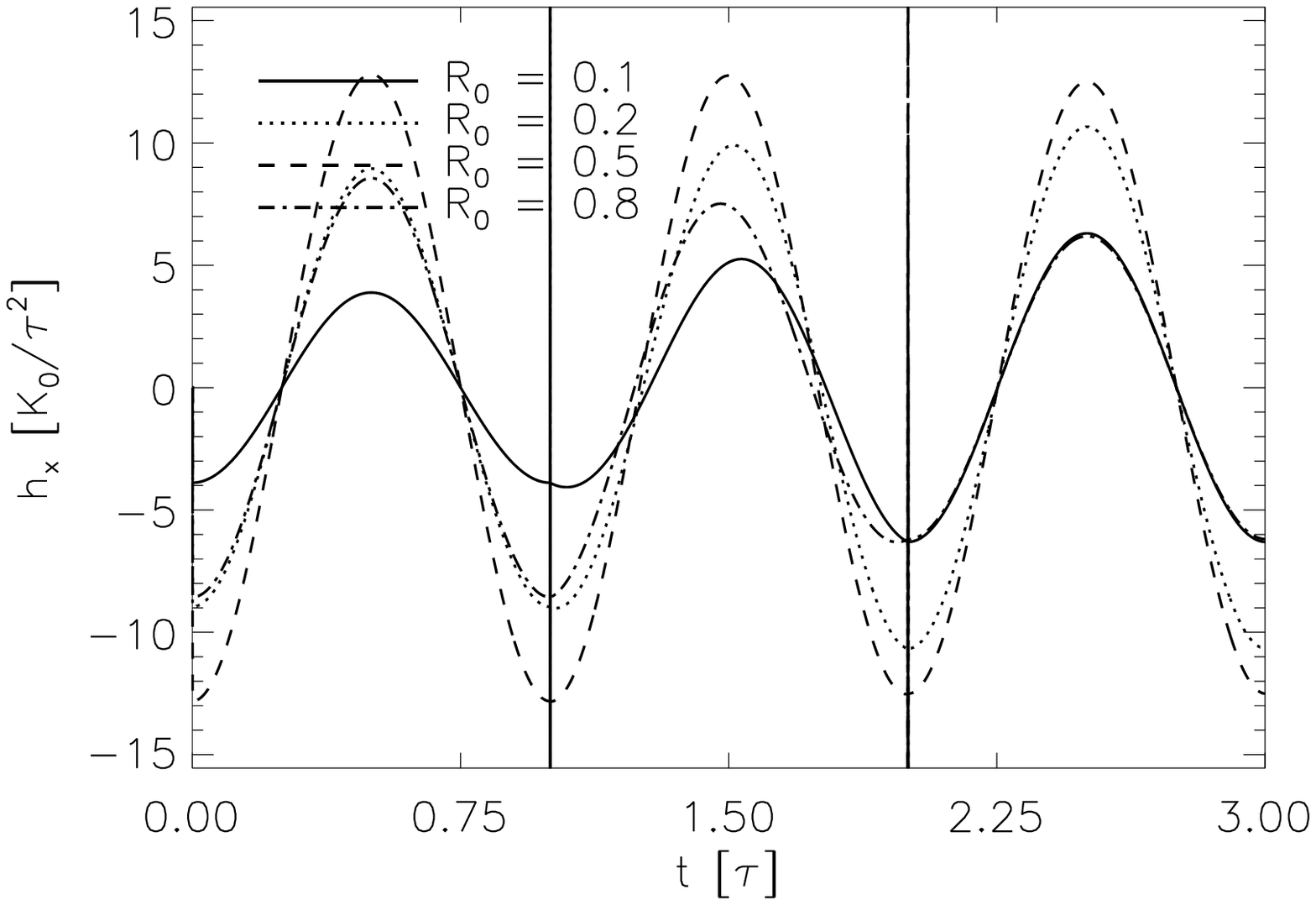}
\includegraphics[scale=0.4]{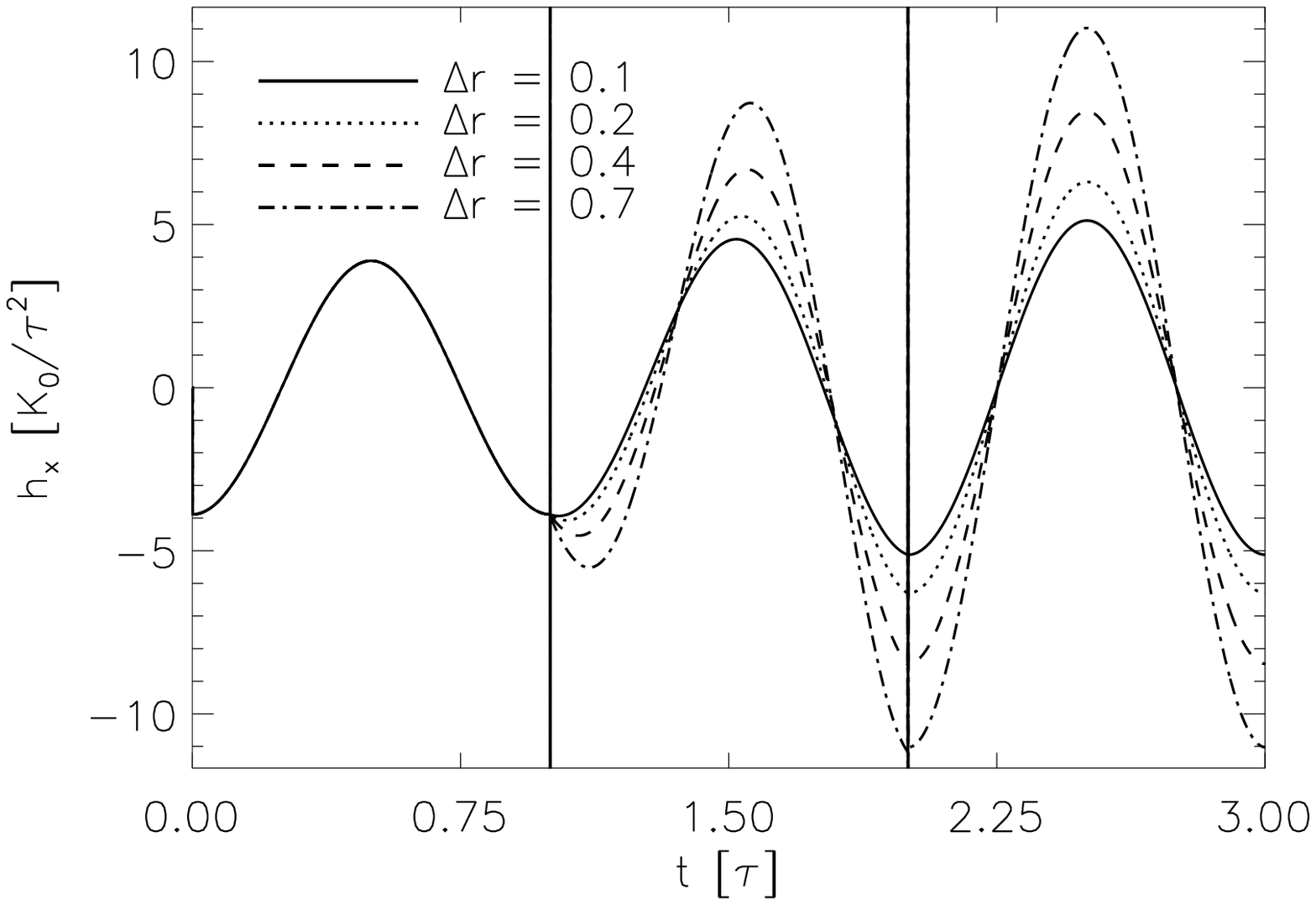}
\includegraphics[scale=0.4]{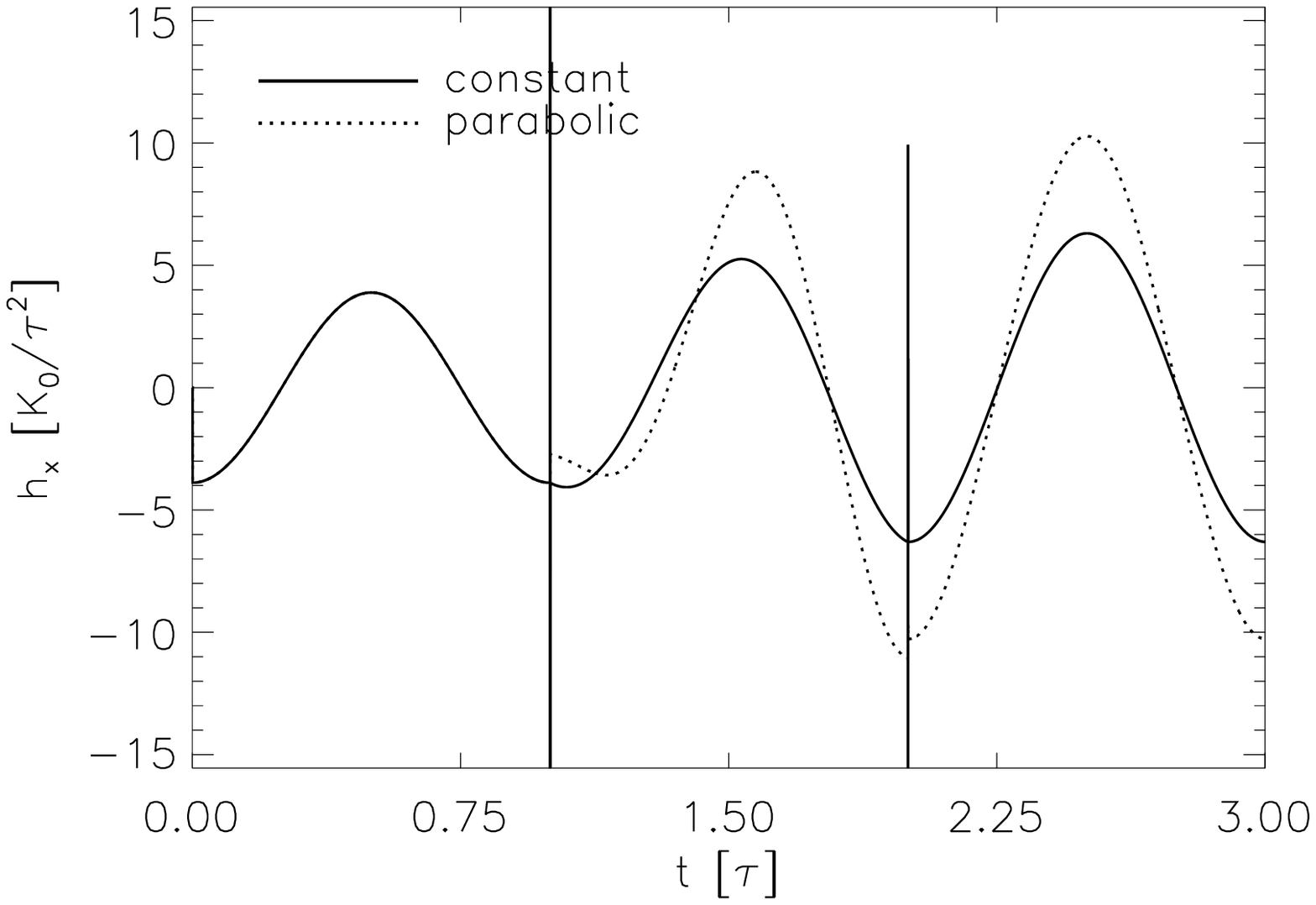}
\end{center}
\caption{Gravitational wave strain in the cross polarisation as a function of time $h_{\times}(t)$ in units of $K_0/\tau^2$ for a single vortex with $\tilde{\omega}/(2\pi)=1.0$ as a function of model parameters.  \emph{Top left}: Initial radius $\tilde{R}_0=0.1$, 0.25, 0.5 and 0.75, with $\Delta\tilde{r}=0.2$.  \emph{Top right}:  Distance moved $\Delta\tilde{r}=0.1$, 0.2, 0.4 and 0.7, with $\tilde{R}_0=0.1$. \emph{Bottom}:  Constant (\emph{solid} curve) and parabolic (\emph{dotted} curve) speed profile.  All panels have $\tilde{t}_{\rm{g}}=1$.}
\label{fig:ch6:CA_single_dr}
\end{figure*}

A glitch occurs when many vortices simultaneously unpin and move outward in response to a trigger.  In this paper, we remain agnostic about the nature of the trigger, a notorious unsolved problem \citep{Anderson:1975p84,CHENG:1988p180,JahanMiri:2005p8678,Melatos:2009p4511,Warszawski:2012PRB}.

Consider $\Delta N_{\rm{v}}$ vortices that unpin from positions originally arranged in an evenly-spaced array [known as an `Abrikosov lattice', which has hexagonal symmetry \citep{Abrikosov:1957}].  A superfluid is irrotational; it only sustains vorticity at vortex singularities where one has 
\begin{equation}
\label{eq:sing}
\mathbf{\nabla}\times\mathbf{v}=\kappa\delta^{\left(2\right)} \left[\mathbf{x}_{\perp}-\mathbf{x}_{\perp\rm{v}}(t)\right]\hat{\mathbf{z}}~,
\end{equation}
for a vortex whose axis runs parallel to the rotation axis and intersects the equatorial plane of the star at $\mathbf{x}_{\perp\rm{v}}(t)$.  Here, $\kappa=h/m=10^{-7}~\rm{m}^2\rm{s}^{-1}$ is the quantum of circulation, where $m$ is twice the neutron mass.  The contribution to $S^{21}$ from a vortex at radius $R_{\rm{v}}(t)$ and azimuth $\phi_{\rm{v}}(t)$ at time $t$ in a fluid of uniform density $\rho$ is therefore calculated by substituting Eq.~(\ref{eq:sing}) into Eq.~(\ref{eq:slm_sph}), with the result
\begin{equation}
 S^{21} = \frac{1}{3}\sqrt{\frac{512\pi}{45}}\rho\kappa e^{-i\phi_{\rm{v}}(t)}R_{\rm{v}}(t) \left[R_{\rm{s}}^2-R_{\rm{v}}(t)^2\right]^{3/2}~.
 \label{eq:S21}
\end{equation}

Let us define dimensionless variables $\tilde{R}_{\rm{v}}=R_{\rm{v}}(t)/R_{\rm{s}}$, $\tilde{t}=t/\tau$, $\tilde{\phi}_{\rm{v}}=\phi_{\rm{v}}(t)/(2\pi)$, and $ \tilde{\omega}=\omega \tau$, where $\tau$ denotes the glitch duration, and $R_{\rm{s}}$ is the stellar radius.   The radial distance travelled by each vortex between unpinning and repinning is denoted by $\Delta\tilde{r}=\Delta r/R_{\rm{s}}$ (dimensionless).  It is taken (arbitrarily) to be the same for all vortices and is a parameter of the model, to be evaluated in the future via Gross-Pitaevskii simulations.  Hence we write $\tilde{R}_{\rm{v}}=\tilde{R}_0+\tilde{d}(\tilde{t})$ and $\tilde{\phi}_{\rm{v}}=\tilde{\phi}_0+\tilde{\omega}\tilde{t}$, where the initial values are chosen from $\tilde{\phi}_0\in [0,1)$ and $\tilde{R}_0\in [0,1-\Delta\tilde{r}]$ respectively.  [Unless indicated, symbols used in plot labels are the dimensionless versions (tildes are omitted in plot labels).]  In these units the single-vortex metric perturbation takes the form,
\begin{eqnarray}
\label{eq:hdless}
 h_{jk}^{TT}(R_0,\phi_0,\omega,t)&=& T^{B2,21}_{jk}\frac{K_0}{\tau^2} e^{-2\pi i\tilde{\phi}_0}\nonumber\\
 & &\times \frac{\partial^2}{\partial \tilde{t}^2}\left[e^{-2\pi i\tilde{\omega}\tilde{t}}\tilde{R}_{\rm{v}}(1-\tilde{R}_{\rm{v}}^2)^{3/2}\right]~.
\end{eqnarray}
Non-glitch pulsar parameters are absorbed in the multiplicative constant
\begin{equation}
\label{eq:M}
 K_0 = \frac{G}{c^5 D}\left(\frac{512\pi}{405}\right)^{1/2}\rho\kappa R_{\rm{s}}^4~,
\end{equation}
except for $\tilde{\omega}$.  The glitch-related parameter $\tau$ (vortex travel time) is kept separate.
  
For a pinned vortex, with $\tilde{d}(\tilde{t})=0$ or $\Delta \tilde{r}$, the wave strain reduces to
\begin{eqnarray}
\label{eq:single_stat}
 h^{TT}_{jk}(\tilde{R}_0,\tilde{\phi}_0,\tilde{\omega} ,\tilde{t})&=&-4\pi^2T^{B2,21}_{jk}\frac{K_0}{\tau^2} e^{-2\pi i (\tilde{\phi}_0 + \tilde{\omega} \tilde{t})}  \nonumber\\
 & & \times\tilde{R}_0 (1 - \tilde{R}_0^2)^{3/2} \tilde{\omega}^2~.
\end{eqnarray}
Equation~(\ref{eq:single_stat}) is therefore the signal emitted by a vortex corotating with the stellar crust.  Additionally, in the regime $\tilde{\omega}\gg 1$, where an unpinned vortex moves much further azimuthally than radially during a glitch, Eq.~(\ref{eq:single_stat}) is a good approximation to the wave strain resulting from the outward motion of a single unpinned vortex too.

\section{Single vortex signal}\label{sec:single}

In this section, we evaluate $h_{\times}(t)$ for a single vortex as it unpins, moves, and repins.  [The results for $h_{+}(t)$ are similar.] We investigate the effect on the wave form of changing $\omega$, $R_0$, $\Delta r$, and $d(t)$.  The radial trajectory is described by
\begin{eqnarray}\label{eq:doft}
 \tilde{d}(\tilde{t}) = \left\{ \begin{array}{cl}
 0 &\mbox{$\tilde{t}<\tilde{t}_{\rm{g}}$} \\
 \int_{\tilde{t}_{\rm{g}}}^{\tilde{t}_{\rm{g}}+1} d\tilde{t}\,\tilde{v}_{\rm{v}}(\tilde{t})&\mbox{$\tilde{t}_{\rm{g}}\leq \tilde{t}\leq \tilde{t}_{\rm{g}}+1$}\\
 \Delta\tilde{r}&\mbox{$\tilde{t}>\tilde{t}_{\rm{g}}+1$}
				\end{array} \right.~;
\end{eqnarray}
the vortex starts moving at $\tilde{t}=\tilde{t}_{\rm{g}}$.  We experiment with both constant and parabolic speed profiles, $\tilde{v}_{\rm{v}}(t)$, in what follows.  The wave strain is calculated by substituting $\tilde{R}_{\rm{v}}=\tilde{R}_0+\tilde{d}(\tilde{t})$ into Eq.~(\ref{eq:hdless}).  

To begin with, we assume that the speed of the vortex whilst unpinned is constant, i.e. $\tilde{v}_{\rm{v}}(\tilde{t})=\Delta \tilde{r}$ ($\tilde{t}_{\rm{g}}\leq\tilde{t}\leq\tilde{t}_{\rm{g}}+1$ and $\tilde{t}_{\rm{g}}=1$).  
Figure~\ref{fig:ch6:CA_single_om} graphs the wave strain in the cross polarisation $h_{\times}(\tilde{t})$ in units of $K_0/\tau^2$ [see Eq.~(\ref{eq:M})] for $\tilde{\omega}/(2\pi)=0.1$, 0.5, 1.0 and 5.0 (\emph{top left}, \emph{top right}, \emph{bottom left} and \emph{bottom right} respectively), corresponding to 0.1, 0.5, 1 and 5 revolutions during the glitch.  In all cases, $\tilde{t}_{\rm{g}}=1$.  The graphs extend one time unit beyond when the vortex repins.  Even while pinned ($\tilde{t}<1$ and $\tilde{t}>2$), a single off-centre vortex produces an oscillatory gravitational wave signal with period $2\pi/\tilde{\omega}$, as its superfluid velocity field is not axisymmetric.  Whilst the vortex remains pinned, the zero-to-peak amplitude of the wave strain, $h_{\rm{max}}$, is proportional to $\tilde{\omega}^2$, as described by Eq.~(\ref{eq:single_stat}).  While the vortex is moving, $h_{\rm{max}}$ changes at a rate $\partial (h_{\rm{max}}/K_0)/\partial t=4\pi^2\tilde{\omega}^2(1-\tilde{R}_{\rm{v}}^2)^{3/2}\left[1-2\tilde{v}_{\rm{v}}/(1-\tilde{R}_{\rm{v}}^2)\right]$. We draw attention to the discontinuous gradient of $h_{\times}(\tilde{t})$ at $\tilde{t}=1$ and $\tilde{t}=2$ (even when the vertical jumps are ignored), which occurs because the acceleration is artificially infinite instantaneously.  Below we introduce a parabolic speed profile, which smoothes the acceleration and removes the spikes in $h_{\times}(\tilde{t}_{\rm{g}})$ and $h_{\times}(\tilde{t}_{\rm{g}}+1)$.

The shape of the signal does not vary greatly with $\tilde{R}_0$.  The \emph{top left} panel of Fig.~\ref{fig:ch6:CA_single_dr} graphs $h_{\times}$ for $0.1\leq\tilde{R}_0\leq 0.75$, with fixed $\tilde{\omega}/(2\pi) =1$ and $\Delta \tilde{r} =0.2$.  The amplitude does not vary monotonically with $\tilde{R}_0$, since the term $\rho \mathbf{v}$, which appears in Eq.~(\ref{eq:slm_sph}), is a function of the volume-weighted angular velocity (a single vortex does not generate rigid-body rotation).  In fact, the amplitude peaks at $\tilde{R}_0=0.5$ followed by $\tilde{R}_0=0.25$, 0.1 and 0.75 (in descending order).  This ordering is preserved whilst the vortex is moving ($1<\tilde{t}<2$); however, for $\tilde{t}>2$, the ordering becomes $\tilde{R}_0=0.25$, 0.1, 0.5, 0.75.  Notably, the time at which the strain is a maximum for different $\tilde{R}_0$ is not identical during the glitch.  Once again, this feature arises because $h_{\times}$ depends nonlinearly on $R_{\rm{v}}$; the sinusoidal signal resulting from the rotation of a pinned vortex is amplified or diminished as the vortex moves radially, resulting in a non-sinusoidal signal.

The phase of the signal also varies with $\Delta \tilde{r}$.  We graph $h_{\times}$ for different $\Delta \tilde{r}$ in the \emph{top right} panel of Fig.~\ref{fig:ch6:CA_single_dr}, for $\tilde{\omega}/(2\pi) =1$ and $\tilde{R}_0 =0.1$ fixed.  For each value of $\Delta \tilde{r}$, traversed in unit time, the in-motion wave strain scales proportional to $\Delta \tilde{r}$.  Once the vortex repins, the wave strain amplitude depends on the new radial position $\tilde{R}_0+\Delta \tilde{r}$, as described by Eq.~(\ref{eq:single_stat}). 

Finally, in the \emph{bottom} panel of Fig.~\ref{fig:ch6:CA_single_dr}, we compare the wave strain for a vortex trajectory with a constant (\emph{solid} curve; see above) and parabolic (\emph{dotted} curve) speed profile, viz.
\begin{equation}
\label{eq:vpara}
\tilde{v}_{\rm{v}}(\tilde{t})=6\Delta \tilde{r}(\tilde{t}-\tilde{t}_{\rm{g}})[1-(\tilde{t}-\tilde{t}_{\rm{g}})]~.
\end{equation} 
If we ignore the initial spike, a parabolic speed profile results in a higher wave strain, because the maximum $\tilde{v}_{\rm{v}}$ is 1.5 times higher than for constant speed.  

To summarise, the strongest gravitational wave signal from the motion of a single vortex is achieved when $\tilde{\omega}$ and the maximum vortex speed are high, $\Delta \tilde{r}$ is large, and $\tilde{R}_0=0.5$.  

\section{Vortex avalanche signal}\label{sec:multi}
\subsection{Unpinning geometry}\label{subsec:geom}

\begin{figure*}
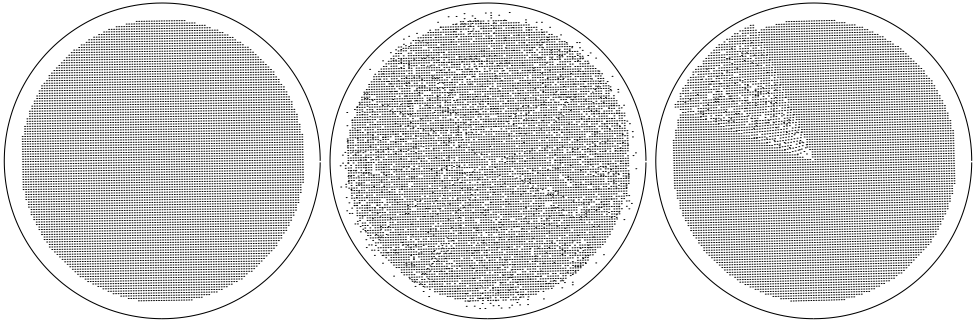

\begin{center}
\includegraphics[scale=0.465,angle=90]{5a.epsi}
\includegraphics[scale=0.465,angle=90]{5b.epsi}
\includegraphics[scale=0.465,angle=90]{5c.epsi}
\end{center}
\caption{\emph{Left}:  Initial vortex positions for $10^4$ vortices in an equatorial cross-section of a neutron star.  Vortices are confined to radii $\tilde{R}<1-\Delta\tilde{r}$ so that unpinned vortices do not leave the star.  \emph{Centre}:  Final vortex positions after a creep-like glitch.  \emph{Right}:  Final vortex positions after an avalanche-like glitch with opening angle $\Delta\tilde{\phi}=1/8$.}
\label{fig:ch6:CA_images}
\end{figure*}

\begin{figure*}
\begin{center}
\includegraphics[scale=0.4]{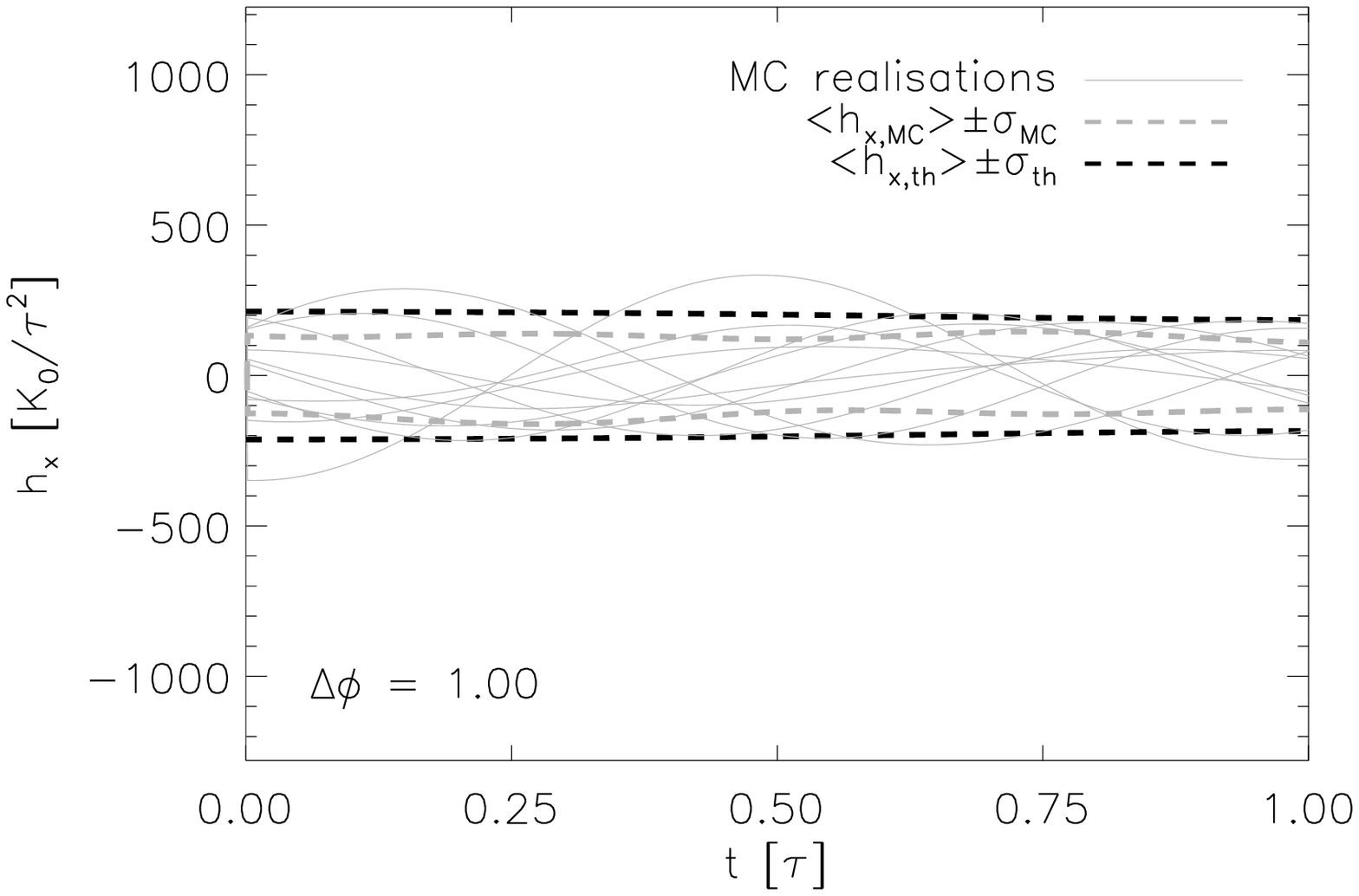}
\includegraphics[scale=0.4]{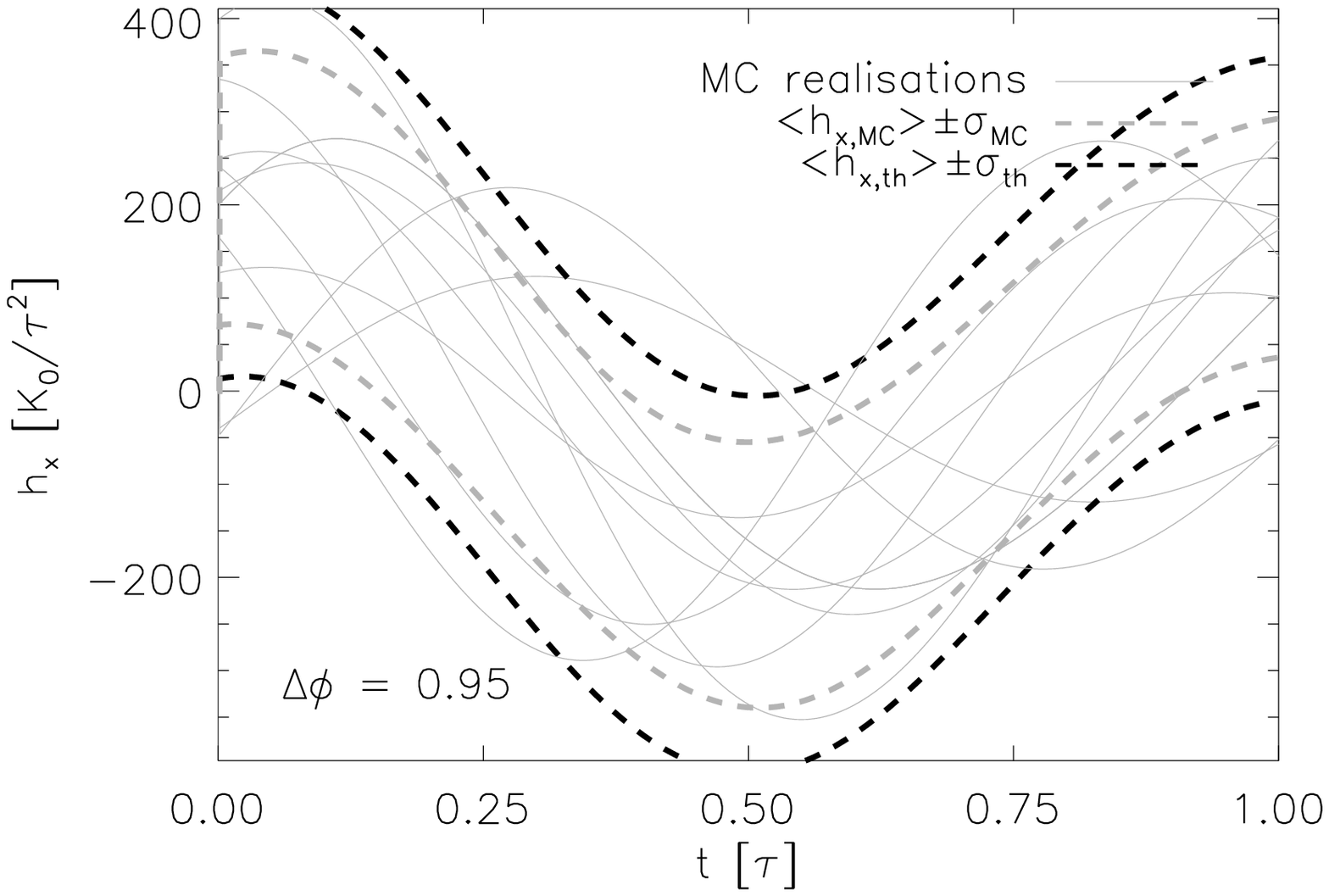}
\includegraphics[scale=0.4]{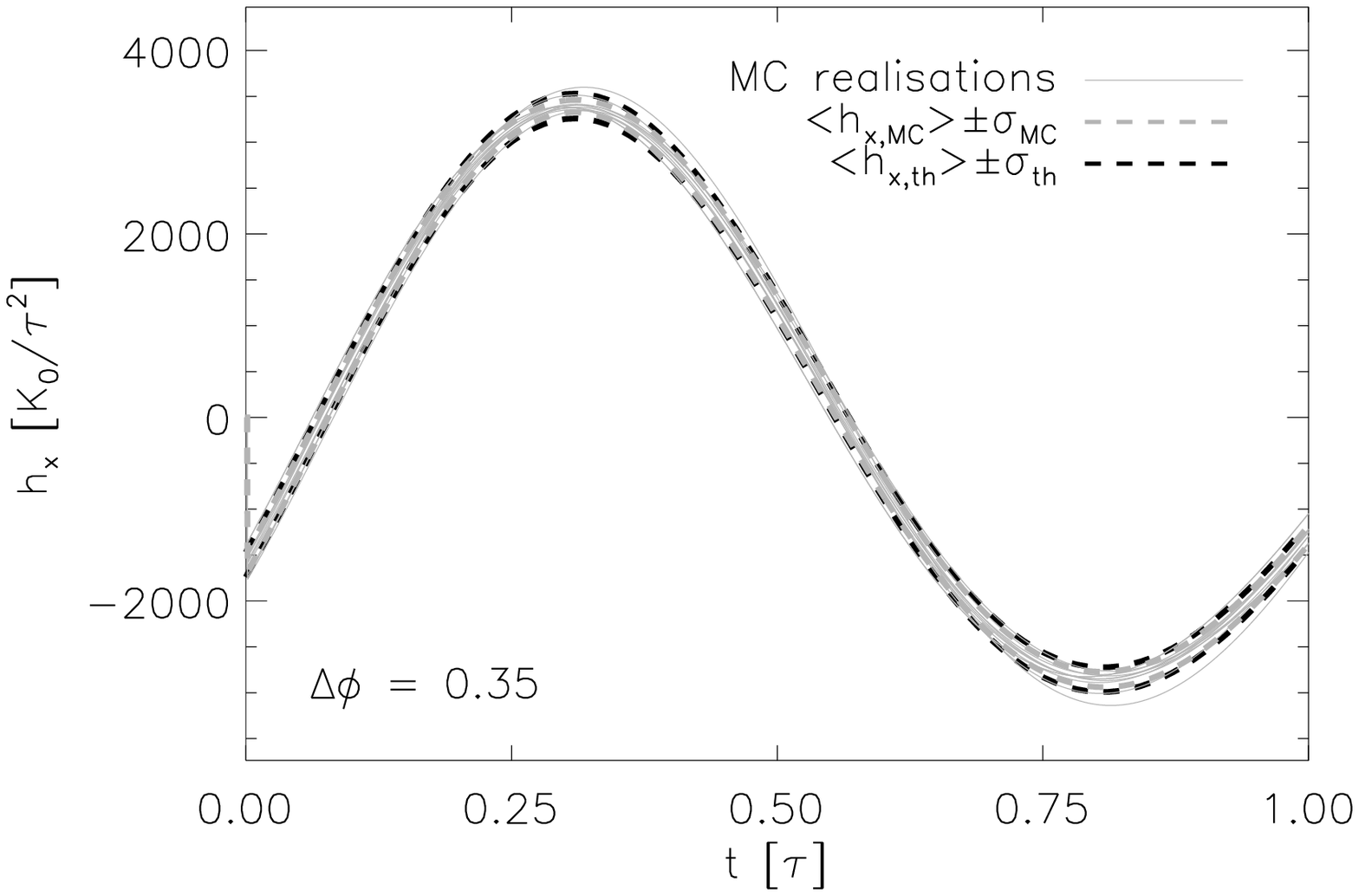}
\end{center}
\caption{Monte-Carlo results for the wave strain in the cross polarisation as a function of time $h_{\times}(t)$ for vortex avalanches with opening angles $\Delta\tilde{\phi}_0=1.0$, 0.95 and 0.35 (\emph{top left}, \emph{top right} and \emph{bottom} panels respectively).  \emph{Solid grey} curves represent $h_{\times}(t)$ for 10 different glitch realisations.  \emph{Dashed grey} curves display the mean plus/minus one standard deviation of 20 (only 10 are plotted) Monte-Carlo realisations.  \emph{Dashed black} curves are theoretical predictions of the mean plus/minus one standard deviation given in Sec.~\ref{subsec:wavestrain}.  Simulation parameters:  $N_{\rm{v}}=5\times 10^4$, $\tilde{\omega}/(2\pi) = 1.0$, $\Delta\tilde{r}=0.1$.}
\label{fig:ch6:h_dphi}
\end{figure*}

\begin{figure*}
\begin{center}
\includegraphics[scale=0.4]{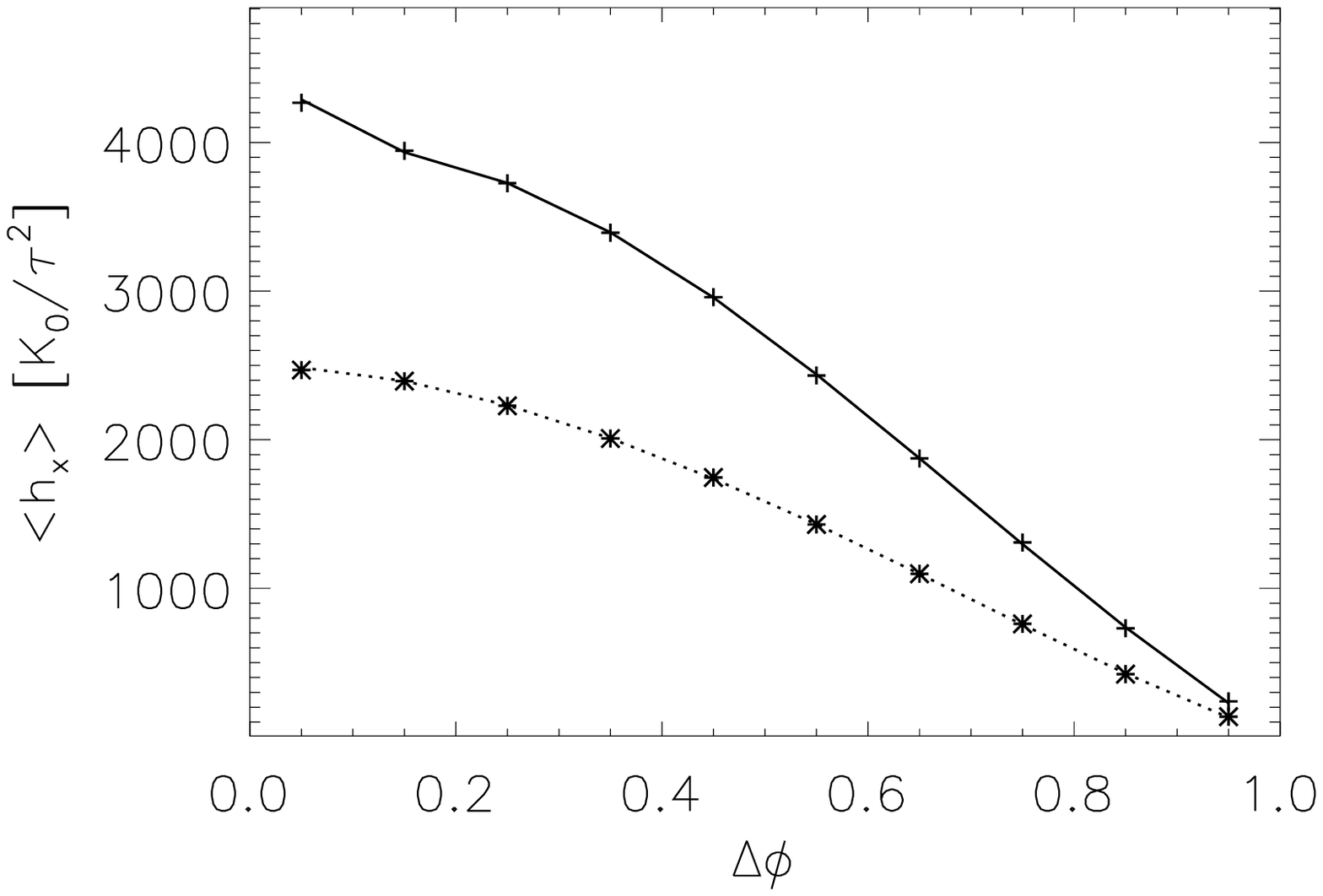}
\includegraphics[scale=0.4]{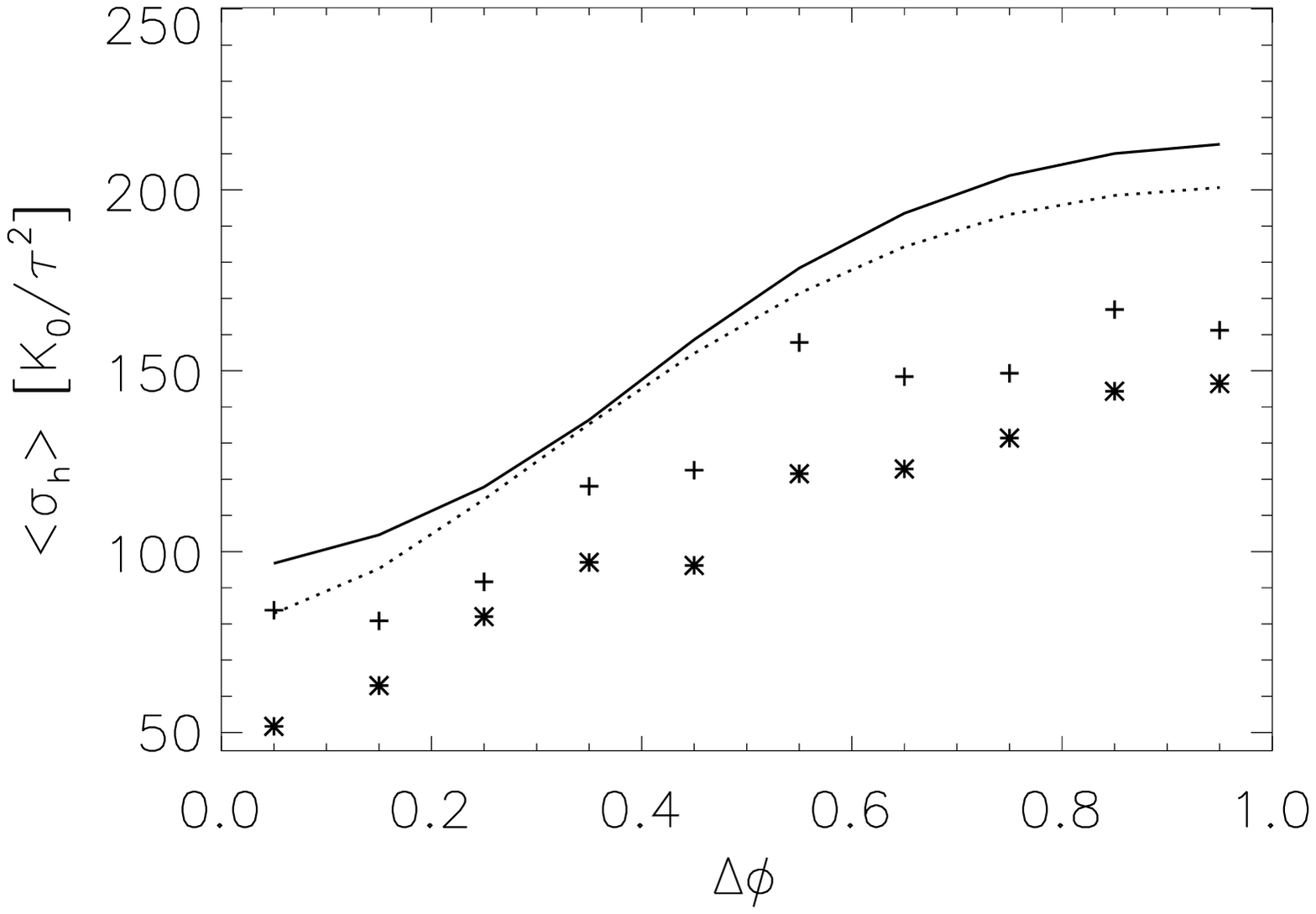}
\includegraphics[scale=0.4]{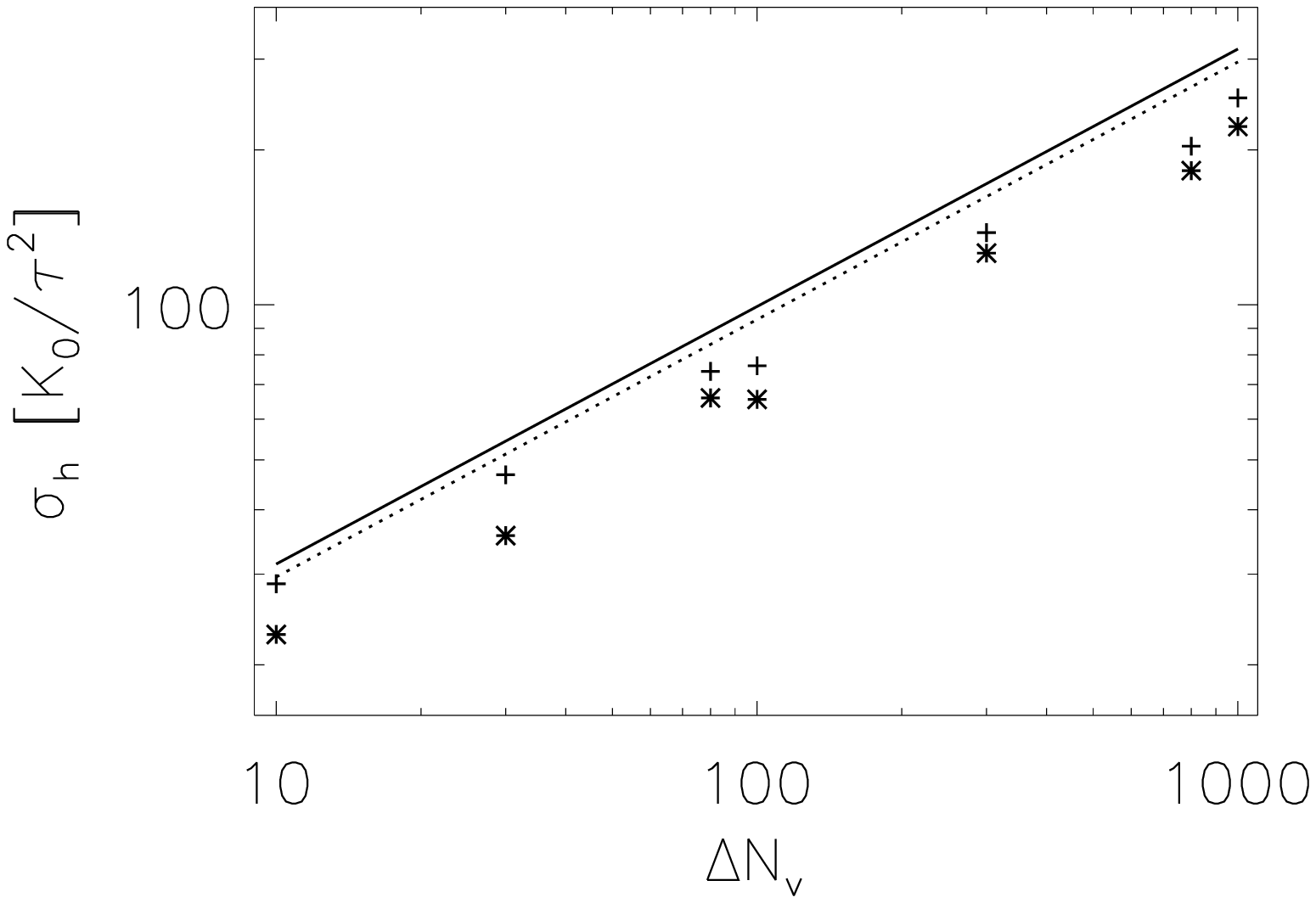}
\end{center}
\caption{Gravitational wave strain statistics from Monte-Carlo simulations in units of $K_0/\tau^2$. \emph{Top left}:  Maximum (\emph{crosses}) and time-averaged (\emph{asterisks}) $\langle h_{\times}\rangle$ (in units of $K_0/\tau^2$) as a function of avalanche opening angle $\Delta\tilde{\phi}_0$. The \emph{solid} and \emph{dotted} curves are the corresponding analytic predictions. \emph{Top right}:  Maximum (\emph{crosses}) and time-averaged (\emph{asterisks}) standard deviation [$\sigma_h=(\langle h_{\times}^2\rangle-\langle h_{\times}\rangle^2)^{1/2}$] as a function of avalanche opening angle $\Delta\tilde{\phi}_0$.  The \emph{solid} and \emph{dotted} curves are the corresponding analytic predictions. \emph{Bottom}:  Maximum (\emph{crosses}) and time-averaged (\emph{asterisks}) standard deviation as a function of number of vortices involved in the avalanche, $\Delta N_{\rm{v}}$.  The \emph{solid} and \emph{dotted} curves are the corresponding analytic predictions.}
\label{fig:ch6:stat_dphi}
\end{figure*}
\begin{table}
\begin{center}
\begin{tabular}{l|l}
\hline Quantity & Value\\\hline
$\tilde{\omega}/(2\pi)$ & 1\\
$\Delta\tilde{\phi}$ & 0.25\\
$\Delta\tilde{r}$ & 0.1\\
$\Delta\tilde{\omega}/\tilde{\omega}$ & 0.001\\
$\Delta N_{\rm{v}}$ & 500\\\hline
\end{tabular}
\caption{Canonical glitch parameters adopted in the simulations in Sec.~\ref{sec:multi}.}
\label{tab:glitch}
\end{center}
\end{table}

Current glitch detection methods based on radio telescope data are sensitive to glitches involving the simultaneous unpinning and outward motion of at least $\Delta N_{\rm{v}}\sim 10^{12}$ vortices.  The following approximate formula relates $\Delta N_{\rm{v}}$ to the fractional frequency jump $\Delta\omega/\omega$ observed:
\begin{equation}
\label{eq:dom}
 \frac{\Delta\omega}{\omega}\approx \Delta\tilde{r}\frac{\Delta N_{\rm{v}}}{N_{\rm{v}}}\frac{I_{\rm{s}}}{I_{\rm{c}}}\,.
\end{equation}
It involves the ratio of the superfluid and crust moments of inertia, $I_{\rm{s}}/I_{\rm{c}}$ (which depends on what fraction of vortices are pinned; $I_{\rm{s}}/I_{\rm{c}}\sim 10^{-2}$) \citep{Alpar:1989p182,Link:1992p20}, and the total number of vortices $N_{\rm{v}}\approx 2\pi\omega R_{\rm{s}}^2/\kappa\approx 10^{15} \,[\omega/(1~\rm{rad~s}^{-1})]$.\footnote{This approximation assumes that each vortex is responsible for an equal fraction of the total angular momentum of the superfluid, which is not accurate when the vortices are spread evenly from the centre to the edge of the star.}  Hence $\Delta N_{\rm{v}}=10^{12}$ corresponds to $\Delta\omega/\omega \sim10^{-11}$, typically the smallest glitch size observed to date \citep{Janssen:2006p78,Melatos:2008p204,Chukwude:2010,Espinoza:2011}.  

The geometry and chronology of a vortex avalanche are vital, yet regrettably unknown, ingredients of the gravitational wave calculation. Do all vortices move simultaneously, from evenly distributed positions throughout the star?  Evenly distributed pinning sites with a broad range of pinning potentials represent one possible configuration that would give rise to this scenario  \citep{Jones:1991p3949,Melatos:2009p4511}.   Or, do vortices unpin in a cascade restricted to a wedge within the star \citep{Alpar:1981p18,Warszawski:2008p4510}?  A branching process, where an unpinned vortex also preferentially unpins its nearest neighbours, leading to a branching tree of activity, would produce this outcome.  We adopt the terms `creep-like' \citep{Alpar:1989p182} and `avalanche-like' \citep[see ][]{Warszawski:2008p4510} to describe the former and latter possibilities respectively. During creep-like glitches, the non-axisymmetric change in velocity field due to one vortex largely cancels a similar change from a vortex on the opposite side of the star.  Hence we have $\langle h_{\times}\rangle=0$ (ensemble average) and the typical strength of the signal in any particular realisation is of order the dispersion $\langle h_{\times}^2\rangle^{1/2}$.  During avalanche-like glitches, cancellation is  incomplete and $\langle h_{\times}\rangle$ is not zero in general.

In what follows, we calculate only the contribution to the wave strain from vortices that unpin and move during a glitch.  In reality, even the static, pinned Abrikosov vortex lattice is not perfectly smooth and axisymmetric: vorticity is carried by point-like vortices rather than in a smooth field.  We ignore the small contribution to the wave strain from pinned vortices.  In addition, we neglect the contribution from persistent asymmetries in the vortex distribution near the stellar core due to incomplete, stratification-limited Ekman spin up \citep{Bennett:2010recovery}.

We perform Monte-Carlo calculations by choosing the initial positions of unpinned vortices at random to investigate the dispersion in $h_{\times}$ as well as its mean.  The \emph{left} panel of Fig.~\ref{fig:ch6:CA_images} shows the initial vortex configuration for $N_{\rm{v}}=10^4$.  Vortices (represented as \emph{dots}) are initially placed in an hexagonal Abrikosov lattice, inside a circle of radius $1-\Delta \tilde{r}$.  The \emph{centre} panel of Fig.~\ref{fig:ch6:CA_images} shows the vortex positions after a creep-like glitch.  The \emph{right} panel shows the positions following an avalanche-like glitch of the same size.  The initial positions ($\tilde{R}_0$ and $\tilde{\phi}_0$) of vortices that unpin are drawn from the following probability distribution functions (PDFs):
\begin{eqnarray}
\label{eq:PDFR0}
 p(\tilde{R}_0) &=&  2\tilde{R}_0~,\\
\label{eq:PDFphi0}
 q(\tilde{\phi}_0) &=& \frac{1}{\Delta \tilde{\phi}_0}[H(\tilde{\phi}_0+\Delta\tilde{\phi}_0/2)-H(\tilde{\phi}_0-\Delta\tilde{\phi}_0/2)]~,
\end{eqnarray}
where $H(\cdot)$ is the Heaviside step function, and we choose $\tilde{\phi}_0=0$ as the bisector of the avalanche without loss of generality.  For creep-like and avalanche-like glitches, we have $\Delta \tilde{\phi_0}=1$ and $\Delta\tilde{\phi}_0<1$ respectively. We adopt the parabolic speed profile from Eq.~(\ref{eq:vpara}), so as to avoid discontinuities leading to artificial spikes in the wave strain.  A parabolic profile is also a close approximation to the spiral motion of a vortex as it repins, observed in the quantum mechanical simulations described in Sec.~\ref{sec:GPE}.

\subsection{Wave strain}\label{subsec:wavestrain}

We now calculate the gravitational wave signal from a vortex avalanche.  We evaluate $h_{\times}(t)$ in two ways:  (1) Monte-Carlo simulations, in which $\Delta N_{\rm{v}}$, $\tilde{R}_0$ and $\tilde{\phi}_0$ are drawn from the avalanche size PDF, $g(\Delta\omega/\omega)$, $p(\tilde{R}_0)$ and $q(\tilde{\phi}_0)$ respectively, using Eq.~(\ref{eq:dom}) with $I_{\rm{s}}/I_{\rm{c}}=1$; and (2) marginalising over the PDFs in Eq.~(\ref{eq:PDFR0}) and (\ref{eq:PDFphi0}) to obtain the moments of $h_{\times}(t)$ for fixed $\Delta\omega/\omega$.  Since the contribution to $h_{\times}$ from each vortex is drawn from the same underlying PDF, and these contributions are summed together, the statistics of the aggregate $h_{\times}(t)$ obey the central limit theorem ($\Delta N_{\rm{v}}\gg 1$).  That is, at each time $t$, $h_{\times}(t)$ is a Gaussian variable with mean $\Delta N_{\rm{v}}\mu_1$ and variance $\Delta N_{\rm{v}}\sigma^2_1$, where $\mu_1$ and $\sigma^2_1$ are the mean and variance respectively of the contribution from a single vortex, given by
\begin{eqnarray}
\label{eq:hmarginmean}
\mu_1(t) &=& \frac{2K_0}{\tau^2\Delta\tilde{\phi}_0}\int_0^{1-\tilde{\Delta} r} d\tilde{R}_0\,\tilde{R}_0\int_{-\Delta\tilde{\phi}/2}^{\Delta\tilde{\phi}/2}\,d\tilde{\phi}_0 h_{jk}^{TT}\\
\label{eq:hmarginvar}
\sigma_1^2(t) &=& \langle h_1^2\rangle-\mu_1^2~,\\
\langle h_1^2\rangle(t) &=&\left(\frac{K_0\sqrt{2}}{\tau^2\Delta\tilde{\phi}_0}\right)^2\int_0^{1-\tilde{\Delta} r}d\tilde{R}_0\,\tilde{R}_0\int_{-\Delta\tilde{\phi}/2}^{\Delta\tilde{\phi}/2}\,d\tilde{\phi}_0 \left|h_{jk}^{TT}\right|^2~.
\end{eqnarray}
Here, $h_{jk}^{TT}$ is calculated from Eq.~(\ref{eq:hdless}).  For $\Delta\tilde{\phi}_0=1$, Eq.~(\ref{eq:hmarginmean}) implies $\mu_1=0$.  However, any particular realisation of the initial vortex positions is not symmetrically distributed around the star, implying $\sigma^2_1> 0$.

In Fig.~\ref{fig:ch6:h_dphi} we graph $h_{\times}(\tilde{t})$  (in units of $K_0/\tau^2$) from 10 glitch realisations (\emph{grey} curves), for $\Delta\tilde{\phi}=1$, 0.95 and 0.35 (\emph{top left}, \emph{top right} and \emph{bottom} panels respectively), overplotted with curves representing one standard deviation above and below the mean (\emph{dashed grey} curves). Unless otherwise stated, we adopt the set of canonical glitch parameters in Table~\ref{tab:glitch}.  We compare these curves to the analytic prediction $\Delta N_{\rm{v}}\mu_1(t)\pm\sqrt{\Delta N_{\rm{v}}} \sigma_1(t)$, given by Eq.~(\ref{eq:hmarginmean}) and (\ref{eq:hmarginvar}) (\emph{dashed black curves}).  Both methods of calculation assume that all $\Delta N_{\rm{v}}$ vortices move simultaneously; in a real avalanche, vortices unpin in a domino chain, but the domino time-scale is rapid compared to the time separation between glitches.  Even for 20 realisations, the standard deviation of the Monte-Carlo results, $\sigma_{\rm{MC}}$, deviates from $\sigma_1\sqrt{\Delta N_{\rm{v}}}$ by less than 1\% for $\Delta\tilde{\phi}=0.35$.  For $\Delta\tilde{\phi}=1$, the realisations yield a mean signal that oscillates about zero [Eq.~(\ref{eq:hmarginmean}) gives precisely zero; our simulations involve a finite number of vortices, resulting in residual dispersion].  

The position of the strain maximum, or phase of the signal, varies by up to half a revolution from glitch to glitch.  The phase shift appears for the same reason discussed in Sec.~\ref{sec:single}; the signal amplitude from each vortex in the avalanche is a non-monotonic function of its radial position.  For a given $\Delta N_{\rm{v}}$, avalanches that are restricted to a thin wedge of the star are more likely than avalanches in larger regions to involve vortices from all radii, effectively smoothing over this effect; for $\Delta\tilde{\phi}\ll1$, the phase difference between different realisations is negligible. 

In order to systematically parametrise the gravitational wave signal, we define four quantities ($\langle \cdot\rangle$ denotes an ensemble average over realisations and $\sigma_h$ is the standard deviation of the glitch signal from many vortices).
\begin{enumerate}
\item The time-averaged mean, $\langle h_{\times}\rangle=\tau^{-1}\int_0^{\tau} dt\langle h_{\times}\rangle(t)$.
\item The maximum amplitude of the mean $\langle h_{\times}\rangle_{\rm{max}}=\rm{max}\{\langle h_{\times}\rangle (t)|t\in[0,\tau]\}$.
\item The time-averaged standard deviation, $\langle \sigma_h\rangle=\tau^{-1}\int_0^{\tau} \sigma_{h}(t) dt$.
\item The maximum amplitude of the standard deviation, $\sigma_{\rm{max}}=\rm{max}\{\sigma_h|t\in[0,\tau]\}$.
\end{enumerate}

The \emph{top} panels of Fig.~\ref{fig:ch6:stat_dphi} graph these parameters as a function of $\Delta\tilde{\phi}$.  For $\langle h_{\times}\rangle$ and $\langle h_{\times}\rangle_{\rm{max}}$, Eq.~(\ref{eq:hmarginmean}) conforms exactly to the Monte-Carlo output (see \emph{left} panel).  The signal strength is a decreasing function of the avalanche opening angle:  as $\Delta\tilde{\phi}$ increases, it is more likely that the signal from the motion of one vortex interferes destructively with that of another vortex on the other side of the star. The \emph{top right} panel shows that, for $\Delta\tilde{\phi}\gtrsim 0.3$,  $\langle h_{\times}\rangle$ decreases linearly with $\Delta\tilde{\phi}$.  Both $\langle\sigma_h\rangle$ and $\sigma_{\rm{max}}$ are increasing functions of $\Delta\tilde{\phi}$ for $\Delta\tilde{\phi}\lesssim 0.5$.  For $\Delta\tilde{\phi}\gtrsim 0.5$, $\langle\sigma_h\rangle$ and $\sigma_{\rm{max}}$ saturate, since the avalanche spans more than half the star, and hence the motion of vortices on one side of the star start to `cancel' with motion of vortices on the opposite side.

The \emph{bottom} panel of Fig.~\ref{fig:ch6:stat_dphi} graphs $\langle\sigma_h\rangle$ and $\sigma_{\rm{max}}$ as functions of $\Delta N_{\rm{v}}$.  As predicted by the central limit theorem, both quantities increase with the square root of $\Delta N_{\rm{v}}$.  For glitches involving a significant fraction of the vortices, this trend breaks down, since most of the vortices unpin during the avalanche, not just a small random sample.

In the following sections, we consider avalanches with $\Delta\tilde{\phi}=0.25$, and explore the change in the mean and standard deviation of the signal as a function of pulsar angular velocity, glitch size, and vortex travel distance.

\subsection{Glitch size}\label{subsec:size}

\begin{figure*}
\begin{center}
\includegraphics[scale=0.4]{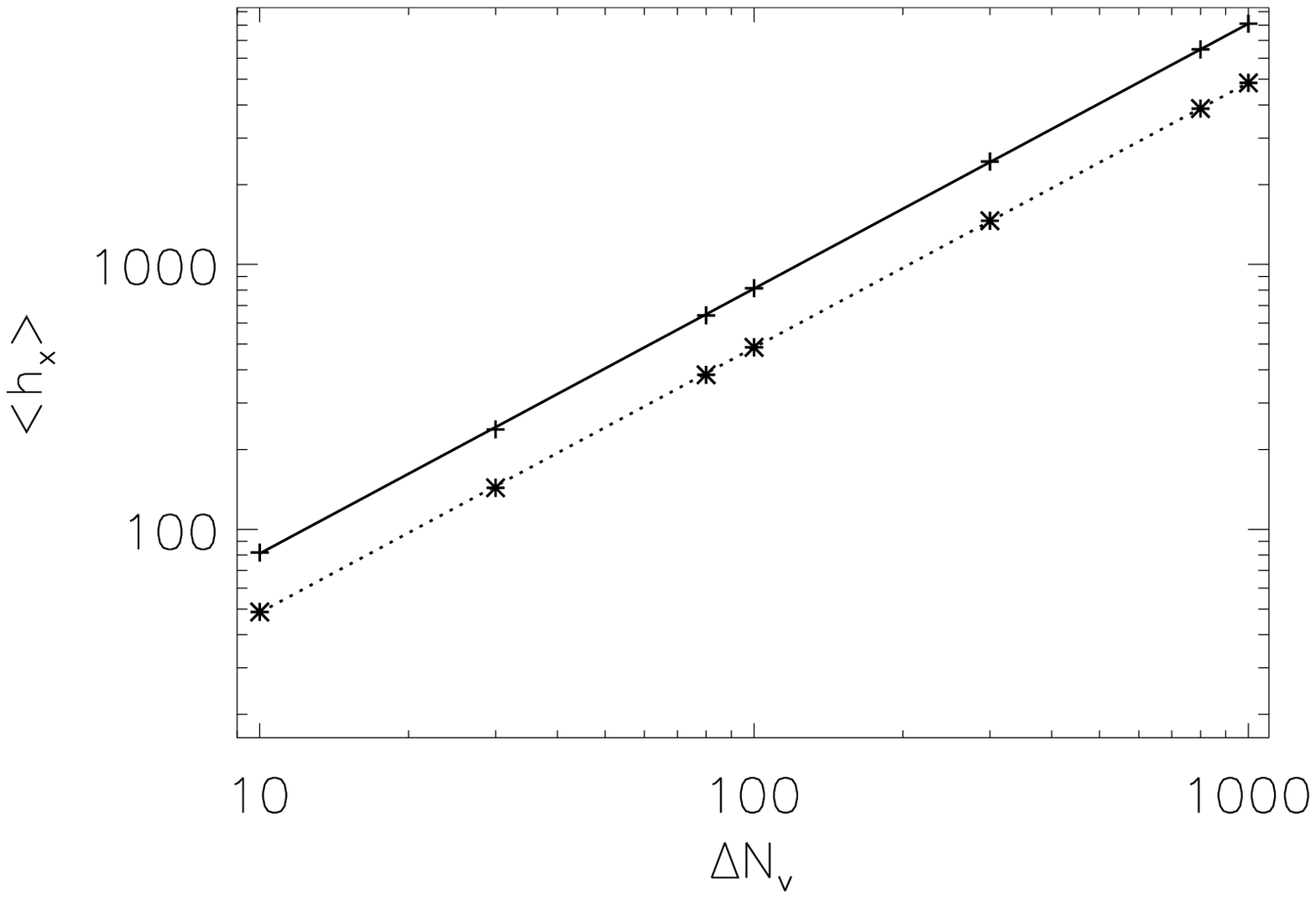}
\includegraphics[scale=0.4]{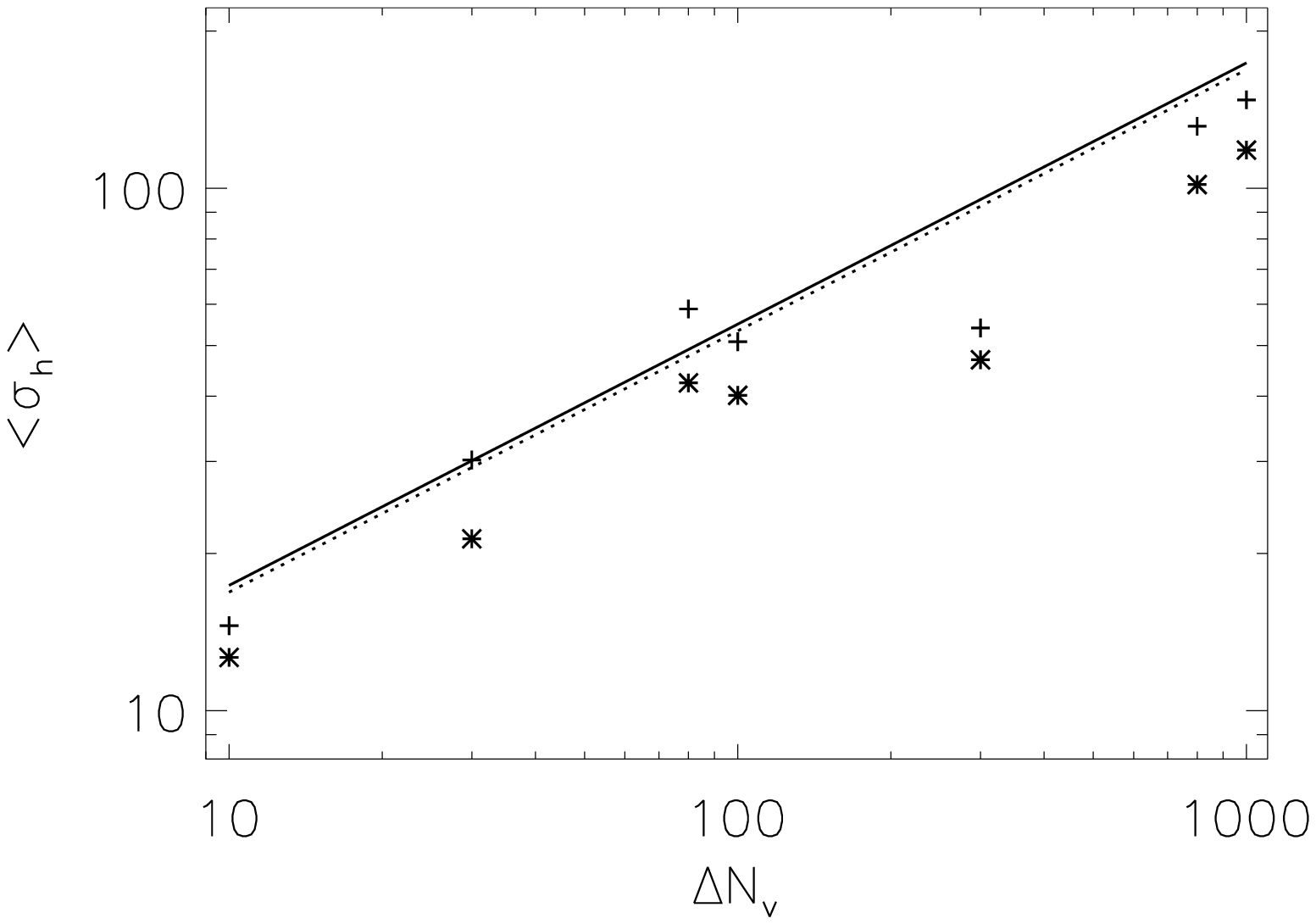}
\includegraphics[scale=0.4]{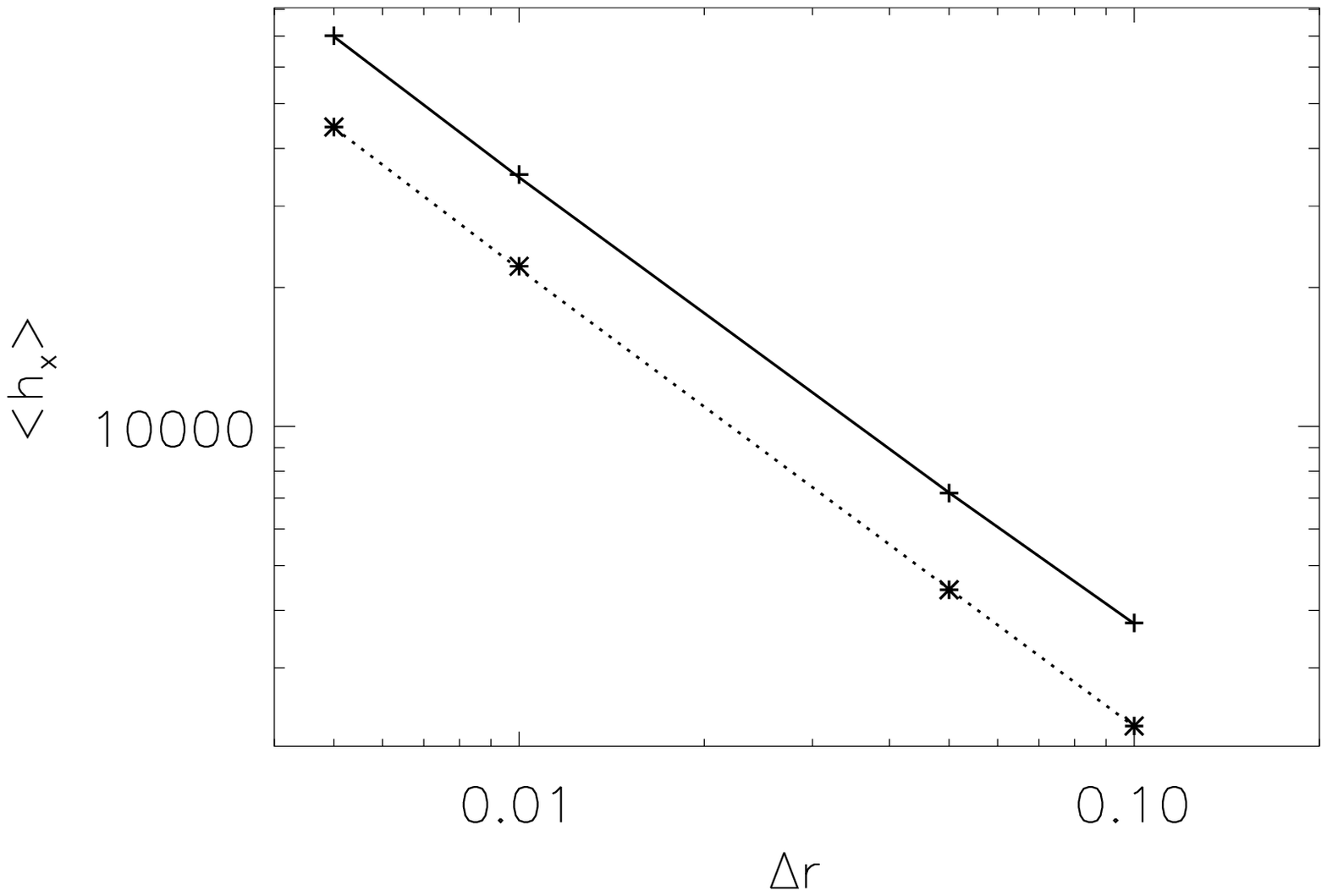}
\includegraphics[scale=0.4]{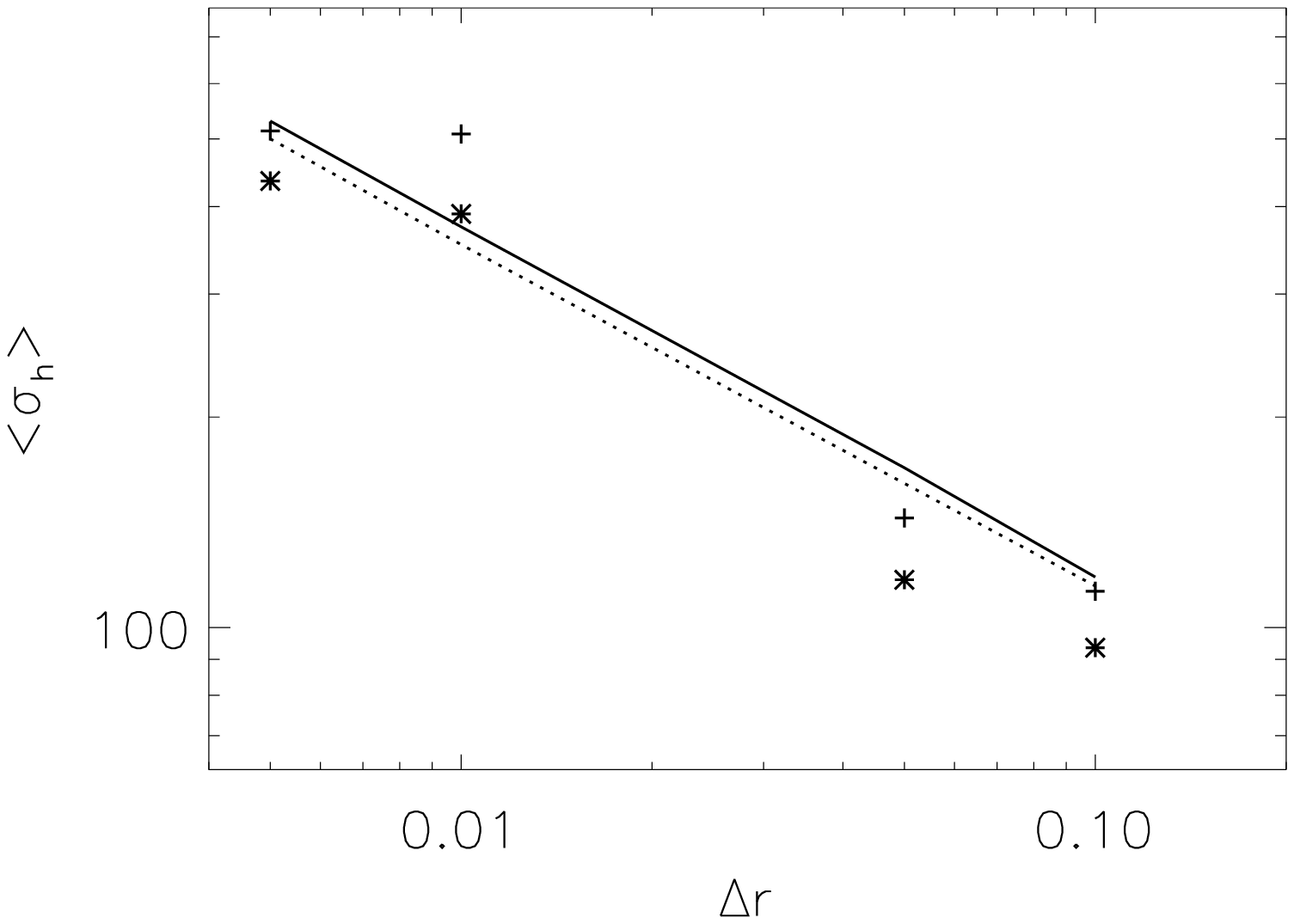}
\includegraphics[scale=0.4]{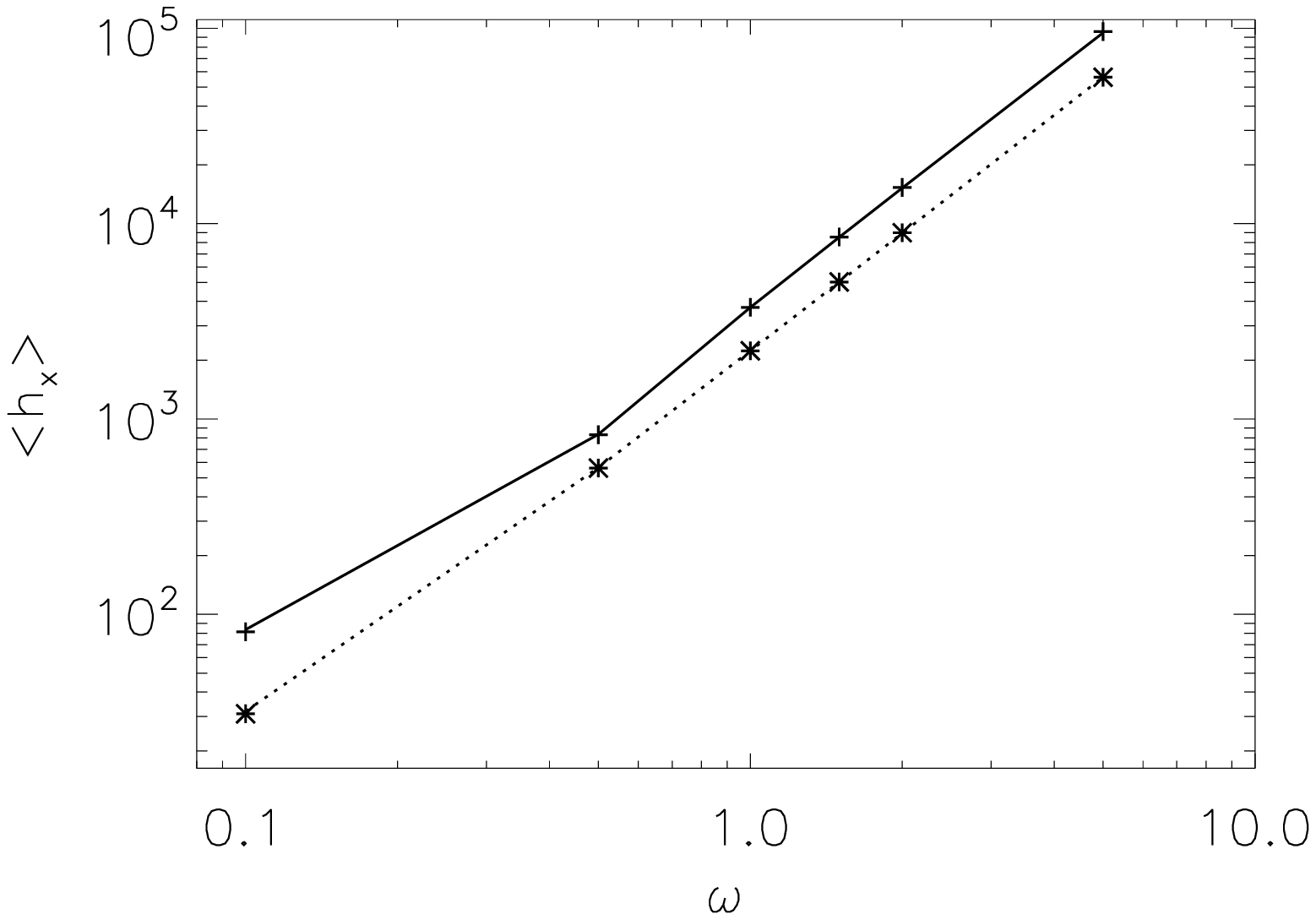}
\includegraphics[scale=0.4]{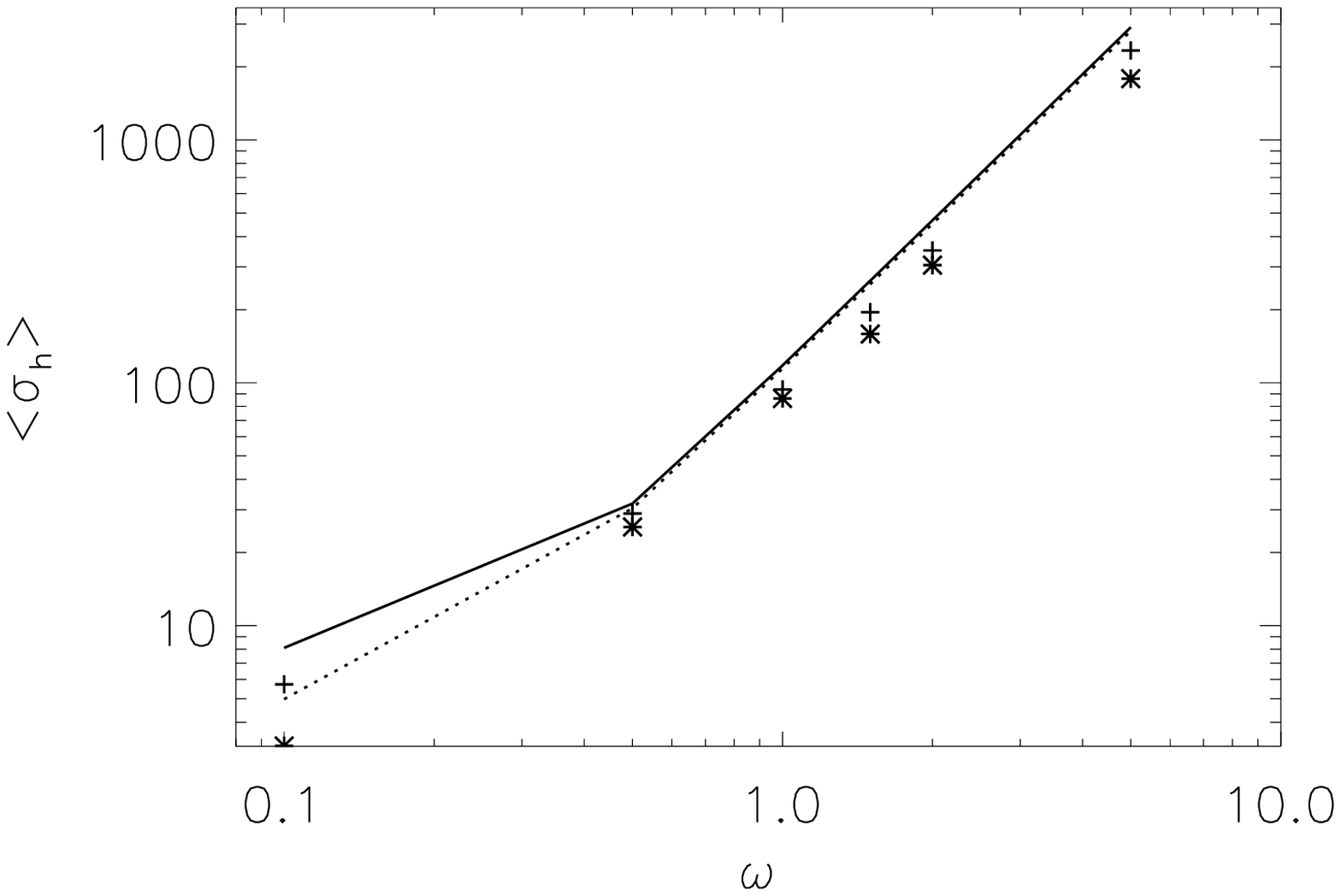}
\end{center}
\caption{Gravitational wave strain statistics from Monte-Carlo simulations of many-vortex glitches in an individual pulsar.  Maximum (\emph{crosses}) and time-averaged (\emph{asterisks}) $\langle h_{\times}\rangle$ (\emph{left}) and $\sigma_{h}$ (\emph{right}). The \emph{solid} and \emph{dotted} curves are the corresponding analytic results.  The mean and standard deviation are given in units of $K_0/\tau^2$.  \emph{Top}:  Variation with the number of vortices that move during the glitch, $\Delta N_{\rm{v}}$.  \emph{Centre}:  Variation with distance travelled by each vortex, $\Delta\tilde{r}$. \emph{Bottom}:  Variation with angular velocity, $\tilde{\omega}$.}
\label{fig:ch6:stat_param}
\end{figure*}

\begin{figure}
\begin{center}
\includegraphics[scale=0.4]{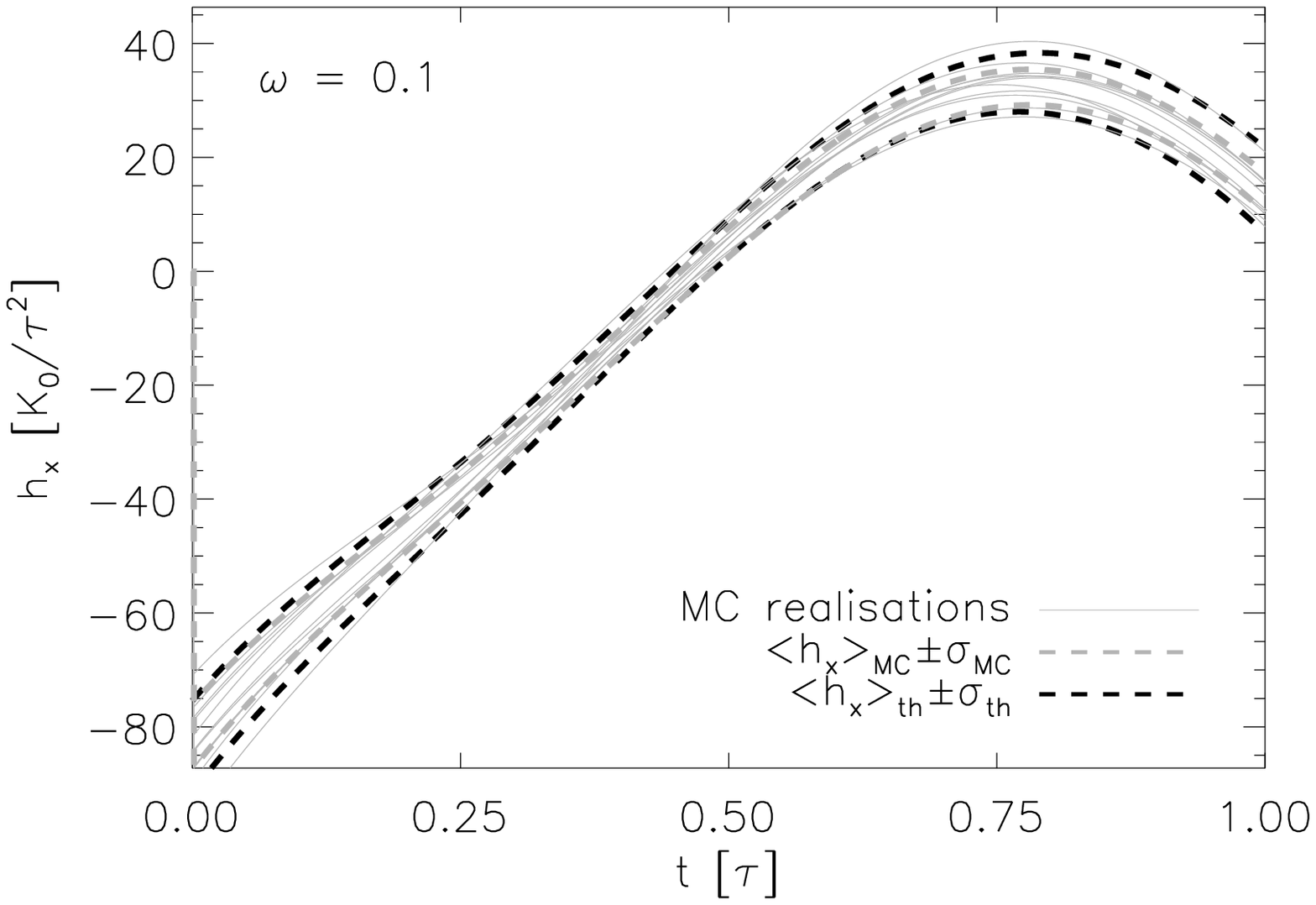}
\includegraphics[scale=0.4]{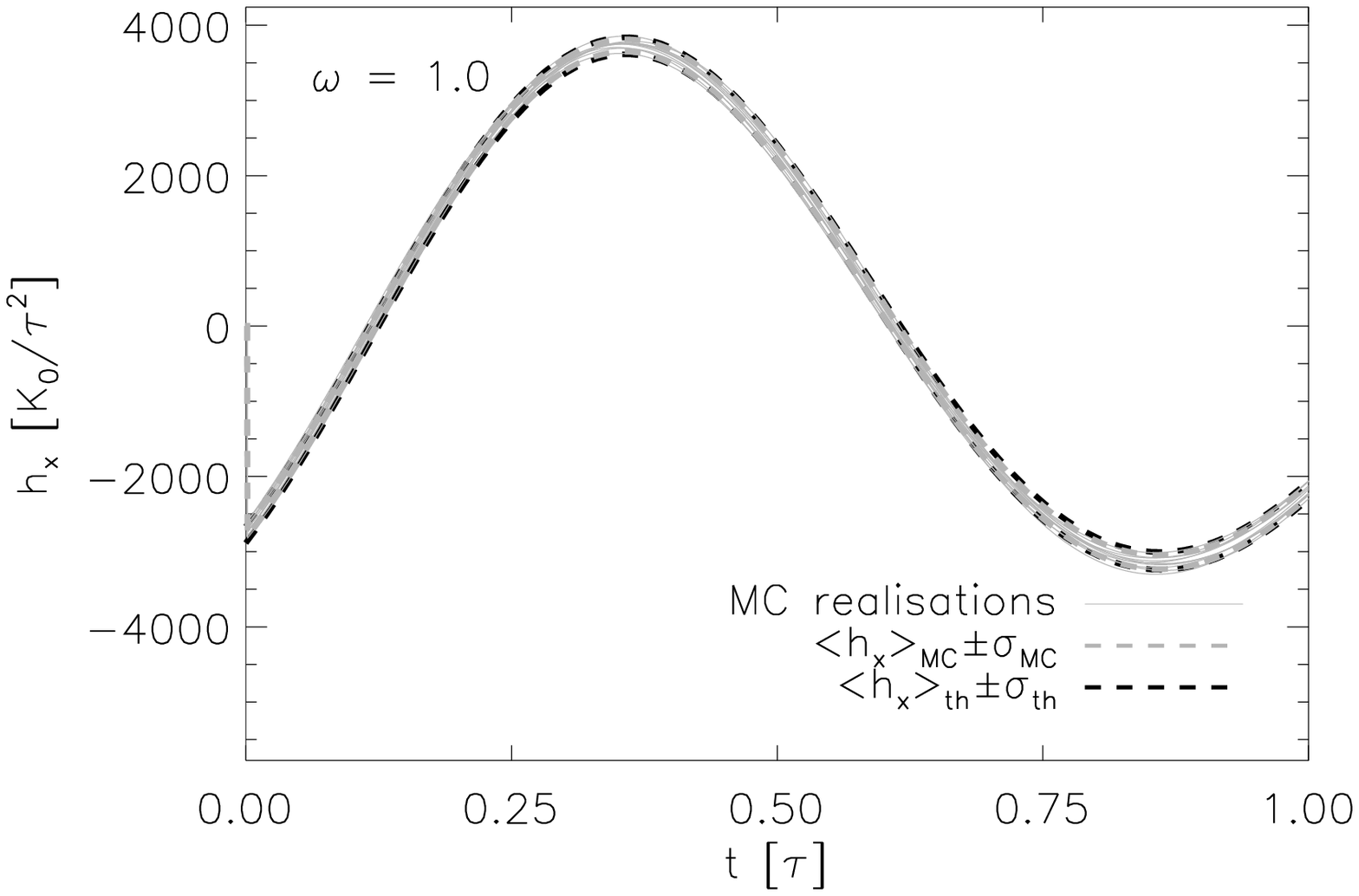}
\includegraphics[scale=0.4]{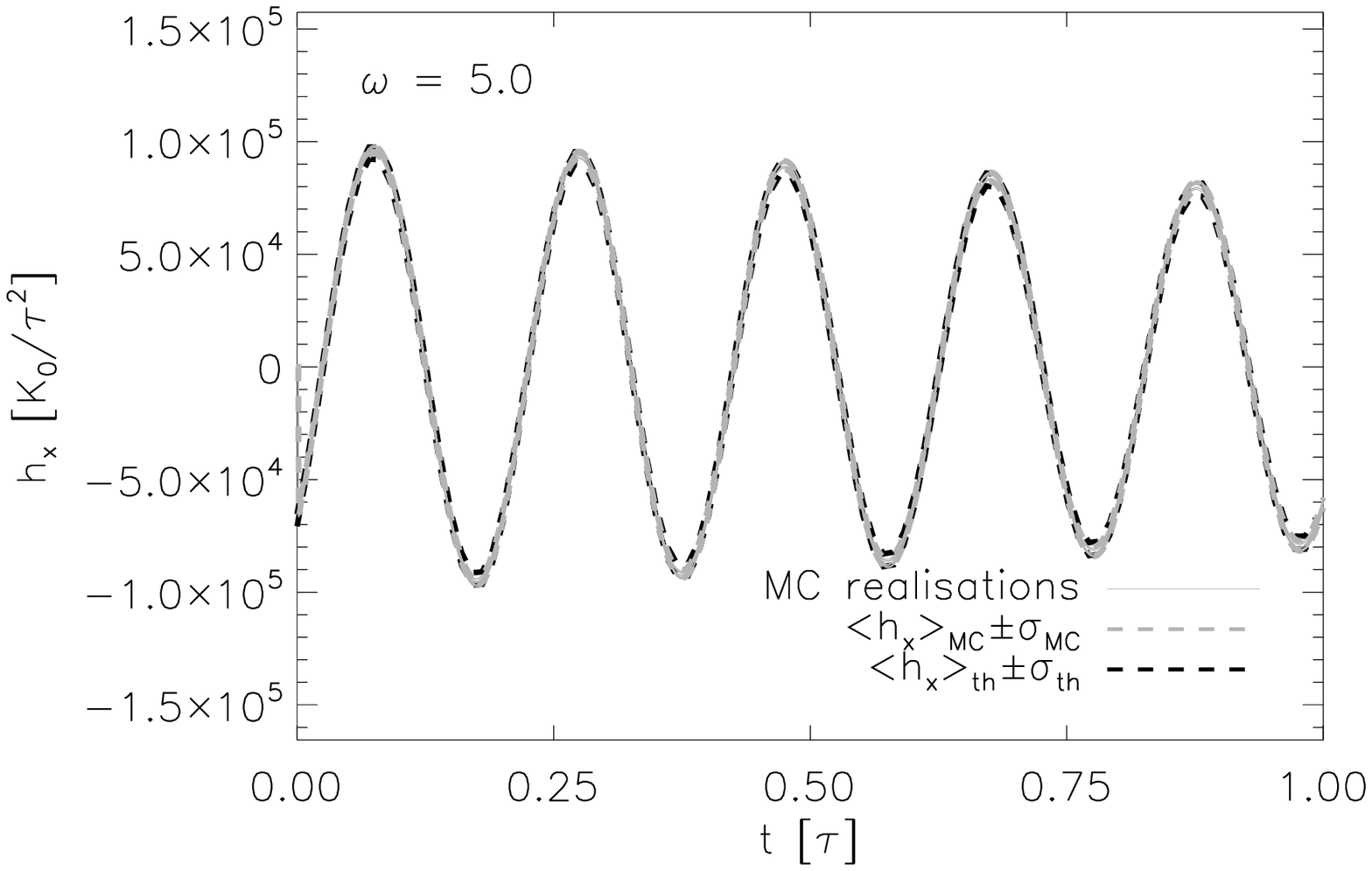}
\end{center}
\caption{Monte-Carlo results for the wave strain in the cross polarisation as a function of time $h_{\times}(t)$ for angular velocity $\tilde{\omega}/(2\pi)=0.1$, 1.0 and 5.0 (\emph{top}, \emph{middle} and \emph{bottom} respectively).  \emph{Solid grey} curves represent $h_{\times}(t)$ for different glitch realisations.  \emph{Dotted grey} and \emph{dashed grey} curves are the mean and the mean plus/minus one standard deviation of 20 (only 10 are plotted) Monte-Carlo realisations.  \emph{Dotted black} and  \emph{dashed black} curves are the corresponding analytical predictions.  Simulation parameters:  $N_{\rm{v}}=5\times 10^{4}$, $\Delta\tilde{\phi}_0 = 0.25$, $\Delta\tilde{r}=0.1$.}
\label{fig:ch6:h_om}
\end{figure}
 
Both avalanche and creep-like glitches generate wave strains that scale linearly with glitch size (and hence $\Delta N_{\rm{v}}$).  In the \emph{top} panels of Fig.~\ref{fig:ch6:stat_param}, we graph the mean (\emph{left} panel) and standard deviation (\emph{right} panel) of the wave strain as a function of glitch size, expressed in terms of $\Delta N_{\rm{v}}$, for avalanche-like glitches with $\Delta\tilde{\phi}=0.25$.  Both measures of the mean (\emph{top left} panel) introduced in Sec.~\ref{subsec:wavestrain} are linearly increasing functions of $\Delta N_{\rm{v}}$.  The Monte-Carlo (\emph{crosses} and \emph{asterisks}) and analytic (\emph{solid} and \emph{dotted} curves) calculations agree with respect to the means.  On the other hand, analytic calculations of $\langle\sigma_h\rangle$ are up to double the Monte-Carlo values, and $\sigma_{\rm{max}}$ does not agree well with Eq.~(\ref{eq:hmarginvar}), although all calculations respect the scaling $\sigma_{\rm{max}}\propto\sqrt{\Delta N_{\rm{v}}}$.  The same scaling is observed for creep-like glitches, as shown in the \emph{bottom panel} of Fig.~\ref{fig:ch6:stat_dphi}.

\subsection{Vortex travel distance}\label{subsec:motionGW}

Glitch size is the product of the number of vortices that unpin and the distance they traverse, according to Eq.~(\ref{eq:dom}).  In other words, for a given glitch size, either many vortices move a short distance, or few vortices move a large distance.  In this section we investigate how the strength of the gravitational wave signal changes with the distance travelled by each vortex, $\Delta\tilde{r}$, to cause a glitch of fixed size. 

The \emph{centre} panels of Fig.~\ref{fig:ch6:stat_param} demonstrate that the mean gravitational wave strength is a decreasing function of $\Delta\tilde{r}$. This result can be understood by noting that increasing the radial distance traversed by each vortex does not make the system more or less asymmetric.  Therefore, as $\Delta\tilde{r}$ increases, and $\Delta N_{\rm{v}}$ decreases in inverse proportion to keep $\Delta\omega/\omega$ constant, we expect the mean and standard deviation to also decrease.  This is confirmed by the Monte-Carlo and analytic results in Fig.~\ref{fig:ch6:stat_param}.  In calculations not graphed here, we also find that this correlation persists for creep-like glitches.

\subsection{Pulsar angular velocity}\label{subsec:omega}

The mean and dispersion of the wave strain increase with $\tilde{\omega}$; $\langle h_{\times}\rangle$, $\langle h_{\times}\rangle_{\rm{max}}$, $\langle \sigma_h\rangle$, and $\sigma_{\rm{max}}$ are plotted versus $\tilde{\omega}$ in the \emph{bottom} panels of Fig.~\ref{fig:ch6:stat_param}.  The quantities are quadratic functions of $\tilde{\omega}$, as for a single vortex, e.g. Eq.~(\ref{eq:single_stat}) in Sec.~\ref{sec:discrete}.

Figure~\ref{fig:ch6:h_om} demonstrates that, for glitches in which azimuthal vortex motion is faster than radial motion, the gravitational wave signal oscillates with the same angular velocity as the pulsar and diminishes with time.  For $\tilde{\omega}/(2\pi)<1$, the signal executes only a fraction of a full oscillation.  The variation in the signal results from both azimuthal vortex motion and the changing parabolic vortex speed profile.  For example, in the \emph{top} panel of Fig.~\ref{fig:ch6:h_om}, the signal executes approximately half an oscillation, and the vortex circulates one tenth of the way around the star [$\tilde{\omega}/(2\pi)=0.1$].  The dependence on $\tilde{\omega}$ is most obvious in the \emph{bottom} panel of Fig.~\ref{fig:ch6:h_om}, which graphs $h_{\times}(\tilde{t})$ for $\tilde{\omega}/(2\pi)=5.0$ [\emph{solid grey} curves are Monte-Carlo realisations, \emph{dashed} curves are the mean plus/minus one standard deviation for the Monte-Carlo (\emph{grey}) and analytic (\emph{black}) calculations].  Over the lifetime of the glitch, the amplitude of $\langle h_{\times}\rangle (t)$ decreases by $20\%$.  

From Sec.~\ref{sec:single} we know that the signal from a single vortex is a function of its radial position.  As a glitch progresses, the mean (taken over all vortices) radial position increases from its initial value $\langle \tilde{R}_{\rm{v}}\rangle=2/3$.  Figure~\ref{fig:ch6:CA_single_dr} demonstrates that, for $\tilde{R}_{\rm{v}}>0.5$, the maximum signal strength is a decreasing function of $\tilde{R}_{\rm{v}}$, and hence we expect the signal from an avalanche to weaken with time.  It is important to note that for a real glitch this effect is expected to be negligible ($\Delta\tilde{r}\ll \langle \tilde{R}_0\rangle$).  As in previous sections, the number of oscillations of $\langle h_{\times}\rangle (t)$ during the avalanche is not precisely $\tilde{\omega}/(2\pi)$, owing to changes in $\langle \tilde{R}_{\rm{v}}\rangle$ with time. 

\subsection{Amplitude scaling}\label{subsec:glitchscaling}
As a guide to how the signal from a single glitch scales with the physical parameters associated with the glitch, we provide a back-of-the-envelope estimate for the maximum wave strain.  There are two regimes, determined by the relative size of $\tilde{\omega}$ and $\Delta\tilde{r}$.  For $\tilde{\omega}\ll \Delta\tilde{r}$, but keeping $\Delta\tilde{r}\ll 1$ (i.e. the vortices still travel across only across a very small fraction of the star), the azimuthal motion of the vortices is negligible compared to their radial motion; this scenario arises when the vortex travel time is much shorter than the neutron star rotation period.  In this case, the zero-to-peak amplitude of $h_{\rm{\times}}$ is obtained by setting $d\tilde{R}_{\rm{v}}/dt\approx\Delta\tilde{r}/\tau$ and expanding Eq.~(\ref{eq:hdless}) to first order in $\Delta\tilde{r}$ and is given by
\begin{eqnarray}
\langle h_{\times,\rm{max}}\rangle&\approx&  3\frac{K_0}{\tau^2}\Delta\tilde{r}\langle \tilde{R}_0\rangle^2 \Delta N_{\rm{v}}\\
  &=& 10^{-28}
 \left(\frac{D}{1\,\rm{kpc}}\right)^{-1}\left(\frac{\tau}{10\,\rm{ms}}\right)^{-2}\nonumber\\
 & &\times\left(\frac{\Delta\omega/\omega}{10^{-7}}\right)\left(\frac{\omega}{10^2 \rm{rad\,s}^{-1}}\right)~.
\label{eq:drapprox}
\end{eqnarray}
$D$ and $\omega$ are the distance to the source and its angular velocity respectively, and they are pulsar, but not glitch, dependent. $\tau$ and $\Delta r$ are the vortex travel time and distance respectively.  We note that both $\tau$ and $\Delta r$ may vary between glitches, and even between individual vortices moving within a single glitch.  However, in general, the faster and further vortices move, the stronger the gravitational wave signal.  Further studies using first-principles simulations, such as those presented in Sec.~\ref{sec:GPE} and \cite{Warszawski:2012PRB}, and an understanding of how vortex behaviour scales with system size, are needed to fix these parameters accurately.

Alternatively, for  $\tilde{\omega}\gg \Delta\tilde{r}$, but $\tilde{\omega}$ still small, the azimuthal vortex motion is a larger contributor to the wave strain than is the radial motion.  The signal amplitude is obtained by expanding Eq.~(\ref{eq:hdless}) to first order in $\tilde{\omega}$, and scales as
\begin{eqnarray}
 \langle h_{\times,\rm{max}}\rangle &\approx&  \frac{K_0}{\tau^2}(2\pi\tilde{\omega})^2 \langle \tilde{R}_0\rangle \Delta N_{\rm{v}}\\
  &=& 10^{-24}
 \left(\frac{D}{1\,\rm{kpc}}\right)^{-1}\left(\frac{\Delta r}{10^{-2}\,\rm{m}}\right)^{-1}\nonumber\\
 & & \times\left(\frac{\Delta\omega/\omega}{10^{-7}}\right)\left(\frac{\omega}{10^2 \rm{rad\,s}^{-1}}\right)^3~.
\label{eq:omapprox}
\end{eqnarray}
In this regime, corotation of pinned vortices with the crust produces a non-zero signal, provided the superfluid flow is non-axisymmetric, whose strength is a function of the non-uniformity of the vortex configuration \citep{Wasserman:2008p70}, e.g. pinning at grain boundaries (inhomogeneous) versus dislocations and lattice sites (homogeneous).  Here, it is important to note that Eq.~(\ref{eq:omapprox}) estimates the strength of the gravitational wave strain from azimuthal vortex motion on the rotation time scale, even though such motion is ongoing, even when the neutron star is not glitching.  To avoid the conclusion that there is a strong, persistent signal from rotation of pinned vortices, we must assume that non-axisymmetry in the vortex lattice is much more pronounced when a large number of vortices are not pinned (i.e. during a glitch) than during inter-glitch intervals.

\subsection{Example: the 2006 glitch in Vela}\label{subsec:ch6:Vela}

We can use Eq.~(\ref{eq:drapprox}) and (\ref{eq:omapprox}) to constrain the glitch duration for the 2006 Vela glitch.  Non-detection of gravitational waves from this glitch \citep{Clark:2010} put an upper limit of $10^{-20}$ on the wave strain.  For $\tilde{\omega}\ll \Delta\tilde{r}$, with Vela-like parameters ($D\approx 300\,\rm{pc}$, $\omega \approx 71\,\rm{rads}^{-1}$, $\Delta\omega/\omega\approx 10^{-6}$ and $\Delta r=10^{-2}\,\rm{m}$), the lower bound on the glitch duration is then $\tau\gtrsim 10^{-4}\,\rm{ms}$.  
Of course, the independence of $\tau$, $\Delta r$ and $\Delta\omega/\omega$ cannot be taken for granted.  For example, if $\tau$ is the time taken for a vortex to travel a distance $\Delta r$, which in turn depends on the superfluid-crust velocity lag (through the vortex velocity), then we cannot hold $\Delta r$, and even $\Delta\omega/\omega$, constant whilst varying $\tau$.  We also remind the reader that underpinning these estimates is the conservative assumption that the pinned vortex distribution is symmetric, apart from those vortices participating in the avalanche.  Otherwise there would be a persistent gravitational wave signal described by Eq.~(\ref{eq:omapprox}).

We also note that the $\tilde{\omega}\gg\Delta\tilde{r}$ approximation returns a wave strain that is much stronger than the persistent signal ($h\sim 10^{-31}$) arising from turbulence in a differentially rotating Crab-like neutron star \citep{Melatos:2010turb}.  The predicted wave strain resulting from the same physical mechanism operating in a nearby ($D\sim 0.01\,\rm{kpc}$), rapidly rotating ($\Omega\sim 3\times 10^3\,\rm{rad\,s}^{-1}$) neutron star is comparable to our $\tilde{\omega}\gg\Delta\tilde{r}$ scenario [Eq.~(\ref{eq:omapprox})] for a glitching pulsar \citep{Melatos:2010turb}.

\section{Discussion}\label{sec:ch6:conc}

This paper estimates the strength of the gravitational wave burst arising from individual pulsar glitches.  Quantum mechanical simulations of rotating, pinned superfluids demonstrate that a broadband burst of gravitational radiation is emitted in the current quadrupole channel by the spasmodic, non-axisymmetric rearrangement of superfluid vortices during a glitch.  Generalising the simulation results, which are restricted to systems with $\sim 10^2$ vortices, we employ a discrete model of vortex motion to calculate analytically the wave strain from vortex avalanches involving realistic numbers of vortices ($\sim 10^{12}$).  

When a single vortex unpins and moves, the wave strain peaks when the angular velocity and the distance travelled are large, the initial position is half the stellar radius, and the vortex speed is high.  Monte-Carlo simulations agree well with analytic calculations.  The wave strain depends strongly on the unpinning geometry. If vortices unpin at random throughout the star (vortex creep), the signal is smaller than when unpinning is confined to a thin wedge (vortex avalanche); the maximum wave strain from a glitch of fixed size varies inversely with $\Delta\tilde{\phi}$.  In fact, for $\Delta\tilde{\phi}=1$, the mean wave strain is formally zero.   

Quantitatively, the maximum strain is inversely proportional to $\Delta\tilde{r}$, quadratic in $\tilde{\omega}$, and linearly proportional to the number of vortices that unpin and move.  The standard deviation of the maximum strain depends on these physical parameters in the same way.  We note, however, that for a glitch of a given size, $\Delta \tilde{r}$ and $\Delta N_{\rm{v}}$ are not independent variables.  That is, for fixed $\Delta \tilde{r}$, the number of vortices that unpin and move during a glitch of size $s$ is $\Delta N_{\rm{v}}=sN_{\rm{v}} I_{\rm{c}}/(I_{\rm{s}}\Delta\tilde{r})$.  We also find that the increase in average radial vortex position as vortices move outward during a glitch damps the amplitude of the glitch gravitational wave signal.  This is most apparent for glitches with large $\omega$; the glitch signal oscillates, and its zero-to-peak amplitude decreases with time.  However, in real glitches, the change in average vortex position during a glitch is so small that this effect is negligible.

The wave strain for a glitch in which a vortex travels further azimuthally than radially ($\tilde{\omega}\gg 1$) has a zero-to-peak amplitude $h\sim 10^{-24}$.  For a glitch dominated by radial motion ($\tilde{\omega}\ll 1$), the zero-to-peak amplitude is $h\sim 10^{-28}$.  These estimates are based on a glitch of size $10^{-7}$.  Convenient scalings of wave strain with glitch size, stellar angular velocity, vortex travel distance, and glitch duration are presented in Eq.~(\ref{eq:drapprox}) and (\ref{eq:omapprox}) respectively.  We remind the reader that we conservatively assume that vortices that do not move during a glitch do not contribute to the gravitational wave signal.  Therefore, when calculating the signal from $\tilde{\omega}$-dominated vortex motion, we implicitly assume that vortices involved in a glitch are asymmetrically distributed.  

Unlike the individual glitch burst signal from a nearby pulsar, which is eminently detectable, the contribution to the stochastic gravitational wave background is arguably weak.  Nevertheless, for completeness, in App.~\ref{sec:stoch}, we explain how the theoretical machinery in Sec.~\ref{sec:current}\,--\,\ref{sec:multi} can be applied to calculate the contribution to the background rigorously.  Estimates of the strength and spectrum of a stochastic background of gravitational waves resulting from sources of cosmological and astrophysical origin are essential to determining the detectability of burst events and continuous wave sources \citep{Allen:1999,Ferrari:1999,Regimbau:2006Apj}.  We find that power-law-distributed glitch sizes (power-law index $-3/2$) from a Galactic population of neutron stars contribute to the stochastic background with a signal-to-noise ratio $\sim 10^{-4}$ (Einstein Telescope).  Importantly, however, the inclusion of extra-Galactic pulsars must boost the signal substantially, something that merits careful consideration in a  future paper.

The calculations presented here are premised on several assumptions about the physics of glitches and the glitch behaviour of the local neutron star population.  Most importantly, our calculation assumes that glitches result from superfluid vortex unpinning avalanches.  Alternative glitch theories, such as fluid instabilities \citep{Andersson:2004} and starquakes \citep{Link:1996p136,Negi:2010}, generate different signal strengths and spectra.  Previous estimates of the maximum achievable signal-to-noise ratio for glitch detections from the energy deposited in $f$-modes give signal-to-noise ratios $0.4\lesssim S/N\lesssim 3\times 10^5$ \citep{Andersson:2001p1942}. The range in $S/N$ is given for a EURO detector including (lower bound) and ignoring (upper bound) shot noise. Larger $S/N$ (up to $\sim 7.4\times 10^5$) are predicted for instability modes in which the proton and neutron fluids in the stellar core are counter-moving.  \cite{Clark:2010} placed Bayesian $90\%$ confidence upper limits on the peak amplitude from the 2006 Vela glitch of $1.4\times 10^{-20}$, corresponding to a maximum energy released through oscillations of the fundamental quadrupole mode of $1.3\times 10^{35}\,\rm{erg}$.

Even within the vortex unpinning paradigm, changes to the unpinning physics may alter the estimated gravitational wave signal strength significantly.  If, as suggested in the pulsar glitch literature \citep{Melatos:2008p204,Warszawski:2012PRB}, glitches result from a domino effect, then contributions from individual vortices should be summed consecutively and coherently, rather than simultaneously.  This is likely to lead to a greater wave strain, as the motion of one vortex is less likely to cancel geometrically with another.

In conclusion, the gravitational wave signal emitted during the spin-up phase of individual glitches resulting from superfluid vortex avalanches is not detectable by current detectors, but it is close as Eq.~(\ref{eq:omapprox}) indicates.  Once a detection is made, other informative experiments are possible, e.g. searching for all-sky correlations between the $+$ and $\times$ polarisations.  These matters are deferred to future work.
\appendix
\section{Stochastic background}\label{sec:stoch}

Gravitational waves from neutron star glitches contribute to a stochastic gravitational wave background composed of many unresolved astrophysical sources \citep{LIGO:2009}.  The Milky Way is populated by $\sim 10^9$ neutron stars, of which $\approx 2\times 10^3$ have been observed electromagnetically so far \citep{atnf}.  Making some simple assumptions about the glitch rate of the neutron star population as a whole (non-pulsating objects should emit gravitational radiation when they glitch too), we estimate the strength of the glitch background. In this first attempt, we ignore the long-term spin-down evolution of pulsar spins during the lifetime of the Galaxy.  We also restrict attention to sources in the Milky Way, as we know nothing about the glitch properties of extra-Galactic neutron stars.  With the latter restriction, we find that the Milky Way glitch background is weaker than other LIGO backgrounds.  However, the theoretical machinery to calculate the signal, which flows directly from the analysis in Sec.~\ref{sec:GPE}-\ref{sec:multi}, is described here for completeness.  Simple order-of-magnitude estimates suggest that the background may be significant when extra-Galactic contributions are included, a topic for a future paper.

\subsection{Total glitch rate}\label{subsec:totalrate}
We estimate the total rate of glitches, $\Theta$, by assuming that glitches are manifestations of a self-organised branching process, characterised by a power-law size distribution \citep{Harris,Daly:2007p8447}
\begin{equation}
\label{eq:g}
g(s)=(-1/2)(s^{-1/2}_+-s^{-1/2}_-)^{-1}s^{-3/2}~,
\end{equation}
where $s_{\rm{min}}\leq s=\Delta\omega/\omega\leq s_+$ is the range of fractional glitch sizes.  This distribution is supported qualitatively by observational data for several individual pulsars \citep{Melatos:2008p204}, although it can be shown by a Kolmogorov-Smirnov test on aggregate data that $g(s)\propto s^{-3/2}$ is not universal \citep{Melatos:2008p204}.  Across the population, one observes $s_+\approx 10^{-4}$ and $s_{\rm{min}}\approx 10^{-11}$ (the observed minimum glitch size) respectively.  The physical lower bound on glitch size, $s_{\rm{-}}$, is given by the fractional change in angular velocity induced by the outward radial motion of a single vortex by a distance characterised by the mean inter-vortex separation $\Delta r\approx[\kappa/(2\omega)]^{1/2}$.  Since the total number of vortices in a pulsar is $N_{\rm{v}}= 2\pi\omega R_{\rm{s}}^2/\kappa$, we obtain
\begin{eqnarray}
s_{\rm{-}}	&=& \left(\frac{\kappa}{2\omega}\right)^{1/2}\frac{\kappa}{2\pi\omega R_{\rm{s}}^3}\frac{I_{\rm{s}}}{I_{\rm{c}}}\\
			&\sim& 10^{-23}\left(\frac{\omega}{10^2~\rm{rad~s}^{-1}}\right)^{-3/2}~.
\end{eqnarray}

To proceed, we ask the question, what fraction of glitches are within the observable range? The integrated probability between $s_{\rm{min}}$ and $s_+$  is 
\begin{equation}
\label{eq:observedglitches}
\int_{s_{\rm{min}}}^{s_+}dsg(s)=(s_{\rm{min}}^{-1/2}-s_+^{-1/2})/(s^{-1/2}_+-s^{-1/2}_-)\sim 10^{-7}~.
\end{equation}
Hence only one in $10^7$ glitches is observed by radio telescopes at current sensitivities and for typical observational duty cycles.  We now make the simplifying assumption that the glitch rate in each individual pulsar, $\lambda$, falls off exponentially with characteristic age [$\lambda(T_{\rm{c}})\propto \exp(-T_{\rm{c}}/T_0)$, where $T_0$ is some reference characteristic age] \citep{Lyne:2000,Melatos:2008p204}, $T_{\rm{c}}=\omega/(2\dot{\omega})$,  and that neutron stars are being born at a constant rate, leading to a flat distribution of neutron star ages in the galaxy, $p(T_{\rm{c}})=(T_{\rm{max}}-T_{\rm{min}})^{-1}$.  A corollary of this first assumption is that the Crab pulsar ($T_{\rm{c}}\approx 10^{3}\,\rm{yr}$) glitches approximately three times more often than the Vela pulsar ($T_{\rm{c}}\approx 10^{4}\,\rm{yr}$), which agrees roughly with observations \citep{McKenna:1990p1874,Melatos:2008p204}.  The PDF of glitch rates is then
\begin{eqnarray}
 p(\lambda) &=& \frac{A}{\lambda^2}(T_{\rm{max}}-T_{\rm{min}})^{-1}~.
\end{eqnarray}
Normalisation gives $A\approx 10^3~\rm{yr}^{-2}$, where $T_{\rm{min}}=10^{3}~\rm{yr}$ and $T_{\rm{max}}=10^{10}~\rm{yr}$ are the minimum and maximum age of local neutron stars respectively.  The mean glitching rate is then 
\begin{eqnarray}
 \langle \lambda\rangle &=& \int_{\lambda_{\rm{min}}}^{\lambda_{\rm{max}}}d\lambda~\lambda p(\lambda)\\
 &=& \frac{A}{T_{\rm{max}}-T_{\rm{min}}}\ln\left(\frac{T_{\rm{max}}}{T_{\rm{min}}}\right)
\end{eqnarray}
which translates into $\langle\lambda\rangle\sim 10^{-13}\rm{s}^{-1}$ for an individual pulsar.  In other words, the mean is dominated by old objects.  Given a population of $N_{\rm{p}}=10^9$ pulsars in the Milky Way, the total glitching rate is then 
\begin{eqnarray}
\Theta &=& 10^{2}\left(\frac{N_{\rm{p}}}{10^9}\right)\left(\frac{\langle\lambda\rangle}{10^{-13}~\rm{s}^{-1}}\right)~\rm{s}^{-1}
\end{eqnarray}
[the factor of $10^7$ comes from Eq.~(\ref{eq:observedglitches}), which accounts for unobserved small glitches].  In the next section we see that the relative size of $\Theta$ and $\tau$ determines the smoothness of the gravitational wave background.  

For simplicity, and due to a lack of information, we exclude glitches in extragalactic pulsars from the above discussion.  However, we show below that the energy density in gravitational waves is proportional to the aggregate glitch rate in the cosmic volume under consideration, $\Theta$, divided by $\langle D\rangle^2$, the square of the average distance between source and detector.  One typically has $\Theta\propto \langle D\rangle^3$ for isotropically distributed sources, suggesting that the contribution from extragalactic pulsars actually dominates.  We defer this important calculation to future work.  To do it properly, one must account for the redshift-dependent source formation rate and input the statistical properties of extra-galactic neutron stars \citep{Regimbau:2006AA,Howell:2010,Marassi:2010}.

\subsection{Temporal structure}

\begin{figure*}
\begin{center}
\includegraphics[scale=0.4]{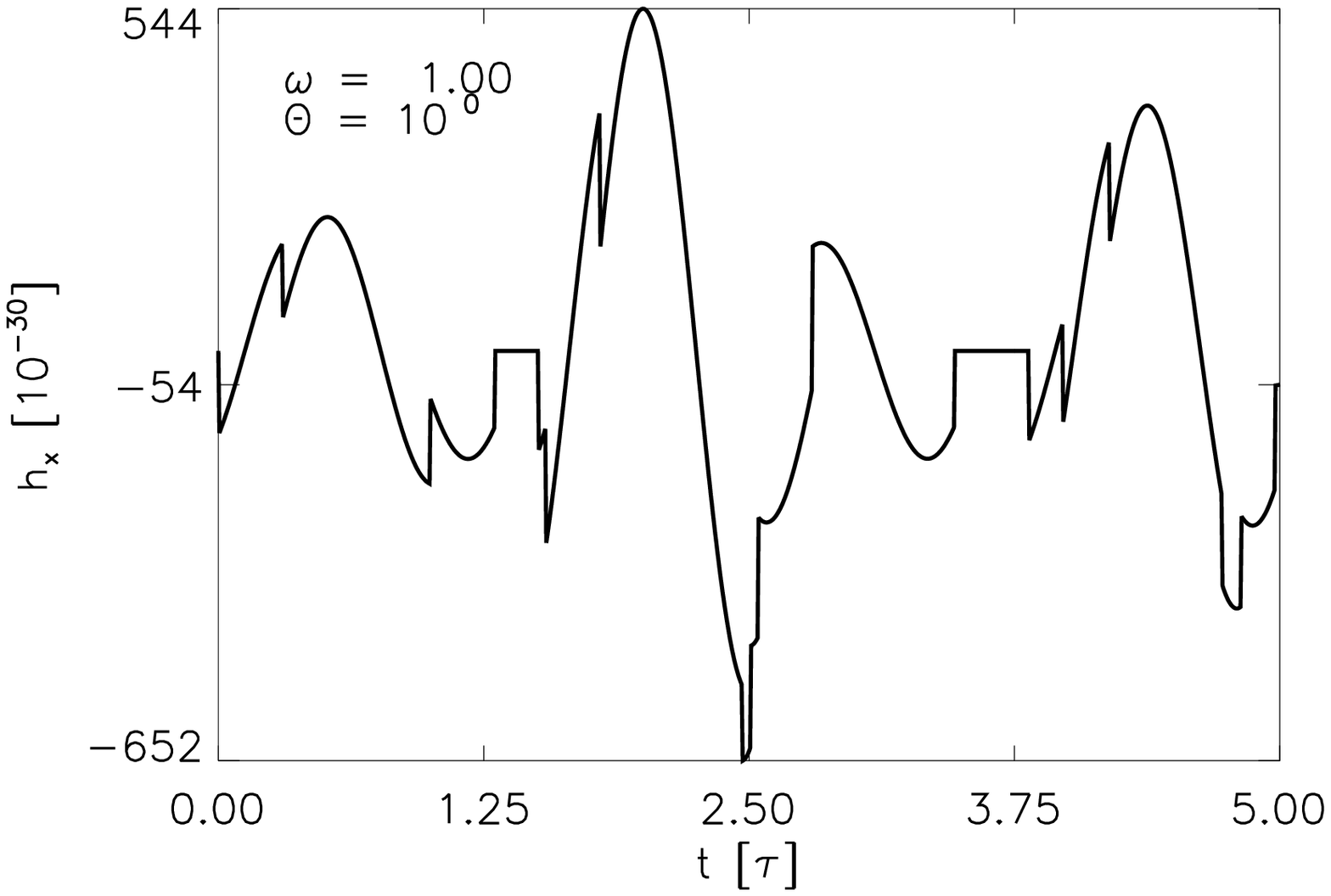}
\includegraphics[scale=0.4]{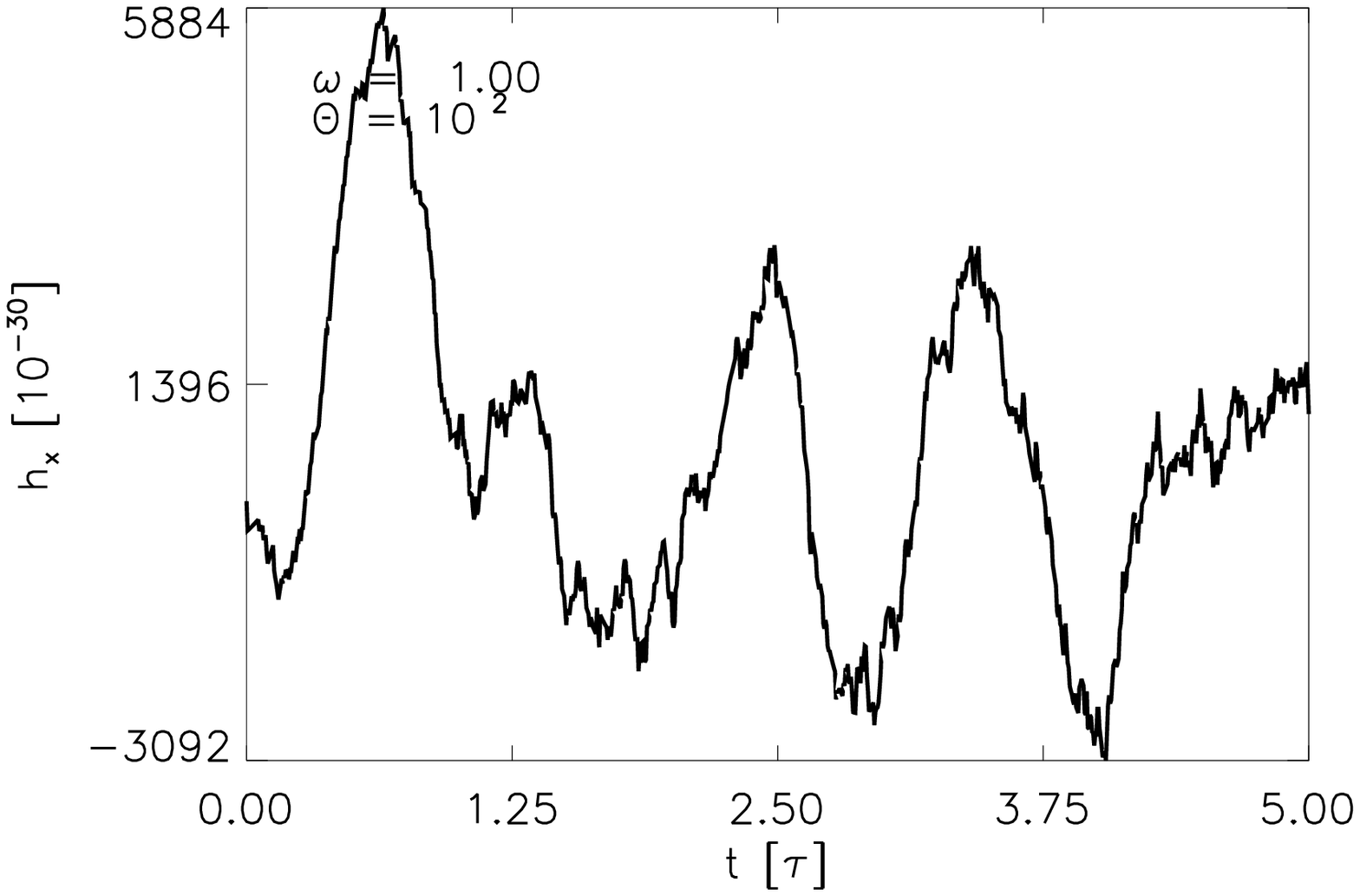}
\end{center}
\caption{\emph{Top}:  Gravitational wave strain in the cross polarisation $h_{\times}(\tilde{t})$ in units of $K_0/\tau^2$ as a function of time from glitches from multiple pulsars.  The time interval is sufficient to bracket five consecutive glitches.  Glitch sizes are drawn from a power-law distribution with minimum and maximum $s_{\rm{min}} = 5\times 10^{-7}$ and $s_{\rm{max}} = 1\times 10^{-6}$ respectively.  The time between glitches is drawn from an exponential distribution with mean rate $\Theta = 1$ (\emph{left}) and $\Theta=10^2$ (\emph{right}).  Parameters:  $\Delta\tilde{r}=10^{-6}$, $\tau = 0.01$, $\tilde{\omega}/(2\pi)=1$.}
\label{fig:ch6:bg_stateg}
\end{figure*}
\begin{figure*}
\begin{center}
\includegraphics[scale=0.4]{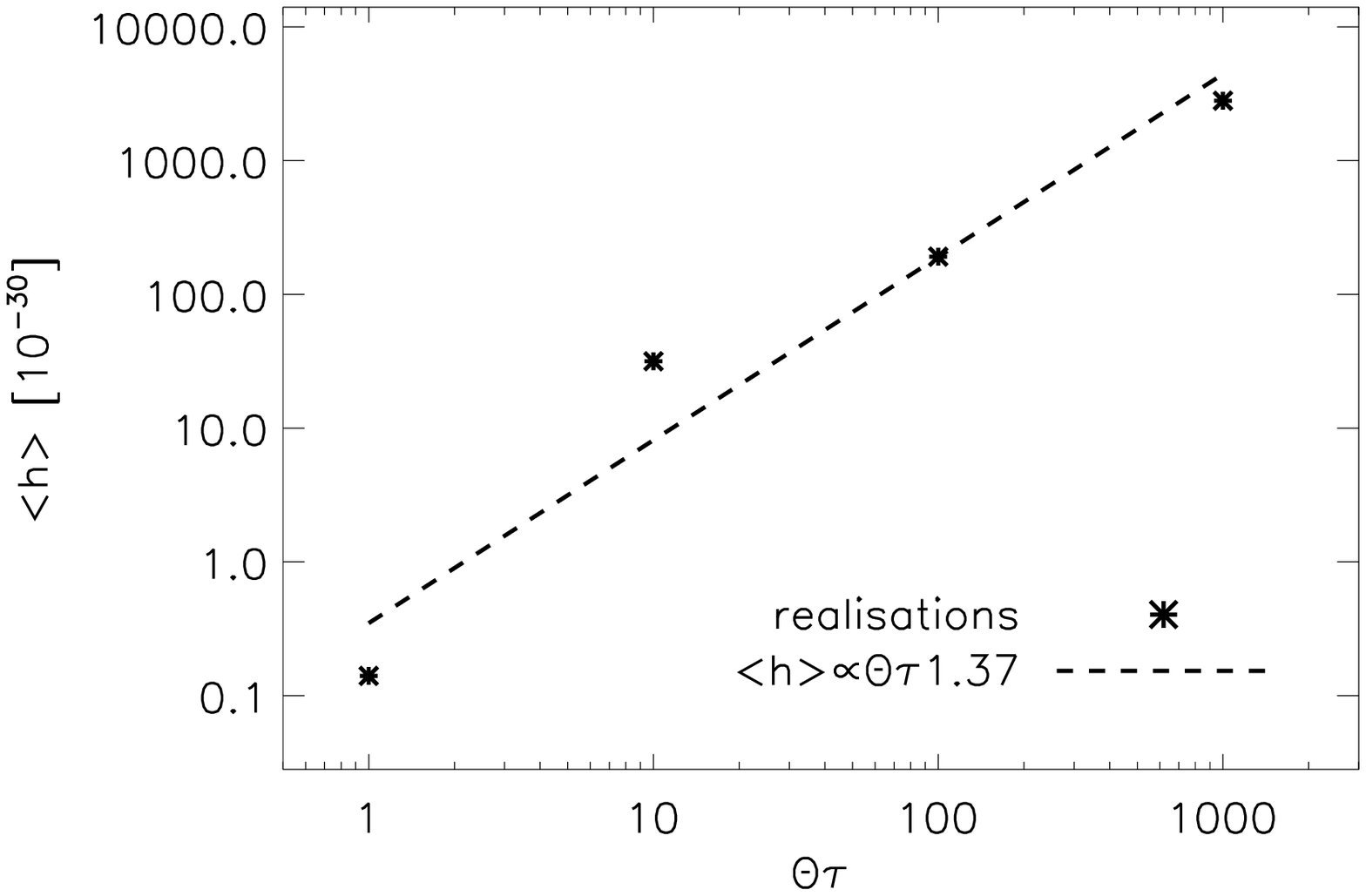}
\includegraphics[scale=0.4]{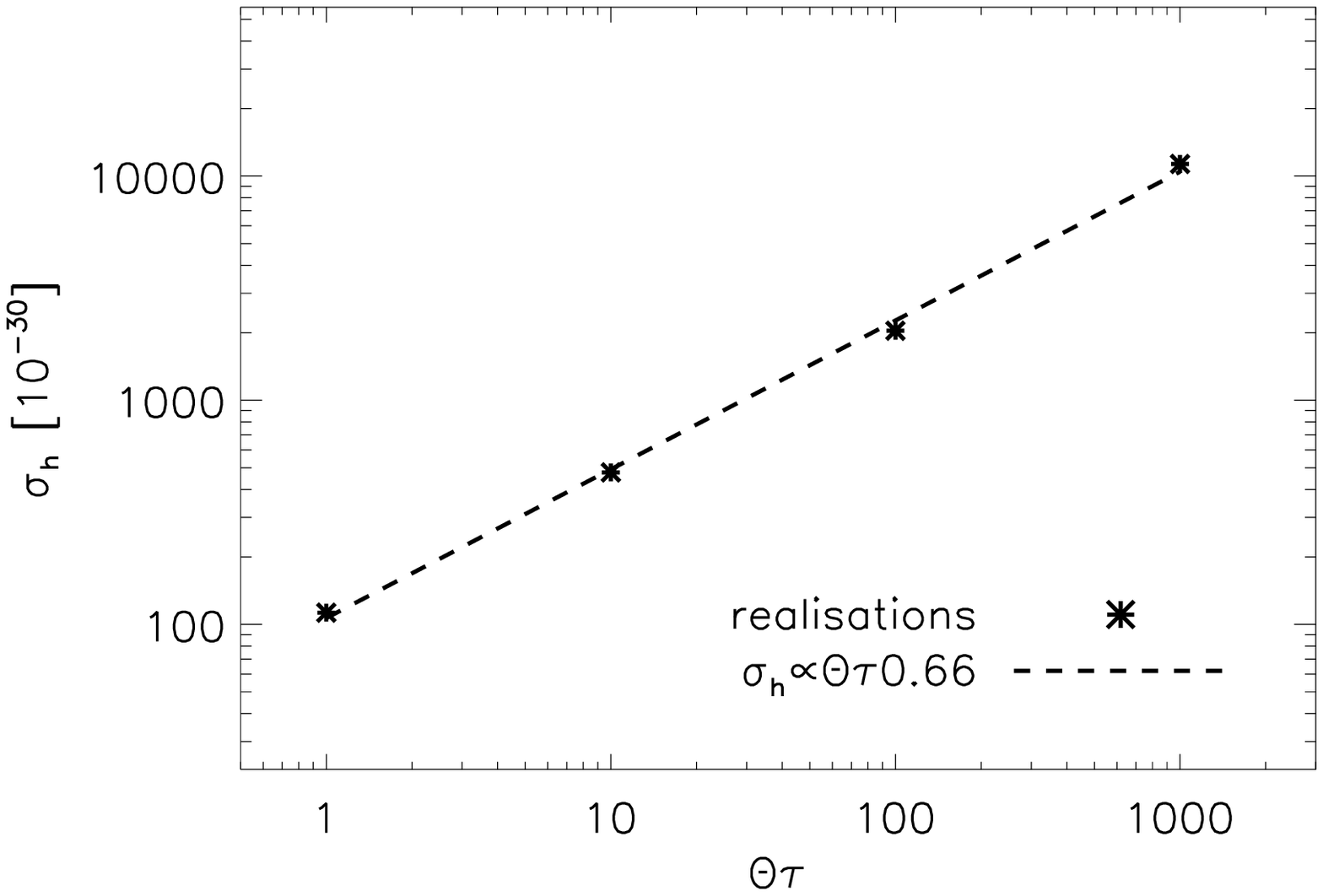}
\end{center}
\caption{Mean (\emph{left}) and standard deviation (\emph{right}) of superposition of glitch signals for different values of $\Theta$.  Parameters:  $\Delta\tilde{r}=10^{-6}$, $\tau = 0.01$, $\tilde{\omega}/(2\pi)=1$.}
\label{fig:ch6:bg_stat}
\end{figure*}
The character of the stochastic gravitational wave background from pulsar glitches depends on the average number of concurrent glitches, given by $\Theta\tau$.  In the regime $\Theta\tau\ll 1$, glitches produce a `shot noise' background of distinguishable burst events.  In Sec.~\ref{subsec:glitchscaling}  we place upper and lower bounds on the strength of such distinguishable events.   In the opposite regime, $\Theta\tau\gg1$, glitches generate a continuous (i.e. never absent) background, whose mean and variance we estimate in App.~\ref{sec:gravitational waveED}.  The intermediate regime $\Theta\tau\approx 1$ is often described as `popcorn' noise \citep{Regimbau:2006AA}.

Let us construct the functional form of the background in a time interval of length $T$, during which $\Theta T$ bursts arrive at Earth.  The wave strain during this interval is
\begin{equation}
h_{jk}^{TT}(t)=h_1(t-t_1)+h_2(t-t_2)+...+h_{\Theta T}(t-t_{\Theta T})~,
\end{equation}
where $t_n$ is the start time of the $n$-th glitch, and $h_n(t-t_n)$ is the gravitational wave signal from the $n$-th glitch.  We make the simplifying assumption that the glitch duration, $\tau$, is the same for all glitches, and that the glitch profiles $h_1(x) =...=h_{\Theta T}(x)$ have the same shape but different amplitudes.  Continuous and shot noise arise from situations with $\langle t_n-t_{n-1}\rangle\ll \tau$ and $\langle t_n-t_{n-1}\rangle\gg\tau$ respectively, where the averages are taken over many inter-glitch intervals.  

For each glitch, we select values for $s$  and $D$ from probability density functions describing the distributions of glitch size and distance from Earth,  $g(s)$ and $\Delta(D)$.  The latter PDF is
\begin{equation}
\label{eq:PDFd}
 \Delta(D) = Q\left(\frac{D+D_1}{D_{\odot}+D_1}\right)^a\exp\left[-b\left(\frac{D-D_{\odot}}{D_{\odot}+D_1}\right)\right]~,
\end{equation}
where $Q$ is a normalisation constant, $a=1.64\pm0.11$, $b=4.01\pm0.24$, $D_1=0.55\pm0.10~\rm{kpc}$, $D_{\odot}$ is the Earth-Sun distance and $D$ is measured from the centre of the galaxy \citep{2006Faucher}.  The inter-glitch time is chosen from an exponential PDF with mean rate $\Theta$ \citep{Melatos:2008p204}.  We use Eq.~(\ref{eq:hmarginmean}) to evaluate $h_n(t-t_n)$.  

Figure~\ref{fig:ch6:bg_stateg} graphs $h_{\times}(\tilde{t})$ for  $\Theta\tau=1$ and $10^2$ (\emph{left} and \emph{right} respectively) and $\tilde{\omega}/(2\pi)=1.0$.  For $\Theta\tau=1$, each glitch is distinguishable.  On average, $\Theta\tau$ glitches are stacked at any given time.  Glitch sizes are drawn from a power law that covers 22 decades.  Feasible Monte Carlo simulations do not sample this distribution comprehensively, so that mean sampled glitch size is many orders of magnitude smaller than $\int ds\,sg(s)$.  However, Monte-Carlo simulations used here to check the analytic theory, not to deliver actual numbers, which is deferred to a future paper.  For testing purposes, a compressed $s$ range is ample.  

In the \emph{left} panel of Fig.~\ref{fig:ch6:bg_stat} we plot the mean wave strain as a function of $\Theta\tau$ for a superposition of glitches whose sizes are in the range $5\times 10^{-7}$ to $1\times 10^{-6}$.  Again, the reduced range ensures that the Monte-Carlo simulation comprehensively samples $g(s)$.  A least-squares power-law fit returns a power-law index of $0.94$.  The mean standard deviation $\langle\sigma_h\rangle$, graphed in the \emph{right} panel of Fig.~\ref{fig:ch6:bg_stat}, is also well described by a power law with index $0.64$, roughly consistent with $\langle\sigma_h\rangle\propto\sqrt{\Delta N_{\rm{v}}}$.

\subsection{Gravitational wave energy density}\label{sec:gravitational waveED}
We now use the analytic expressions derived in Sec.~\ref{sec:multi} to calculate the glitch contribution to the dimensionless, cosmological, energy density parameter $\Omega_{\rm{gw}}$.  We choose this quantity to make contact with other studies of stochastic gravitational-wave backgrounds, including the recent LIGO bound on theories of cosmic strings \citep{LIGO:2009}.  The calculation is relevant to the regime $\Theta \tau\gg1$ \citep{Howell:2010}, in which the background is continuous in time.  Following \cite{Ferrari:1999} and \cite{Maggiore:2000}, we write $\Omega_{\rm{gw}}$ as a function of wave frequency, $\nu$, by relating it to the spectral energy density, $d^2E_{\rm{gw}}/(d\nu dS)$, through the expression
\begin{equation}
\label{eq:omgw}
 \Omega_{\rm{gw}}(\nu) = \frac{\nu\Theta}{c^3\rho_{\rm{cr}}}\frac{d^2E_{\rm{gw}}}{d\nu dS}~,
\end{equation}
where $\rho_{\rm{cr}}=3H_0^2/(8\pi G)$ is the critical density to close the universe, $H_0$ is Hubble's constant, $\Theta$ is the total event rate from Sec.~\ref{subsec:totalrate}, $dS$ is the area element on the sky, and $E_{\rm{gw}}(\nu,\mathbf{n})$ is the total energy radiated in the direction $\mathbf{n}$ at frequency $\nu$.  

We begin the task of calculating $d^2E_{\rm{gw}}/(d\nu dS)$ by writing the energy flux from a source as \citep{Kokkotas:2001}
\begin{equation}
 \frac{d^2E}{dtdS} = \frac{c^3}{16\pi^2 G} \left|\frac{\partial h^{\rm{TT}}_{jk}(t)}{\partial t}\frac{\partial h^{\rm{TT}}_{jk}(t)}{\partial t}\right|~.
\end{equation}
The total radiated energy per unit area per glitch is then
\begin{eqnarray}
\label{eq:ch6:dEdW}
 \frac{dE_{\rm{gw}}}{dS}&=&\int_0^{\tau} dt \frac{d^2E_{\rm{gw}}}{dtdS}\nonumber\\
   &=& \frac{c^3}{16\pi^2 G}\int_{0}^{\infty}d\nu\left|\mathcal{F}\left[\frac{\partial h_{jk}^{TT}}{\partial t}\right]\mathcal{F}\left[\frac{\partial h_{jk}^{TT}}{\partial t}\right]\right|~,
\end{eqnarray}
where the second equality is due to Parseval's theorem and $\mathcal{F}[\cdot]$ denotes the Fourier transform.  From Eq.~(\ref{eq:ch6:dEdW}), we identify
\begin{eqnarray}
\label{eq:dEdomdS}
 \frac{d^2E_{\rm{gw}}}{d\nu dS}&=&\frac{c^3}{16\pi^2 G}\left|\mathcal{F}\left[\frac{\partial h_{jk}^{TT}}{\partial t}\right]\mathcal{F}\left[\frac{\partial h_{jk}^{TT}}{\partial t}\right]\right|
\end{eqnarray}
as the time-averaged energy flux per unit frequency. The beam pattern, $T^{B2,21}_{jk}$, which depends on the observer's position through the angles $\phi$ and $\iota$ \citep{Jaranowski:1998}, is ensconced in $h_{jk}^{TT}$.  Averaging over line-of-sight orientation by integrating the right-hand side of Eq.~(\ref{eq:dEdomdS}) with respect to $\cos\phi$ and $\iota$ \citep{Jaranowski:1998}, we can write 
\begin{equation}
\label{eq:omgwfull}
 \Omega_{\rm{gw}}(\tilde{\nu}) = \frac{f_{TB2}\Theta\tilde{\nu}K_0^2}{16\pi^2G\rho_{\rm{cr}}\tau^5}\left|\mathcal{F}\left[\frac{\partial \tilde{h}^{\rm{TT}}_{jk}}{\partial t}\right]\mathcal{F}\left[\frac{\partial \tilde{h}^{\rm{TT}}_{jk}}{\partial t}\right]\right|~,
\end{equation}
with $\tilde{\nu}=\nu\tau$, $f_{TB2} = \int_{0}^{2\pi}d\phi|T^{jk,B2}T^{jk,B2}|$ and $\tilde{h}^{\rm{TT}}_{jk}=(\tau^2/K_0)h^{\rm{TT}}_{jk}$.

We now calculate $|\mathcal{F}[\partial h^{\rm{TT}}_{jk}/\partial t]\mathcal{F}[\partial h^{\rm{TT}}_{jk}/\partial t]|$ for glitches arriving from an ensemble of neutron stars.  This is a stochastic quantity, because $h_{jk}^{TT}$ depends on the random variables $R_0$ and $\phi_0$.   The mean and variance of $\Omega_{\rm{gw}}$ are found in two stages.  First, we calculate $\mathcal{F}[\partial h^{\rm{TT}}_{jk}/\partial t]$ for an individual glitch by applying the central limit theorem to add up the contributions from the $\Delta N_{\rm{v}}$ vortices involved.  Second, we use properties of the $\chi^2$ distribution to add up the signals from multiple glitches in multiple pulsars.

According to the central limit theorem, the mean and variance of $|\mathcal{F}[\partial \tilde{h}^{\rm{TT}}_{jk}/\partial t]\mathcal{F}[\partial \tilde{h}^{\rm{TT}}_{jk}/\partial t]|$, for a glitch involving $\Delta N_{\rm{v}}=sN_{\rm{v}}R_{\rm{s}}/\Delta r$ vortices, are $\Delta N_{\rm{v}}\mu_1$ and $\Delta N_{\rm{v}}\sigma_{1}^2$ respectively [$\mu_1$ and $\sigma_1^2$ are the Fourier transforms of Eq.~(\ref{eq:hmarginmean}) and Eq.~(\ref{eq:hmarginvar})].  The single-vortex moments $\mu_1$ and $\sigma_1$ are calculated by marginalising over the PDFs for initial vortex position in Eq.~(\ref{eq:PDFR0}) and Eq.~(\ref{eq:PDFphi0}), the glitch size distribution $g(s)$, and the source distance distribution $\Delta (D)$.  

The total $|\mathcal{F}[\partial \tilde{h}^{\rm{TT}}_{jk}/\partial t]\mathcal{F}[\partial \tilde{h}^{\rm{TT}}_{jk}/\partial t]|$ is the sum of contributions from $\Theta\tau$ simultaneous glitches.  Its statistics obey the non-central $\chi^2$ distribution, with non-centrality parameter $\lambda_{\chi^2} =\Theta\tau \Delta N_{\rm{v}}\mu_1^2/\sigma_{1}^2$, and mean and variance given by
\begin{eqnarray}
\label{eq:momentsom1}
 \mu_{\chi} &=& \Theta\tau\sigma_1^2\Delta N_{\rm{v}}\left(1+\frac{\Delta N_{\rm{v}}\mu_1^2}{\sigma_1^2}\right)~,\\
\label{eq:momentsom2}
 \sigma_{\chi}^2&=&
  2\Theta\tau\sigma_1^4\Delta N_{\rm{v}}^2\left(1+\frac{2\Delta N_{\rm{v}}\mu_1^2}{\sigma_1^2}\right)~.
\end{eqnarray}

\begin{figure*}
\begin{center}
\includegraphics[scale=0.375]{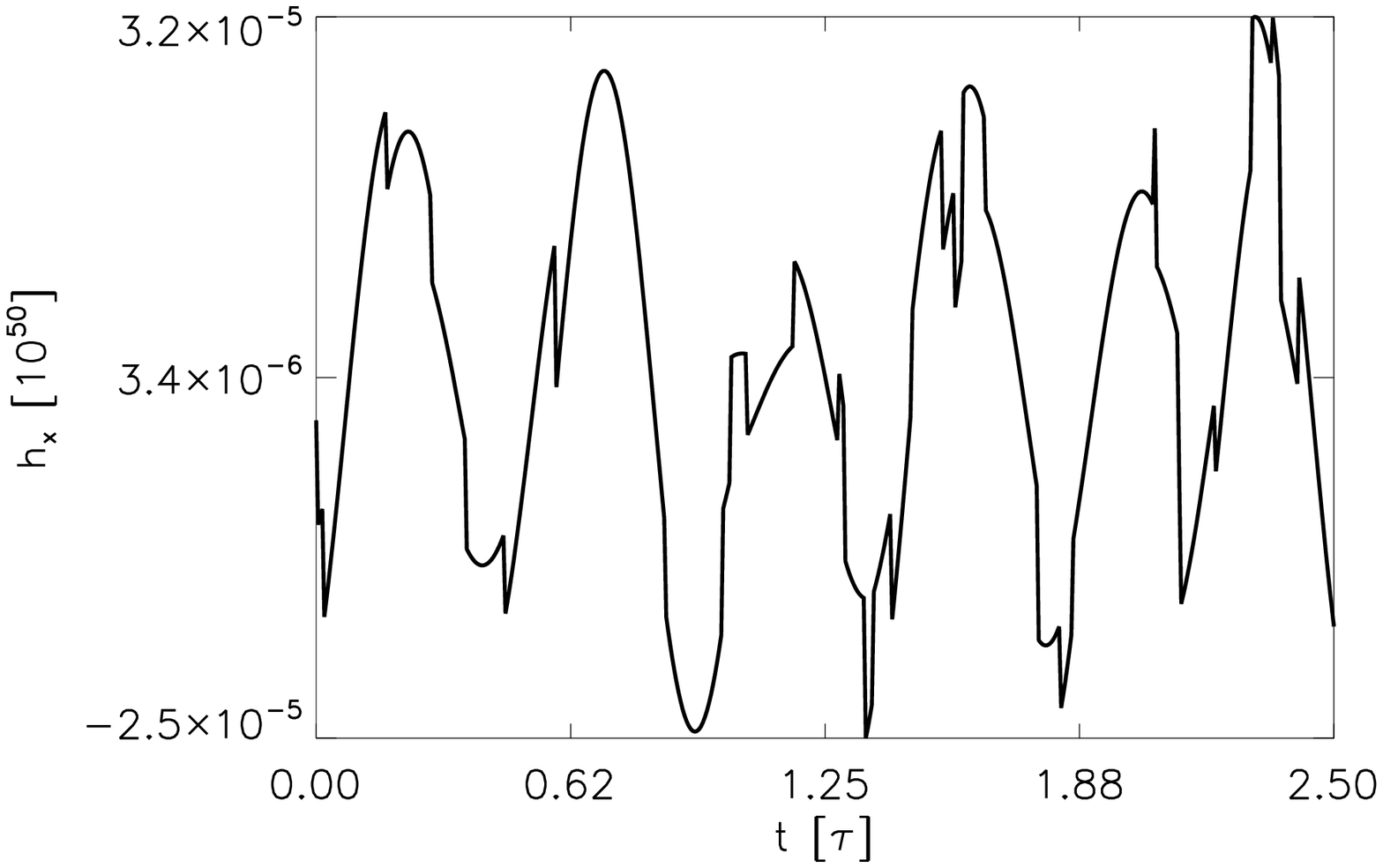}
\includegraphics[scale=0.375]{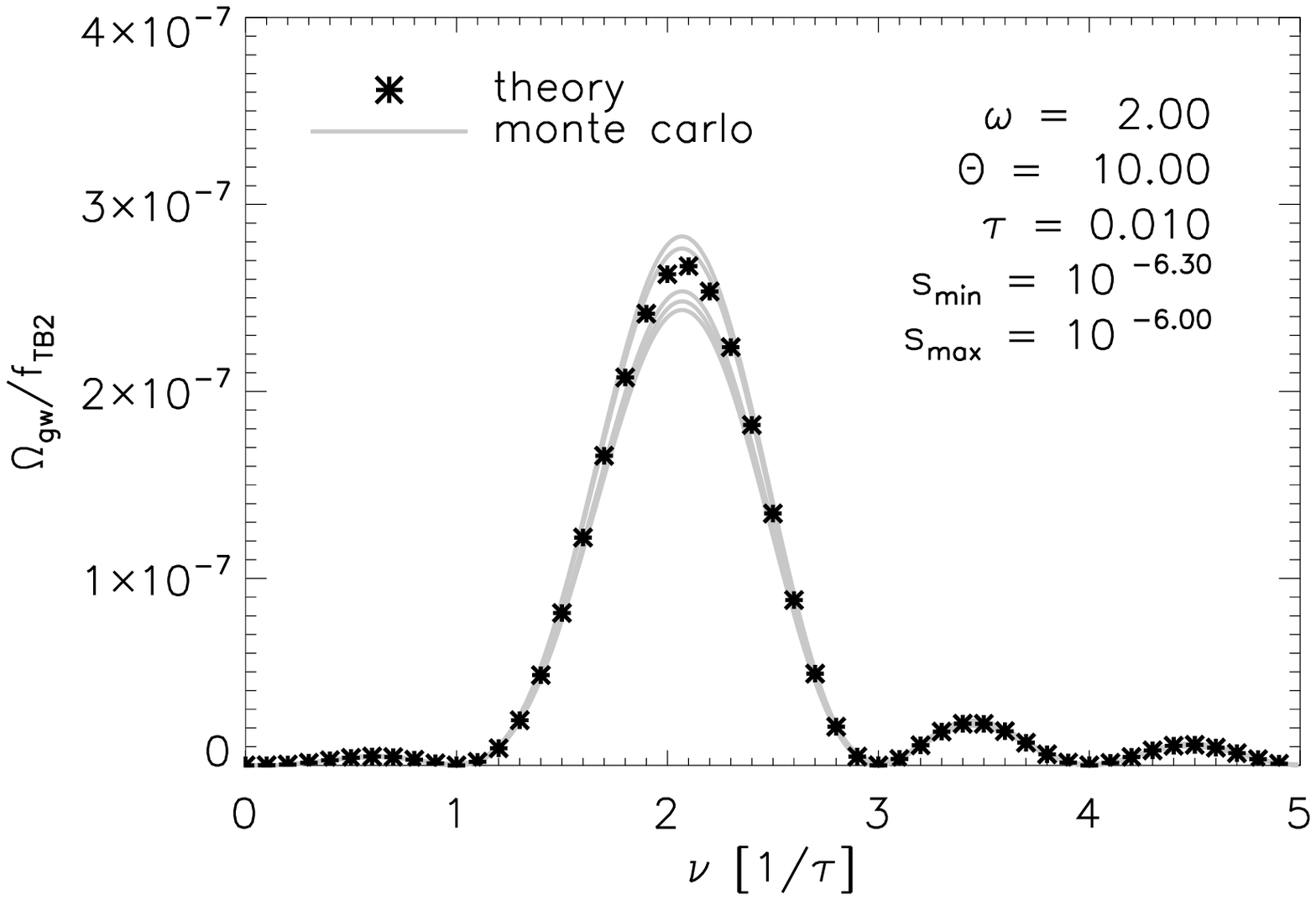}
\includegraphics[scale=0.375]{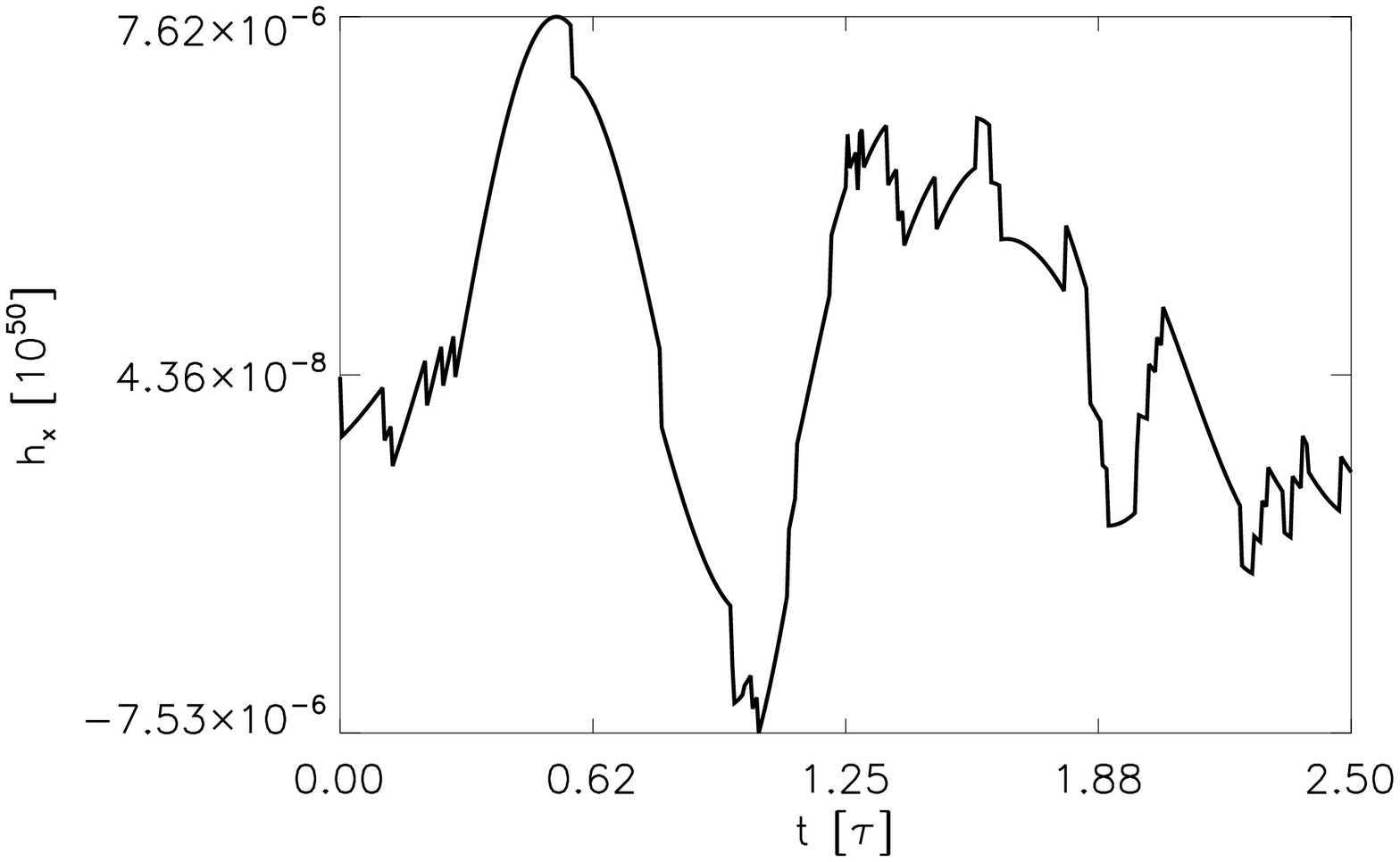}
\includegraphics[scale=0.375]{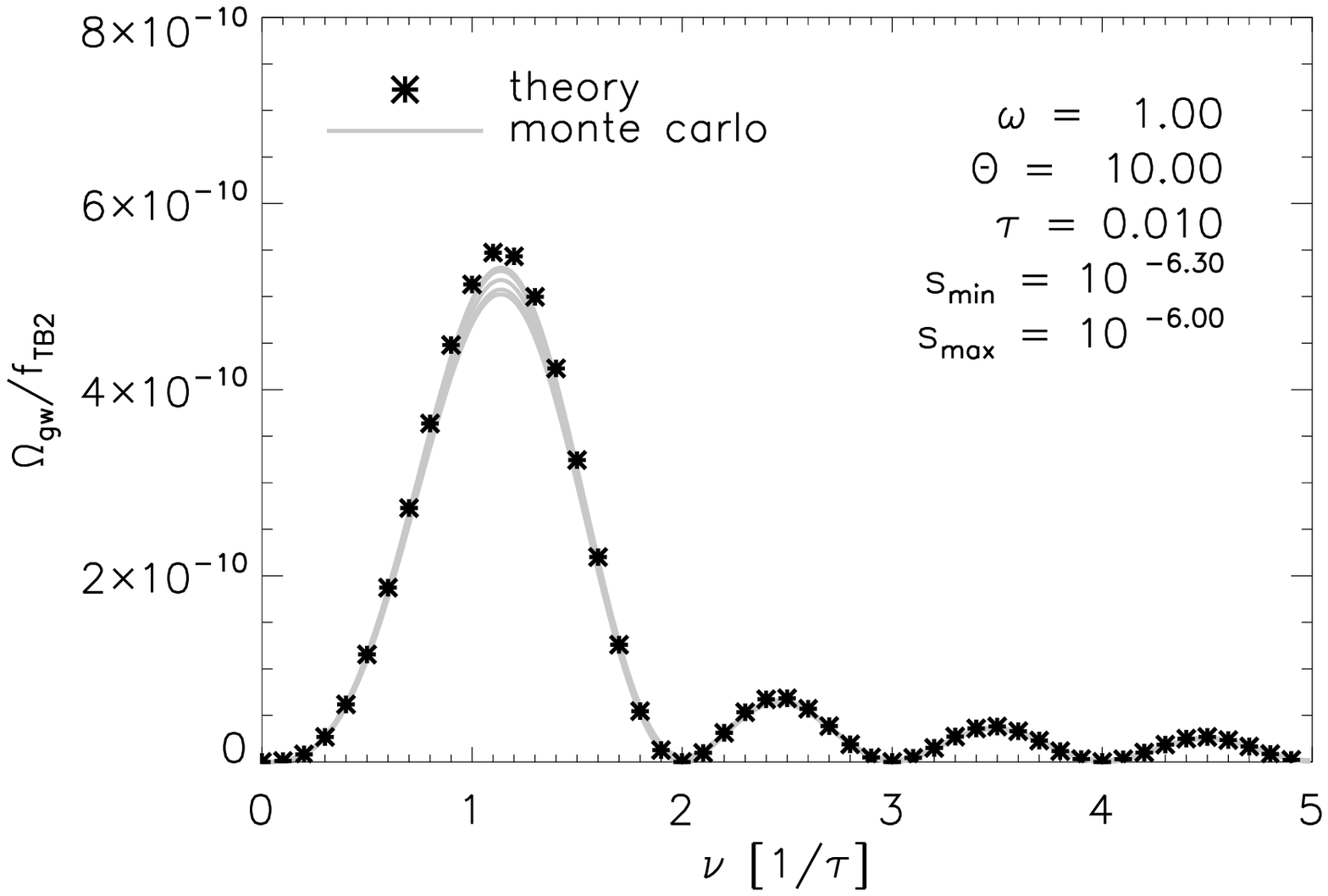}
\includegraphics[scale=0.375]{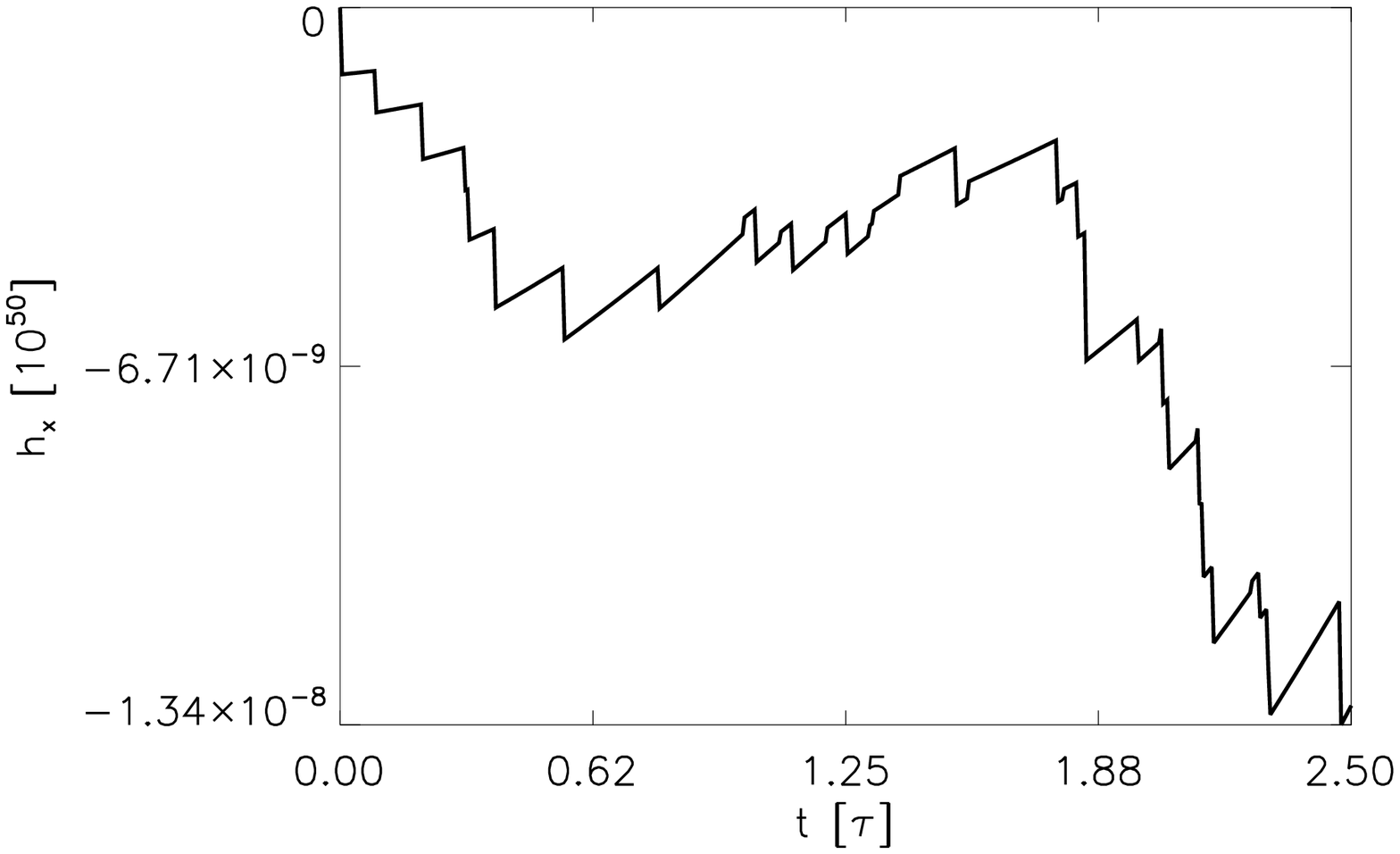}
\includegraphics[scale=0.375]{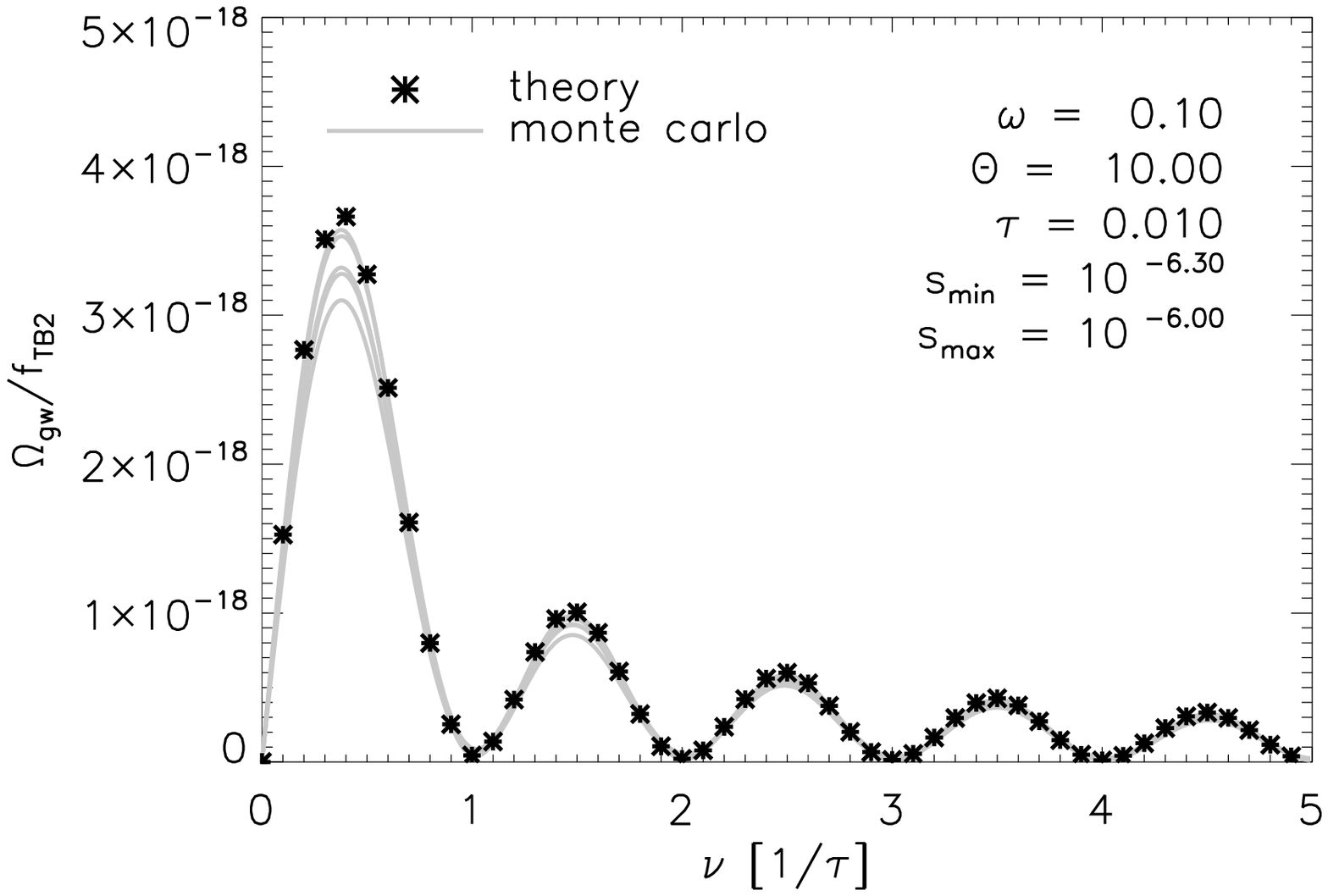}
\includegraphics[scale=0.375]{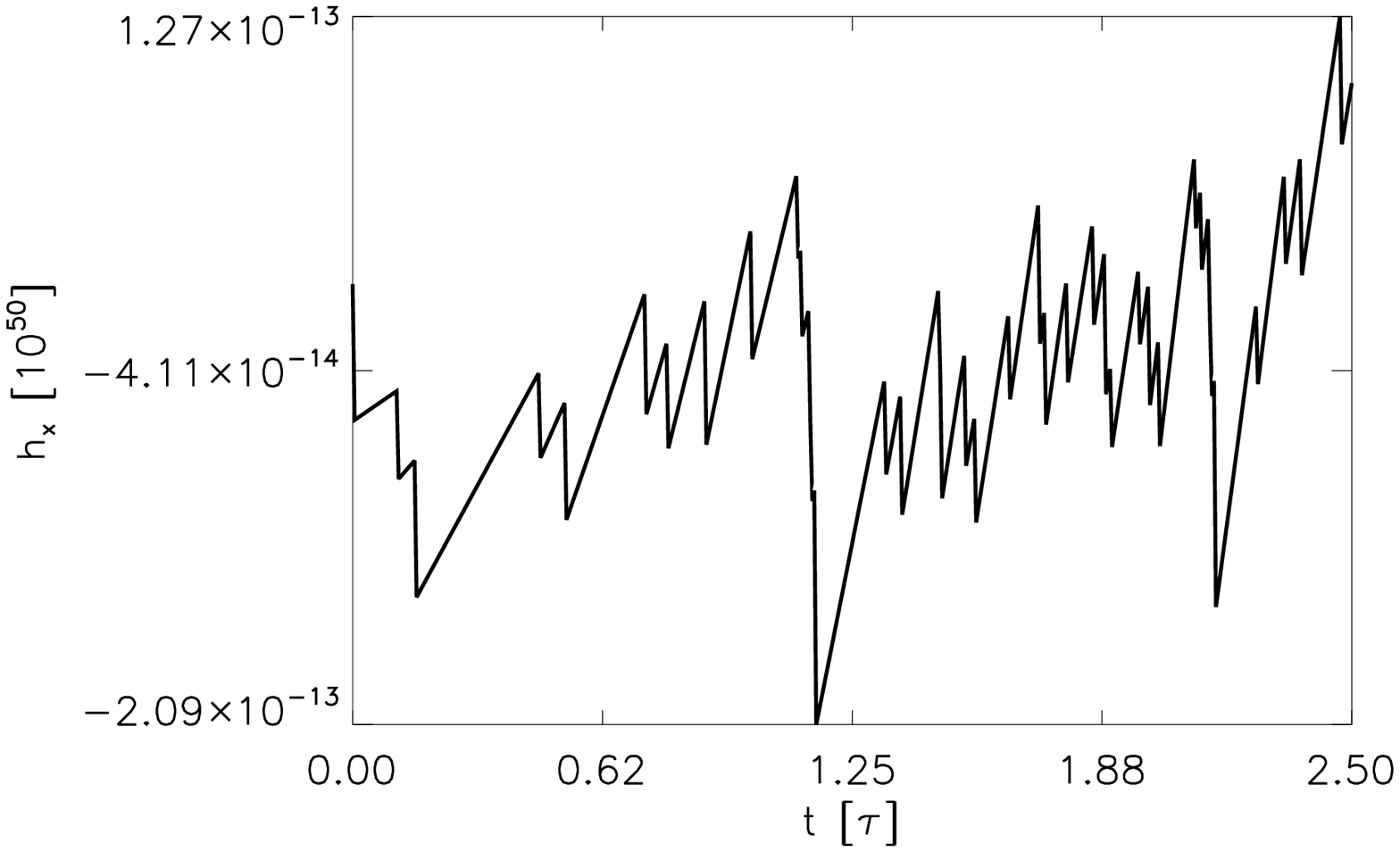}
\includegraphics[scale=0.375]{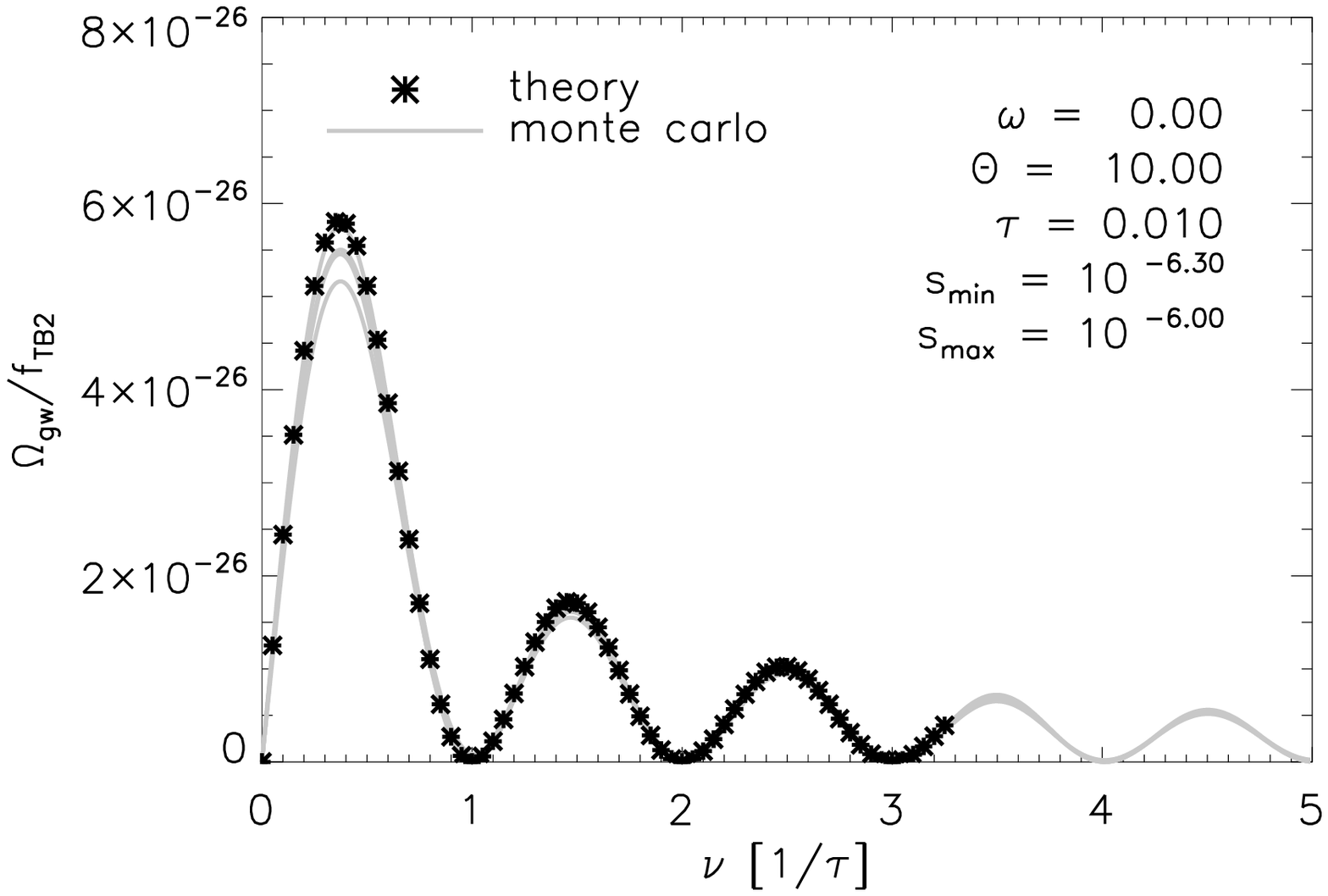}
\end{center}
\caption{Superposition of gravitational wave signals in the cross polarisation $h_{\times}(t)$ (in units of $K_0/\tau^2$; \emph{left column}) and gravitational wave energy density parameter, $\Omega_{\rm{gw}}(\nu)$ (\emph{right column}), from a population of glitching pulsars with mean aggregate glitch rate $\Theta = 10$, for $\tilde{\omega}/(2\pi)=2.0$, 1.0, 0.1 and 0 (\emph{top} to \emph{bottom} respectively).  The \emph{grey} curves in the \emph{right} column are different realisations of the underlying glitch size and distance distributions.  The \emph{asterisks} correspond to the theoretical calculation described in Sec.~\ref{sec:stoch}.  Minimum and maximum glitch sizes are $5\times 10^{-7}$ and $10^{-6}$ respectively.  Unless otherwise stated, the pulsar parameters are given in Table.~\ref{tab:glitch}.}
\label{fig:ch6:testomgw} 
\end{figure*}

We note that, strictly, Eq.~(\ref{eq:momentsom1}) and (\ref{eq:momentsom2}) apply only when $\Delta N_{\rm{v}}$ is constant.  The correct expressions contain sums over different glitch sizes.  However, comparison between Monte-Carlo simulations and Eq.~(\ref{eq:momentsom1}) and (\ref{eq:momentsom2}) confirm that the approximation is reasonable.  For simplicity, we also assume $\omega$ to be the same for all pulsars; the observed population of glitching pulsars is dominated by young pulsars with $10\lesssim \omega/(2\pi)\lesssim 100~\rm{Hz}$.  

In the \emph{left} panel of Fig.~\ref{fig:ch6:testomgw} we graph $h_{\times}(t)$ from Monte-Carlo simulations for $\tilde{\omega}/(2\pi)>1$ (\emph{top}), $\tilde{\omega}/(2\pi)=1$ (\emph{top centre}) and $\tilde{\omega}/(2\pi)<1$ (\emph{bottom centre}), representing vortex motions that are shorter, the same, and longer than one rotation period respectively.  In the \emph{right} panel, we graph the corresponding $\Omega_{\rm{gw}}(\tilde{\nu})$ (\emph{grey} curves), overplotted with analytic predictions of the mean (\emph{asterisks}) and standard deviation (\emph{error bars}).  The error bars are too small to be seen.  The \emph{bottom} panels display results for $\tilde{\omega}/(2\pi)=0$, where radial vortex motion is the only contributor to the gravitational wave signal.

A fair comparison between simulation and theory is achieved by restricting the glitch size to range between $5\times10^{-7}$ and $1\times 10^{-6}$, to ensure that the Monte-Carlo algorithm comprehensively samples $g(s)$.  In all four rows in Fig.~\ref{fig:ch6:testomgw}, the analytic mean is an excellent match to the Monte-Carlo results, but the analytic standard deviation underestimates the simulated value by a factor of $\sim 100$.  For $\tilde{\omega}/(2\pi)>1$, $\Omega_{\rm{gw}}(\tilde{\nu})$ peaks at $\tilde{\nu}=\tilde{\omega}/(2\pi)$ and is oscillatory, with nodes at integer multiples of $\tilde{\nu}$, except at $\tilde{\nu}_{\rm{max}}=\tilde{\omega}$.  For $\tilde{\omega}/(2\pi)\lesssim1$, $\Omega_{\rm{gw}}(\tilde{\nu})$ peaks at $\tilde{\nu}_{\rm{max}}\approx  0.375$.  The latter result holds for all $\tilde{\omega}/(2\pi)\ll1$, because the canonical parameters listed in Table~\ref{tab:glitch} describe a glitch in which vortex motion is dominated by azimuthal rather than radial motion.  The peak at $\tilde{\nu}_{\rm{max}}=\tilde{\omega}/(2\pi)$ is roughly five times higher than the next highest peak.  

We estimate $\Omega_{\rm{gw}}$ as
\begin{eqnarray}
&\Omega_{\rm{gw}} &=\frac{K_0^2}{48\pi^2 G\rho_{\rm{cr}}\tau^4}\left(\frac{2\pi\omega R_{\rm{S}}^2}{\kappa}\right)^{3/2}\tilde{\nu}\mu_1^2\\
&= & \left(\frac{D}{10^{20}\,\rm{m}^{-1}}\right)^{-2}
		\left(\frac{\Theta}{10^{2}\,\rm{s}^{-1}}\right)^{2}\nonumber\\
&\times & \left\{\begin{array}{ll}
			10^{-17}
			\left(\frac{\tau}{10^{-2}\,\rm{s}}\right)^{-3}
			\left(\frac{\omega}{10^{2}\,\rm{rad\,s}^{-1}}\right)^{5/2}
			\left(\frac{\Delta r}{10^{-2}\,\rm{m}}\right)^{2}
			&\mbox{,$\tilde{\omega}\gg\Delta\tilde{r}$}\\	
			10^{-23}
			\left(\frac{\tau}{10^{-2}\,\rm{s}}\right)
			\left(\frac{\omega}{10^{2}\,\rm{rad\,s}^{-1}}\right)^{13/2}
			&\mbox{,$\tilde{\omega}\ll\Delta\tilde{r}$}\\
				\end{array} \right.~,	
   \label{eq:om_est}
\end{eqnarray}
where we use $s_+ =\Delta\tilde{r}(I_{\rm{s}}/I_{\rm{c}})$ and $s_- =\Delta\tilde{r}/N_{\rm{v}}$.
These results should be compared to estimates of the background from sources of cosmological origin \citep{Maggiore:2000}, such as cosmic strings [$\Omega_{\rm{gw}}\sim 10^{-10}$ in the frequency range 1.5-2.5\,kHz \citep{Ferrari:1999}], inflation [$\Omega_{\rm{gw}}\sim 10^{-15}$ \citep{Turner:1997}] and core-collapse supernovae [$10^{-14}\lesssim\Omega_{\rm{gw}}\lesssim 10^{-12}$ at 100\,Hz \citep{Buonanno:2005}].  

Although the background looks relatively weak at first blush, it could be much larger than the canonical numbers in Eq.~(\ref{eq:om_est}) if (1) $\tau$ is smaller than $10^{-3}\,\rm{s}$, which is eminently possible, and (2) extragalactic pulsars are included, since one has $\Theta\propto\langle D\rangle^3$ as discussed previously.  These possibilities will be explored in future work.  

\subsection{Signal to noise}\label{subsec:sn}

The optimised signal-to-noise ratio ($S/N$) for an integration time $T$ is given by \citep{Allen:1999} 
\begin{eqnarray}
 \frac{S}{N} &=& \left[\frac{9H_0^4T}{50\pi^4}\int_0^{\infty}d\nu\frac{\gamma^2(\nu)\Omega_{\rm{gw}}^2(\nu)}{\nu^6P_1(\nu)P_2(\nu)}\right]^{1/2} ~,
\end{eqnarray}
where $P_1(\nu)$ and $P_2(\nu)$ are the power spectral noise densities for the two detectors, and $\gamma$ is the normalised overlap reduction function, which characterises the loss of sensitivity due to the separation and relative orientation of the detectors.  The minimum observable $\Omega_{\rm{gw}}$ of two detectors with the sensitivity of the Einstein Telescope (with a signal-to-noise ratio of 2.56) is $\Omega_{\rm{min}}=1.13\times 10^{-11}$ for an integration time of one year, a false alarm rate of $\alpha=10\%$, and a detection rate of $90\%$ \citep{Marassi:2010}.  For $\omega/(2\pi) = 200~\rm{Hz}$, and assuming a Milky Way distance distribution given by Eq.~(\ref{eq:PDFd}), with an aggregate glitch rate of $\Theta =10^2~\rm{s}^{-1}$, and glitches lasting $\tau=0.01~\rm{s}$, we obtain $S/N = 10^{-4}$ for Milky Way glitches (with $\gamma=1$ for simplicity). Very crudely, assuming there are $\sim 10^{10}$ Milky-Way type galaxies, evenly distributed throughout the universe, $S/N\propto \Omega_{\rm{gw}}\propto\langle D\rangle$ becomes $\sim 1$.

\bibliographystyle{mn2e}
\bibliography{GW_ms}

\end{document}